\documentclass[twocolumn]{aa}  
\usepackage{graphicx}
\usepackage{amsmath,amsfonts,amssymb}
\usepackage{natbib}
\usepackage{tabularx}
\usepackage{collcell}
\usepackage{array}
\usepackage{booktabs}
\usepackage{subfigure}
\usepackage{txfonts}
\usepackage{xcolor}
\usepackage{blindtext}
\usepackage{float}
\usepackage{dblfloatfix}
\usepackage{afterpage}
\usepackage{ifthen}
\usepackage[morefloats=12]{morefloats}
\usepackage{tabularx}
\usepackage{placeins}
\usepackage{multicol}
\bibpunct{(}{)}{;}{a}{}{,}
\usepackage[switch]{lineno}
\definecolor{linkcolor}{rgb}{0.6,0,0}
\definecolor{citecolor}{rgb}{0,0,0.75}
\definecolor{urlcolor}{rgb}{0.12,0.46,0.7}
\usepackage[breaklinks, colorlinks, urlcolor=urlcolor,
linkcolor=linkcolor,citecolor=citecolor,pdfencoding=auto]{hyperref}
\hypersetup{linktocpage}
\usepackage{bold-extra}
\usepackage{ifthen}

\def\setsymbol#1#2{\expandafter\def\csname #1\endcsname{#2}}
\def\getsymbol#1{\csname #1\endcsname}

\def\Planck{\textit{Planck}}

\newbox\tablebox    \newdimen\tablewidth
\def\leaderfil{\leaders\hbox to 5pt{\hss.\hss}\hfil}
\def\endPlancktable{\tablewidth=\columnwidth 
    $$\hss\copy\tablebox\hss$$
    \vskip-\lastskip\vskip -2pt}

\def\tablenote#1 #2\par{\begingroup \parindent=0.8em
    \abovedisplayshortskip=0pt\belowdisplayshortskip=0pt
    \noindent
    $$\hss\vbox{\hsize\tablewidth \hangindent=\parindent \hangafter=1 \noindent
    \hbox to \parindent{$^#1$\hss}\strut#2\strut\par}\hss$$
    \endgroup}
\def\doubleline{\vskip 3pt\hrule \vskip 1.5pt \hrule \vskip 5pt}

\def\L2{\ifmmode L_2\else $L_2$\fi}

\def\DeltaT{\ifmmode \Delta T\else $\Delta T$\fi}
\def\deltat{\ifmmode \Delta t\else $\Delta t$\fi}
\def\fknee{\ifmmode f_{\rm knee}\else $f_{\rm knee}$\fi}
\def\Fmax{\ifmmode F_{\rm max}\else $F_{\rm max}$\fi}
\def\solar{\ifmmode{\rm M}_{\mathord\odot}\else${\rm M}_{\mathord\odot}$\fi}
\def\Msolar{\ifmmode{\rm M}_{\mathord\odot}\else${\rm M}_{\mathord\odot}$\fi}
\def\Lsolar{\ifmmode{\rm L}_{\mathord\odot}\else${\rm L}_{\mathord\odot}$\fi}
\def\inv{\ifmmode^{-1}\else$^{-1}$\fi}
\def\mo{\ifmmode^{-1}\else$^{-1}$\fi}
\def\sup#1{\ifmmode ^{\rm #1}\else $^{\rm #1}$\fi}
\def\expo#1{\ifmmode \times 10^{#1}\else $\times 10^{#1}$\fi}
\def\,{\thinspace}
\def\lsim{\mathrel{\raise .4ex\hbox{\rlap{$<$}\lower 1.2ex\hbox{$\sim$}}}}
\def\gsim{\mathrel{\raise .4ex\hbox{\rlap{$>$}\lower 1.2ex\hbox{$\sim$}}}}

\def\simprop{\mathrel{\raise .4ex\hbox{\rlap{$\propto$}\lower 1.2ex\hbox{$\sim$}}}}
\def\deg{\ifmmode^\circ\else$^\circ$\fi}
\def\pdeg{\ifmmode $\setbox0=\hbox{$^{\circ}$}\rlap{\hskip.11\wd0 .}$^{\circ}
          \else \setbox0=\hbox{$^{\circ}$}\rlap{\hskip.11\wd0 .}$^{\circ}$\fi}
\def\arcs{\ifmmode {^{\scriptstyle\prime\prime}}
          \else $^{\scriptstyle\prime\prime}$\fi}
\def\arcm{\ifmmode {^{\scriptstyle\prime}}
          \else $^{\scriptstyle\prime}$\fi}
\newdimen\sa  \newdimen\sb
\def\parcs{\sa=.07em \sb=.03em
     \ifmmode \hbox{\rlap{.}}^{\scriptstyle\prime\kern -\sb\prime}\hbox{\kern -\sa}
     \else \rlap{.}$^{\scriptstyle\prime\kern -\sb\prime}$\kern -\sa\fi}
\def\parcm{\sa=.08em \sb=.03em
     \ifmmode \hbox{\rlap{.}\kern\sa}^{\scriptstyle\prime}\hbox{\kern-\sb}
     \else \rlap{.}\kern\sa$^{\scriptstyle\prime}$\kern-\sb\fi}
\def\ra[#1 #2 #3.#4]{#1\sup{h}#2\sup{m}#3\sup{s}\llap.#4}
\def\dec[#1 #2 #3.#4]{#1\deg#2\arcm#3\arcs\llap.#4}
\def\deco[#1 #2 #3]{#1\deg#2\arcm#3\arcs}
\def\rra[#1 #2]{#1\sup{h}#2\sup{m}}

\def\dots{\relax\ifmmode \ldots\else $\ldots$\fi}
\def\WHzsr{\ifmmode $W\,Hz\mo\,sr\mo$\else W\,Hz\mo\,sr\mo\fi}
\def\mHz{\ifmmode $\,mHz$\else \,mHz\fi}
\def\GHz{\ifmmode $\,GHz$\else \,GHz\fi}
\def\mKs{\ifmmode $\,mK\,s$^{1/2}\else \,mK\,s$^{1/2}$\fi}
\def\muKs{\ifmmode \,\mu$K\,s$^{1/2}\else \,$\mu$K\,s$^{1/2}$\fi}
\def\muKRJs{\ifmmode \,\mu$K$_{\rm RJ}$\,s$^{1/2}\else \,$\mu$K$_{\rm RJ}$\,s$^{1/2}$\fi}
\def\muKHz{\ifmmode \,\mu$K\,Hz$^{-1/2}\else \,$\mu$K\,Hz$^{-1/2}$\fi}
\def\MJysr{\ifmmode \,$MJy\,sr\mo$\else \,MJy\,sr\mo\fi}
\def\MJysrmK{\ifmmode \,$MJy\,sr\mo$\,mK$_{\rm CMB}\mo\else \,MJy\,sr\mo\,mK$_{\rm CMB}\mo$\fi}
\def\microns{\ifmmode \,\mu$m$\else \,$\mu$m\fi}

\def\muK{\ifmmode \,\mu$K$\else \,$\mu$\hbox{K}\fi}
\def\microK{\ifmmode \,\mu$K$\else \,$\mu$\hbox{K}\fi}
\def\muW{\ifmmode \,\mu$W$\else \,$\mu$\hbox{W}\fi}
\def\kms{\ifmmode $\,km\,s$^{-1}\else \,km\,s$^{-1}$\fi}
\def\kmsMpc{\ifmmode $\,\kms\,Mpc\mo$\else \,\kms\,Mpc\mo\fi}

\providecommand{\sorthelp}[1]{}

\def\Cosmoglobe{\textsc{Cosmoglobe}}
\def\commander{\texttt{Commander}}
\def\commanderthree{\texttt{Commander3}}

\def\Planck{\textit{Planck}}
\def\WMAP{\textit{WMAP}}

\def\GAIA{\textit{Gaia}}
\def\CIBER{\textit{CIBER}}
\def\IRAS{\textit{IRAS}}
\def\AKARI{\textit{AKARI}}

\newcommand{\dv}[0]{\vec{d}}

\newcommand{\B}[0]{\tens{B}}
\newcommand{\G}[0]{\tens{G}}

\newcommand{\n}[0]{\vec{n}}

\newcommand{\s}[0]{\vec{s}}
\renewcommand{\a}[0]{\vec{a}}

\renewcommand{\L}[0]{\tens{L}}

\newcommand{\N}[0]{\tens{N}}
\newcommand{\M}[0]{\tens{M}}

\renewcommand{\r}[0]{\vec{r}}

\renewcommand{\P}[0]{\tens{P}}

\newcommand{\Te}[0]{T_{\rm e}}

\newcommand{\BP}{\textsc{BeyondPlanck}}

\newcommand{\cosmoglobe}{\textsc{Cosmoglobe}}

\def\Tcmb{\ifmmode T_\mathrm{CMB}\else $T_{\mathrm{CMB}}$\fi}
\def\Tcold{\ifmmode T_\mathrm{c}\else $T_{\mathrm{c}}$\fi}
\def\Thot{\ifmmode T_\mathrm{h}\else $T_{\mathrm{h}}$\fi}
\def\Tnear{\ifmmode T_\mathrm{n}\else $T_{\mathrm{n}}$\fi}
\def\scmb{\ifmmode s_\mathrm{CMB}\else $s_{\mathrm{CMB}}$\fi}
\def\squad{\ifmmode s_\mathrm{quad}\else $s_{\mathrm{quad}}$\fi}
\def\ssynch{\ifmmode s_\mathrm{s}\else $s_\mathrm{s}$\fi}
\def\sdust{\ifmmode s_\mathrm{d}\else $s_{\mathrm{d}}$\fi}
\def\ssdust{\ifmmode s_\mathrm{sd}\else $s_{\mathrm{sd}}$\fi}
\def\same{\ifmmode s_\mathrm{AME}\else $s_{\mathrm{AME}}$\fi}
\def\ssrc{\ifmmode s_\mathrm{src}\else $s_{\mathrm{src}}$\fi}
\def\sco{\ifmmode s_\mathrm{CO}\else $s_{\mathrm{CO}}$\fi}
\def\sff{\ifmmode s_\mathrm{ff}\else $s_{\mathrm{ff}}$\fi}
\def\gff{\ifmmode g_\mathrm{ff}\else $g_{\mathrm{ff}}$\fi}
\def\fsynch{\ifmmode f_\mathrm{s}\else $f_{\mathrm{s}}$\fi}
\def\fsd{\ifmmode f_\mathrm{sd}\else $f_{\mathrm{sd}}$\fi}
\def\fame{\ifmmode f_\mathrm{AME}\else $f_{\mathrm{AME}}$\fi}
\def\alphasrc{\ifmmode \alpha_\mathrm{src}\else $\alpha_{\mathrm{src}}$\fi}
\def\bcold{\ifmmode \beta_\mathrm{c}\else $\beta_{\mathrm{c}}$\fi}
\def\bhot{\ifmmode \beta_\mathrm{h}\else $\beta_{\mathrm{h}}$\fi}
\def\bnear{\ifmmode \beta_\mathrm{n}\else $\beta_{\mathrm{n}}$\fi}
\def\bsynch{\ifmmode \beta_\mathrm{s}\else $\beta_{\mathrm{s}}$\fi} 
\def\bsun{\ifmmode \beta_\mathrm{sun}\else $\beta_{\mathrm{sun}}$\fi} 
\def\nuzeros{\ifmmode \nu_{0,\mathrm{s}}\else $\nu_{0,\mathrm{s}}$\fi} 
\def\nuzeroff{\ifmmode \nu_{0,\mathrm{ff}}\else $\nu_{0,\mathrm{ff}}$\fi} 
\def\nuzerocold{\ifmmode \nu_{0,\mathrm{c}}\else $\nu_{0,\mathrm{c}}$\fi}
\def\nuzerohot{\ifmmode \nu_{0,\mathrm{h}}\else $\nu_{0,\mathrm{h}}$\fi}
\def\nuzeronear{\ifmmode \nu_{0,\mathrm{n}}\else $\nu_{0,\mathrm{n}}$\fi} 
\def\nuzeroame{\ifmmode \nu_{0,\mathrm{AME}}\else $\nu_{0,\mathrm{AME}}$\fi} 
\def\nuzerosd{\ifmmode \nu_{0,\mathrm{}}\else $\nu_{0,\mathrm{sd}}$\fi} 
\def\nuzerosrc{\ifmmode \nu_{0,\mathrm{src}}\else $\nu_{0,\mathrm{src}}$\fi} 
\def\nup{\ifmmode \nu_{\mathrm{p}}\else $\nu_{\mathrm{p}}$\fi} 
\def\alphasd{\ifmmode \alpha_{\mathrm{sd}}\else $\alpha_{\mathrm{sd}}$\fi} 
\def\Te{\ifmmode T_{\mathrm{e}}\else $T_{\mathrm{e}}$\fi} 
\def\kB{\ifmmode k_\mathrm{B}\else $k_{\mathrm{B}}$\fi}

\begin{document} 
   \title{\bfseries{\Cosmoglobe\ DR2. III. Improved modeling of
       zodiacal light with \textit{COBE}-DIRBE through global Bayesian analysis}}

   \newcommand{\oslo}[0]{1}
\newcommand{\milano}[0]{2}
\newcommand{\ijclab}[0]{3}
\newcommand{\gothenberg}[0]{4}
\newcommand{\milanoinfn}[0]{5}
\newcommand{\trento}[0]{6}

\author{\small
M.~San\inst{\oslo}\thanks{Corresponding author: M.~San; \url{metin.san@astro.uio.no}}
\and
A.~Bonato\inst{\milano}\thanks{Corresponding author: A.~Bonato; \url{angela.bonato1@studenti.unimi.it}}
\and
M.~Galloway\inst{\oslo}
\and
E.~Gjerl\o w\inst{\oslo}
\and
D.~J.~Watts\inst{\oslo}
\and
R.~Aurvik\inst{\oslo}
\and
A.~Basyrov\inst{\oslo}
\and
L.~A.~Bianchi\inst{\oslo}
\and
M.~Brilenkov\inst{\oslo}
\and
H.~K.~Eriksen\inst{\oslo}
\and
U.~Fuskeland\inst{\oslo}
\and
K.~A.~Glasscock\inst{\oslo}
\and
L.~T.~Hergt\inst{\ijclab}
\and
D.~Herman\inst{\oslo}
\and
J.~G.~S.~Lunde\inst{\oslo}
\and
A.~I.~Silva Martins\inst{\oslo}
\and
D.~Sponseller\inst{\gothenberg}
\and
N.-O.~Stutzer\inst{\oslo}
\and
R.~M.~Sullivan\inst{\oslo}
\and
H.~Thommesen\inst{\oslo}
\and
V.~Vikenes\inst{\oslo}
\and
I.~K.~Wehus\inst{\oslo}
\and
L.~Zapelli\inst{\milano, \milanoinfn,\trento}
}

\institute{\small
Institute of Theoretical Astrophysics, University of Oslo, Blindern, Oslo, Norway\goodbreak
\and
Dipartimento di Fisica, Universit\`{a} degli Studi di Milano, Via Celoria, 16, Milano, Italy\goodbreak
\and
Laboratoire de Physique des 2 infinis -- Irène Joliot Curie (IJCLab), Orsay, France
\and
Department of Space, Earth and Environment, Chalmers University of Technology, Gothenburg, Sweden\goodbreak
\and
Università di Trento, Università degli Studi di Milano, CUP E66E23000110001\goodbreak
\and
INFN sezione di Milano, 20133 Milano, Italy\goodbreak
}

   \titlerunning{\Cosmoglobe: Interplanetary dust}
   \authorrunning{M.~San et al.}

   \date{\today}

   \abstract{We present an improved zodiacal light (ZL) model for \textit{COBE}-DIRBE, 
   derived through global Bayesian analysis within the \cosmoglobe\ Data Release~2 (DR2) 
   framework. The parametric form of the ZL model is inspired by the original DIRBE model by Kelsall et al.\ (K98), but the specific best-fit parameter values are re-derived 
   using a combination of DIRBE Calibrated Individual Observations, \Planck\ HFI 
   sky maps, and WISE and \textit{Gaia} compact object catalogs. Furthermore, the ZL 
   parameters are fitted jointly with astrophysical parameters, such as thermal dust and 
   starlight emission, and the new model takes into account excess radiation that appears 
   stationary in solar-centric coordinates, as reported in a companion paper. The relative 
   differences between the predicted signals from K98 and our new model are $\lesssim\,3\,\%$ in 
   the 12 and 25$\,\mu\mathrm{m}$ channels over the full sky. The zero-levels of the cleaned DR2 
   maps are lower than those of the K98 Zodiacal light Subtracted Mission Average (ZSMA) maps 
   by $\sim$\,30\,kJy/sr at 1.25--3.5\,$\mu\mathrm{m}$, which is larger than the entire predicted 
   contribution from high-redshift galaxies to the Cosmic Infrared Background (CIB) at the same 
   wavelengths. At high Galactic latitudes, the total root-mean square of each DR2 map is lower than the corresponding 
   DIRBE ZSMA map of $\sim$\,30\,\% at wavelengths up to and including $3.5\,\mu\mathrm{m}$, and
   $\sim$\,80\,\% at wavelengths 4.9--25\,$\mu\mathrm{m}$.
   The cleaned DR2 maps at 4.9 and 12\,$\mu\mathrm{m}$ are now, for the first time, 
   visually dominated by Galactic signal at high latitudes rather than by ZL residuals. Even the 
   100$\,\mu\mathrm{m}$ channel, which has served as a cornerstone for Galactic studies for three 
   decades, appears cleaner in the current processing. Still, obvious ZL residuals 
   can be seen in several of the DR2 maps, and further work is required to mitigate these. Joint 
   analysis with existing and future high-resolution full-sky surveys such as \textit{AKARI}, \textit{IRAS}, \textit{Planck} 
   HFI, and SPHEREx will be essential both to break key degeneracies in the current model and to 
   determine whether the reported solar-centric excess radiation has a ZL or instrumental origin. 
   On the algorithmic side, more efficient methods for probing massively multi-peaked likelihoods 
   should be explored and implemented. Thus, while the results presented in this paper do redefine 
   the state of the art for DIRBE modeling, it also only represents the first among many steps 
   toward a future optimal Bayesian ZL model.}


   \maketitle

\section{Introduction}
Zodiacal light (ZL, sometimes zodiacal emission or interplanetary dust
emission) is the primary source of diffuse radiation observed in the
infrared sky between 1--100 $\mu$m (see, e.g., \citealp{Leinert1998}
and references therein). This radiation comes from scattering and
re-emission of sunlight from interplanetary dust (IPD) grains, and was
first mapped in detail by the \IRAS\ satellite
\citep{neugebauer:1984}.

The inner Solar System is embedded in a Sun centered cloud of IPD,
with a symmetry axis tilted slightly with respect to the Ecliptic,
known as the zodiacal cloud. The ZL is seasonal, and its appearance in
the sky changes as the Earth moves through the IPD distribution. The
most common way to model the observer-position-dependent ZL is to
evaluate a line-of-sight integral for each observation directly in the
time-ordered domain. The time-varying and three-dimensional nature of
the ZL makes it one of the most challenging foregrounds to model in
astrophysical and cosmological studies of the infrared sky. The lack
of a high-accuracy ZL model has left a large part of the
electromagnetic spectrum inaccessible to cosmological analysis
attempting to measure the Cosmic Infrared Background (CIB;
\citealp{partridge1967}). The main scientific motivation of the \textit{COBE}-DIRBE 
experiment \citep{hauser1998,hauser:2001} was to detect
and characterize the CIB spectrum.

One of the most widely used ZL models in the field of cosmology is the
\textit{COBE}-DIRBE model by \citet{kelsall1998}, often simply
referred to as the K98 model. This is a parametric model that describes 
the three-dimensional distribution and radiative 
properties of IPD using time-dependent measurements from
the \textit{COBE}-DIRBE instrument \citep{hauser1998}. There have been 
extensions to this model proposed since this original work, primarily to 
explain residual monopoles seen in comparisons to other data \citep{sano, Korngut,  skysurf}. 
Additionally, there have been several direct measurements of the IPD density 
\citep{newhorizons, parker}, some of which suggest that the IPD densities may 
be greater than previously predicted \citep{pioneer}. Despite these potential 
shortcomings, the Kelsall model remains the de-facto model in the CMB field today.

Since DIRBE, our
understanding of the infrared sky has improved with new observational
data from experiments like WISE \citep{wright:2010}, \Planck\ HFI
\citep{planck2016-l03}, and \GAIA\ \citep{gaia:2016}. However, these
experiments have largely been analyzed individually, and little or no
coordinated effort has been made toward combining the data from these
experiments into one overall state-of-the-art model of the infrared
sky. The main goal of the current work, summarized in a series of
papers collectively denoted \cosmoglobe\ Data Release 2 (DR2), is to make
the first step toward such a concordance model by leveraging recent
computational advances in Bayesian cosmological data analysis by the
\BP\ \citep[][and references therein]{BP2023} and
\cosmoglobe\ \citep{Watts2023} collaborations. The computer code
implementation is called \commander\ \citep{eriksen:2004, seljebotn:2019, bp03}, which
is a Bayesian Gibbs sampler that was originally designed for component separation, but has now evolved into a complete framework for
end-to-end analysis of cosmic microwave background (CMB) experiments,
in particular \Planck\ LFI \citep{planck2016-l02} and \WMAP\
\citep{bennett2012}. However, as demonstrated in the current work, the
same algorithms are, after relatively minor modifications, directly
applicable to infrared measurements.

One of the most important generalizations required for application of
\commanderthree\ to the infrared sky is the implementation of an
accurate ZL model to remove the time-varying ZL. In this paper, we therefore implement 
support in \texttt{Commander3} for the K98 model, combined with the scattering phase function described 
by \citet{Hong}, and we apply this to the time-domain DIRBE data. 
This new code implementation is based on ZodiPy \citep{San2024}, which is an Astropy-affiliated 
Python package for ZL simulations.\footnote{\url{https://cosmoglobe.github.io/zodipy/}} As an early application of this framework, \cite{San2022} 
demonstrated the removal of ZL from the DIRBE time-ordered data (TOD) with ZodiPy using the 
K98 model. 

ZL is found to be polarized in the near-infrared in both the DIRBE and 
\CIBER\ data \citep{Takimoto2022,Takimoto2023}. We make no attempts at 
modeling polarized ZL in this analysis, but a natural next step in this 
analysis would be to include the polarization data from the 1.25, 2.2, 
and 3.5 $\mu$m DIRBE bands.

The rest of the paper is organized as follows. In
Sect.~\ref{sect:zodi-model}, we introduce our ZL model and discuss
implementation and optimization aspects. Next, in 
Sect.~\ref{sect:param-estimation}, we describe the algorithms and methods
used to fit the ZL parameters within the \cosmoglobe\  framework.
In Sect.~\ref{sect:data}, we describe the data used in the current
analysis. Next, in Sect.~\ref{sect:improved-model}, we present our updated 
ZL model and compare this with the K98 model. Finally, we
conclude in Sect.~\ref{sec:conclusions}.

\section{Zodiacal light modeling}\label{sect:zodi-model}
ZL is commonly modeled sample-by-sample in time-domain by performing
line-of-sight integration at each observation through a parametric
three-dimensional model of the IPD distribution. We adopt a modified version of the 
general parametrization as introduced by K98 for the current
\commanderthree\ implementation, and, in this section, we briefly review
the main aspects of this model, aiming to build intuition about the
underlying parametric model. For full
details of this model, we refer the interested reader to
\citet{kelsall1998} and \citet{Hong}.\footnote{We have also spent large 
resources on implementing and exploring the parametric ZL model presented in 
Appendix~1 of \citet{wright:1998}, but we have unfortunately been unsuccessful 
with achieving goodness-of-fit statistics that are competitive with the current 
K98-based analysis. We are uncertain whether this is due to the intrinsic lower 
flexibility of that model or an undiscovered bug in our implementation, and we 
therefore do not discuss this alternative model further at this time.} 
We also consider a few numerical
approximations that reduce the overall computational cost.

\subsection{Parameterization of interplanetary dust}
IPD in the zodiacal cloud is overall smooth and stable 
\citep{Leinert1989}, and most of the dust may be accounted for by a 
diffuse cloud-like component \citep{kelsall1998}. The origin of IPD 
is debris mainly from comets \citep{Liou1995, ipatov, rigley}, asteroids \citep{Dermott1984} and meteoroids \citep{dikarev}, 
with potential additions from planets \citep{Jorgensen2021}, 
the Kuiper belt \citep{Mann2009}, and interstellar dust passing 
through the Solar System \citep{Robinson2013}. The relative contribution to the overall 
IPD density from these sources is not well known, but many models predict the main 
source to be low-eccentricity Jupiter family comets that rapidly 
disintegrate due to frequent trips through the inner Solar System.
Within the mostly smooth zodiacal cloud, fine structures exist near the Ecliptic plane as a result 
of collisions and fragmentation in the asteroid belt and gravitational 
resonance and disturbance in the orbit of the planets \citep{Low1984, Dermott1984, Dermott1994, Reach1997}. 
There are also structures found in the orbits of other Solar System planets, \citep{kennedy, jones_venus, Stenborg}, 
but these are potential extensions to the Kelsall model.  

We model the IPD distribution as a combination of several zodiacal
components, each described by a number density $n_c(x,y,z)$, where $c$
indicates components and $x$, $y$ and $z$ indicate heliocentric
Ecliptic coordinates. Each zodiacal component is allowed to have a
heliocentric offset $(x_{0,c}, y_{0,c}, z_{0,c})$, such that the
component-centric coordinates become
\begin{equation}    
    \begin{aligned}
        x_c&= x - x_{0,c}\\
        y_c&= y - y_{0,c}\\
        z_c&= z - z_{0,c}.
    \end{aligned}
\end{equation}
Additionally, each zodiacal component is allowed to have a plane of
symmetry that is different from the Ecliptic, which is defined by an
inclination $i_c$ and an ascending node $\Omega_c$. Components that
happen to be azimuthally symmetric are then fully described by a
radial distance $r_c$ from the origin and the height above the
symmetry plane $Z_c$:
\begin{align}
    r_c &= \sqrt{x_c^2 + y_c^2 + z_c^2},\\
    Z_c &= x_c\sin{\Omega_c}\sin{i_c} - y_c \cos{\Omega_c}\sin{i_c} + z_c \cos{i_c},\\
    \zeta_c &= \frac{|Z_c|}{r_c},
\end{align}
with $\zeta_c$ being the radial height above the symmetry plane.

\subsection{Zodiacal components}

With these definitions in hand, we now define parametric models for
each IPD component type.

\subsubsection{Smooth cloud}
By far the most important component is typically referred to as the
``zodiacal cloud''. This component represents the smooth IPD
distribution in the inner Solar System. Its number density is
modeled as
\begin{equation}
    n_\mathrm{C}(x,y,z)=n_{0,
      \mathrm{C}}r_\mathrm{C}^{-\alpha}f(\zeta_\mathrm{C}),
    \label{eq:cloud}
\end{equation}
where $n_{0, \mathrm{C}}$ is the number density at 1\,AU, $\alpha$ is a 
power-law exponent, $f(\zeta_\mathrm{C})$ is the fan-like vertical 
distribution given as 
\begin{equation}
    f(\zeta_\mathrm{C}) = \exp {\left[-\beta g(\zeta_\mathrm{C})^\gamma \right]},
\end{equation}
with
\begin{equation}
    g(\zeta_\mathrm{C}) = \begin{cases}
        \zeta_\mathrm{C}^2/2\mu & \mathrm{for}\; \zeta_\mathrm{C} < \mu,\\
        \zeta_\mathrm{C} - \mu/2 & \mathrm{for}\; \zeta_\mathrm{C} \geq \mu,
    \end{cases}
\end{equation}
where $\beta$, $\gamma$ and $\mu$ are shape parameters.

\subsubsection{Dust bands}
Next, three dust bands are included in the model to represent the
observed shoulder-like structure in the \IRAS\ scans across the Ecliptic
plane.  These bands appear at Ecliptic latitudes of approximately
$\pm1.4^\circ$, $\pm10^\circ$, and $\pm15^\circ$, and are associated with
a blend of the Themis and Koronis, the Eos, and the Io/Maria asteroid families, 
respectively \citep{Reach1997}. Each dust band, indicated by $B_i$, is 
modeled as
\begin{align}
    n_{\mathrm{B}_i}(x,y,z) &= \frac{3 n_{0, \mathrm{B}_i}}{r_{\mathrm{B}_i}} \exp \left[-\left(\frac{\zeta_{\mathrm{B}_i}}{\delta_{\zeta_{\mathrm{B}_i}}}\right)^{6}\right]\left[1 + \left(\frac{\zeta_{\mathrm{B}_i}}{\delta_{\zeta_{\mathrm{B}_i}}}\right)^{p_i}v_i^{-1}\right] \nonumber\\
    &\times\left\{1-\exp \left[-\left(\frac{r_{\mathrm{B}_i}}{\delta_{r_{\mathrm{B}_i}}}\right)^{20}\right]\right\},
\label{eq:band}
\end{align}
where $n_{0, \mathrm{B}_i}$ is the number density of band $\mathrm{B}_i$ 
at 3\,AU, $\delta_{r_{\mathrm{B}_i}}$ is the inner radial cut-off, and 
$p_i$, $v_i$ and $\delta_{\zeta_{\mathrm{B}_i}}$ are shape parameters.

\subsubsection{Circumsolar ring and Earth-trailing feature}
\label{sec:ring}
Finally, a circumsolar ring (denoted ``SR'') component is included in the
model to represent dust that has accumulated in Earth's orbit due to
gravitational effects \citep{Dermott1994}. This component also
includes an enhancement to the IPD distribution at Earth's wake, known
as the ``Earth-trailing feature'' (denoted ``TF''). The composite ring component (denoted ``R'') is then modeled as
\begin{align}
    n_\mathrm{R}(x, y, z, \theta)&=n_{0, \mathrm{SR}} \exp \left[-\frac{\left(r_\mathrm{R}-r_{0, \mathrm{SR}}\right)^2}{\sigma_{R,\mathrm{SR}} ^2}-\frac{\left| Z_\mathrm{R} \right|}{\sigma_{Z, \mathrm{SR}}}\right]\nonumber\\
   &+ n_{0, \mathrm{TF}} \exp \left[-\frac{\left(r_\mathrm{R}-r_{0,
          \mathrm{TF}}\right)^{2}}{\sigma_{R,
          \mathrm{TF}}^{2}}-\frac{\left|Z_\mathrm{F}\right|}{\sigma_{Z,
          \mathrm{TF}}}-\frac{\left(\theta-\theta_{0,
          \mathrm{TF}}\right)^{2}}{\sigma_{\theta,\mathrm{TF}}^{2}}\right],
    \label{eq:ring}
\end{align}
where $\theta$ is the heliocentric longitude of the Earth, and the 
radial locations $r_{0, \mathrm{SR}}$, $r_{0, \mathrm{TF}}$ specify
the distances to the peak densities $n_{0, \mathrm{SR}}$, 
$n_{0, \mathrm{TF}}$. The $\sigma$ parameters are length scales for the 
$r$, $Z$ and $\theta$ parameters, respectively. We note that the 
Earth-trailing feature depends on the position of the Earth and does not 
have a plane symmetry like the other zodiacal components.

\subsection{Radiative and scattering properties}
Equations~\eqref{eq:cloud}--\eqref{eq:ring} define the number density
of each component. However, the signal actually measured with an
infrared detector is defined by an intensity, $I_{\nu}$, typically
measured in units of MJy$\,\mathrm{sr}^{-1}$ or
nW$\,\mathrm{m}^{-2}\,\mathrm{Hz}^{-1}\,\mathrm{sr}^{-1}$. The connection between the
number density and this thermal emission is modeled in terms
of a blackbody modified by an emissivity factor $E_{c,
  \lambda}$,\footnote{In these expressions, $\lambda$ denotes
wavelength channel, and may refer interchangeably to both the physical
wavelength and the DIRBE channel ID, e.g., channel 1 correponds to
1.25\,$\mu$m; see \citet{hauser1998} for a full definition. }
\begin{equation}
    I^\mathrm{Thermal}_{c,\lambda} = E_{c,\lambda} B_\lambda(T),
\end{equation}
where $B_\lambda$ is the Planck function at a wavelength $\lambda$
\citep{kelsall1998}. A key parameter in this equation is the IPD
temperature $T$, which is assumed to fall off with radial distance $r$ (in AU)
from the Sun as
\begin{equation}
    T(r) = T_0 r^{-\delta},
\end{equation}
where $T_0$ is the temperature of IPD at 1\,AU and $\delta$ is a power-law 
exponent which is expected to be $\sim 0.5$ for gray dust. 

In addition to emitting thermally, IPD grains also scatter
sunlight in near-infrared wavelengths. The contribution to the total
signal from scattering reads
\begin{equation}\label{eq: scat_term}
    I^\mathrm{Scattering}_{c, \lambda} = A_{c, \lambda} F_\lambda^\odot(r) \Phi(\Theta),
\end{equation}
where $A_{c, \lambda}$ is the albedo (or reflectivity) of the IPD,
$F_\lambda^\odot(r)$ the solar flux at a radial distance from the Sun,
and $\Phi(\Theta)$ is the so-called phase function for scattering angles
$\Theta$, which describes the angular distribution of the scattered
light. In particular, 
our ZL model features a Henyey-Greenstein phase function as defined by 
\citet{Hong}:
\begin{equation}
  \Phi(\Theta)=\sum_\mathrm{k=1}^3 w_\mathrm{k} \frac{1}{4\pi}\frac{1-g_\mathrm{k}^2}{(1+g_\mathrm{k}^2-2g_\mathrm{k}\cos(\Theta))^{3/2}},
\label{eq:phase_function}
\end{equation}
where $g_1$, $g_2$, $g_3$, $w_2$, and $w_3$ are free parameters, while \mbox{$w_1=1-(w_2+w_3)$.}

\begin{figure}
  \includegraphics[width=0.85\columnwidth]{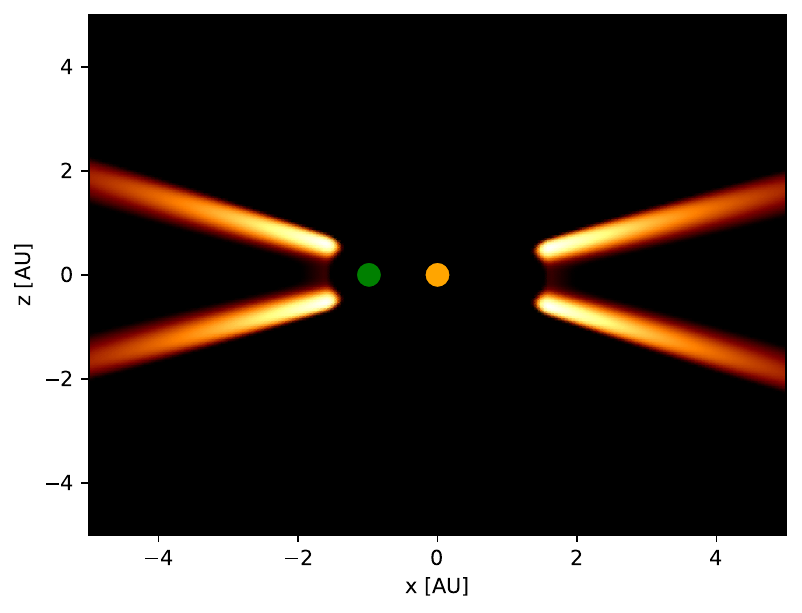}\\
  \includegraphics[width=0.9\columnwidth]{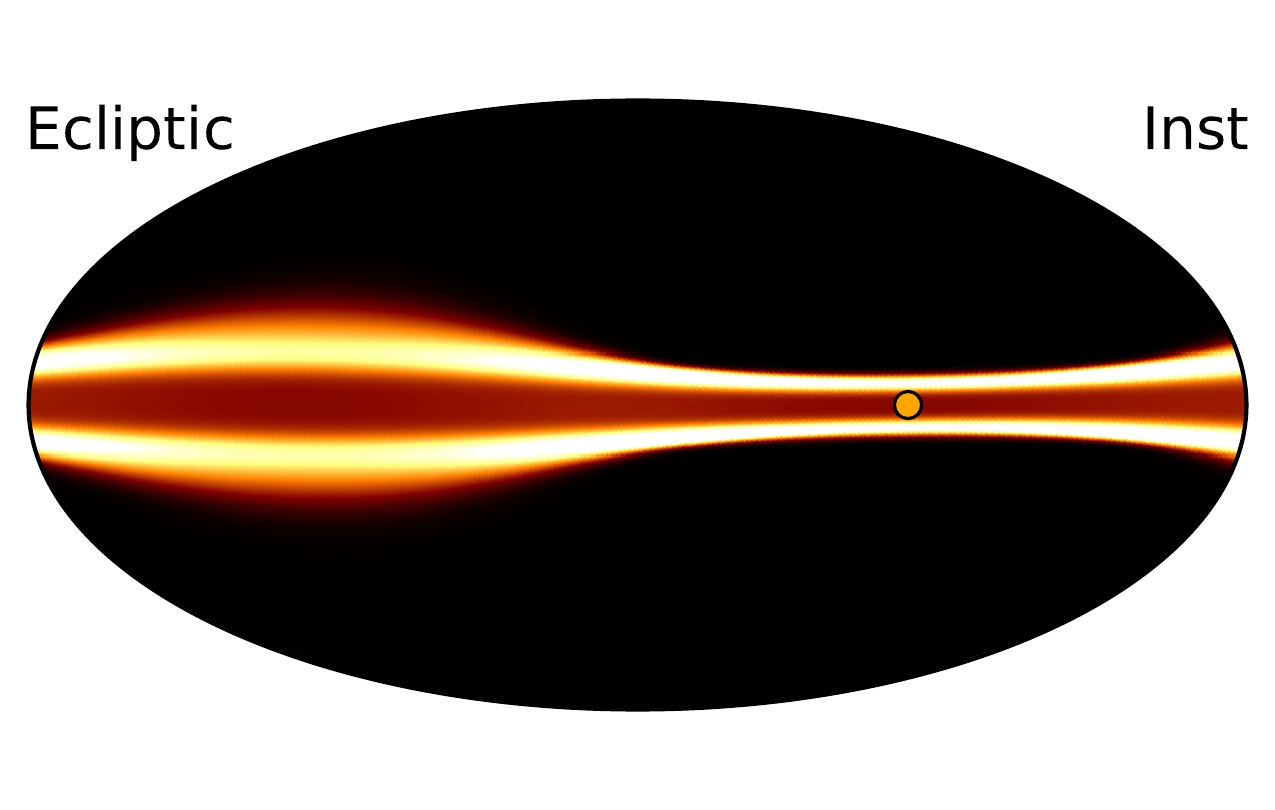}\\
  \includegraphics[width=0.9\columnwidth]{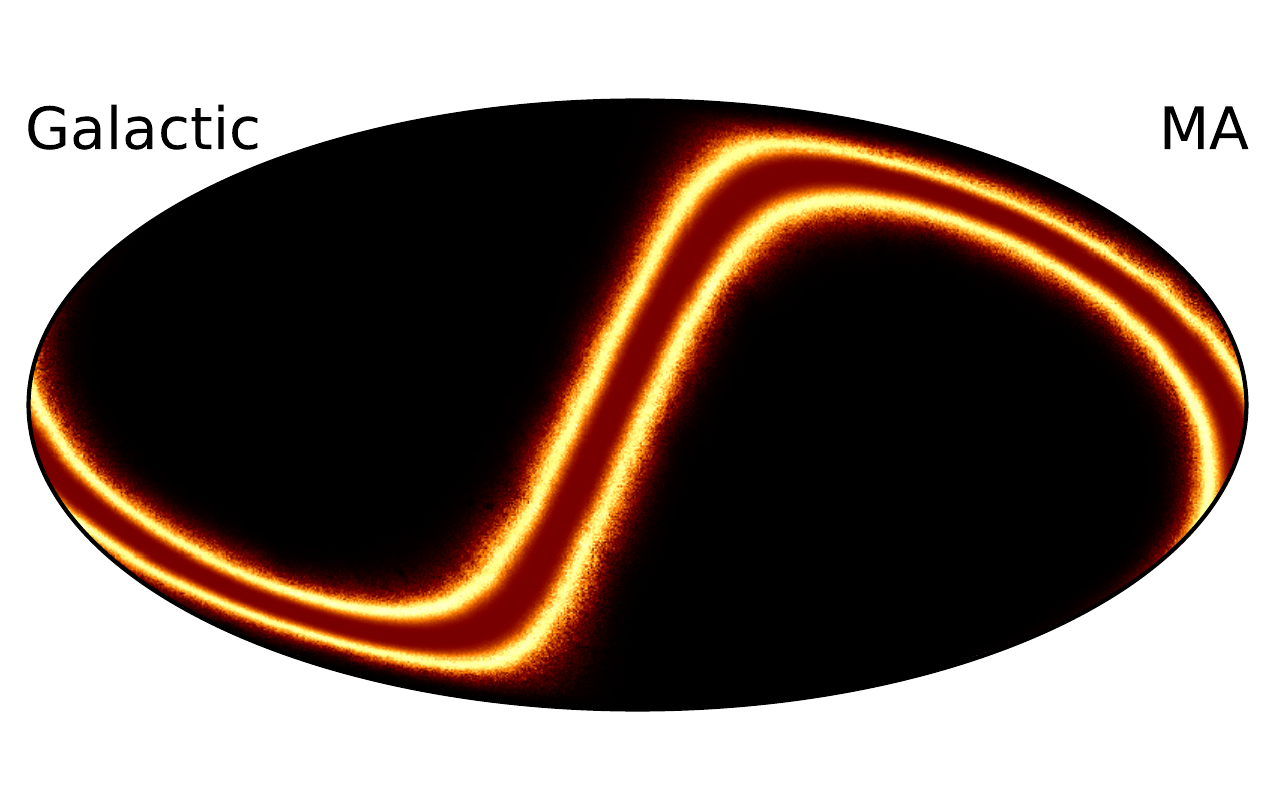}\\
  \includegraphics[width=0.9\columnwidth]{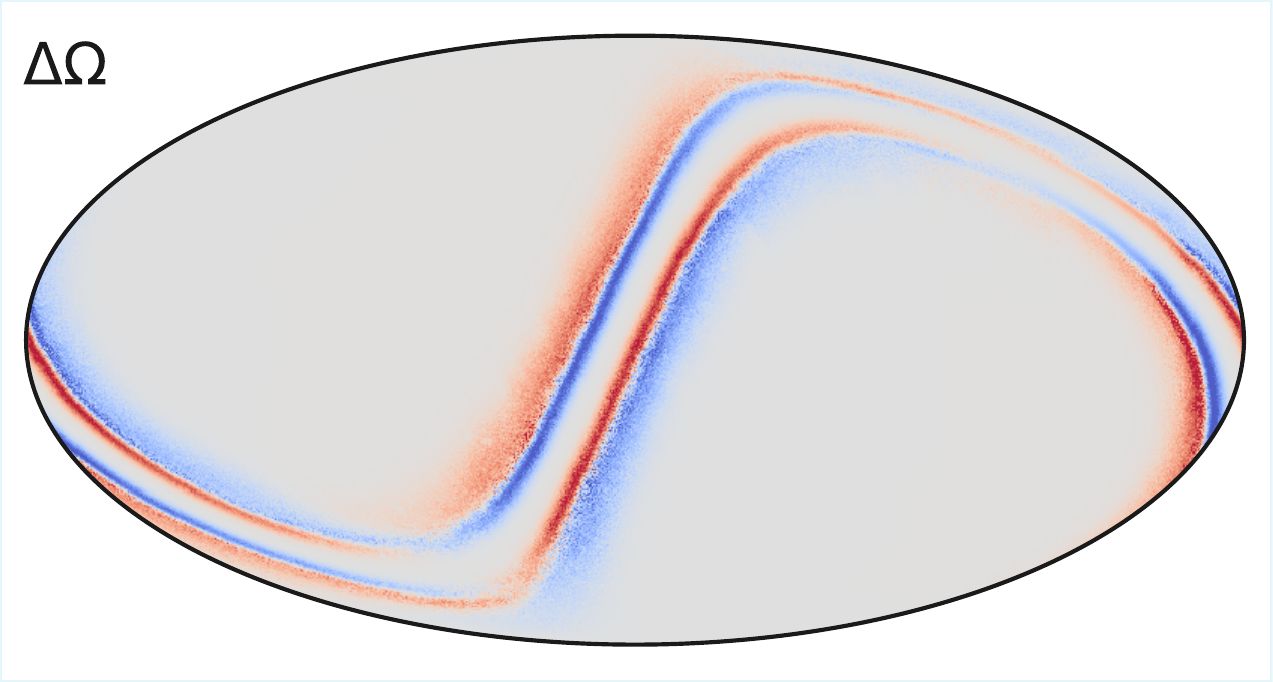}
  \caption{Geometry of the first asteroidal dust band. (\textit{First row}:) 
  Slice through the $x$--$z$ plane of the number density, $n_{0}$, 
  in heliocentric coordinates. The positions of the Sun and Earth 
  are marked by orange and green dots, respectively. (\textit{Second 
  row}:) Observed instantaneous intensity plotted in Ecliptic 
  coordinates, obtained by integrating the above figure along each 
  line-of-sight. (\textit{Third row}:) Same as above, but plotted in
  Galactic coordinates and mission averaged (MA) over nearly a full year of
  observations corresponding to the DIRBE scanning strategy. 
  (\textit{Fourth row}:) Difference between observed
  intensities as defined in the third row after changing the value of
  the ascending node, $\Omega$, by 5\,\%. Similar plots for all
  components and parameters are provided in 
  Appendices~\ref{sec:zodi-comps} and \ref{sec:param-atlas}.}
  \label{fig:band1}
\end{figure}

The total intensity from a single IPD grain is then
\begin{align}\label{eq:I_tot}
    I^\mathrm{Total}_{c, \lambda} &= I^\mathrm{Scattering}_{c,\lambda} + I^\mathrm{Thermal}_{c,\lambda}\nonumber\\
    &= A_{c, \lambda} F_\lambda^\odot \Phi + (1 - A_{c, \lambda})E_{c,\lambda} B_\lambda,
\end{align}
where we have also taken into account that reflective material have lower thermal emission
by introducing a factor $(1 - A_{c, \lambda})$ to the thermal term.
The total ZL signal may then be evaluated by summing up the intensity
from all dust grains, which in practice means evaluating a line-of-sight 
integral for each observation,
\begin{equation}\label{eq:los}
    I_{p,t, \lambda} = \sum_c \int n_c \left[  A_{c, \lambda} F_\lambda^\odot \Phi + \left( 1 - A_{c, \lambda} \right) E_{c,\lambda} B_\lambda \right]\,ds.
\end{equation}
Here, $p$ represents an observed pixel or direction in the sky, $t$ is
the time of observation; $n_c$ is the number density of component $c$
in the line-of-sight; and $ds$ is a small distance along the
line-of-sight $s$ from the observer and toward $p$. Due to the limited signal-to-noise ratios, we follow
\citet{kelsall1998} and only fit one overall albedo for each
high-frequency channel, as well as only one common emissivity for all
three asteroidal bands.

\subsection{Model intuition}

As described above, our model has only $\mathcal{O}(10^2)$ free
parameters, which we collectively denote
$\zeta_{\mathrm{z}}$. Clearly, this is in reality far too few to fully
capture the true complex nature of ZL across many decades in
wavelength. However, even with such a limited number of parameters,
the model is still severely under-constrained when fitted to the DIRBE
data, and the corresponding posterior distribution exhibits many
strong degeneracies. Consequently, most currently available parameter
estimation algorithms are prone to getting trapped in local posterior
maxima, and this then will result in significant residuals in the
final ZL cleaned maps.

In order to interpret such residuals, and potentially define better
starting points for the non-linear optimization algorithm, it is
useful to build up human visual intuition regarding the impact of each
free parameter. Fig.~\ref{fig:band1} shows one specific (and
arbitrary) example of this. First, the top panel shows a
$x$--$z$-plane slice through the three-dimensional IPD number density
distribution for the first dust band, $B_1$. In this figure, the
orange dot marks the Sun's position, while the green dot marks the
observer's (or Earth's) position. Here it is worth noting that this
component is azimuthally symmetric about the Sun, and the full 3D
structure may therefore be visualized by rotating this figure about
the vertical $z$-axis. In this space, it is quite straightforward to
visualize the effect of each free parameter defined by
Eq.~\eqref{eq:band}. For instance, the position of the inner radial
cut-off can be changed by modifying $\delta_{r}$, while the angle
between the $x$-axis and the peak densities may be changed through
$\delta_{\zeta}$. If we modify the $x_0$ offset, the entire density
field will shift left or right.

The second panel in Fig.~\ref{fig:band1} shows the corresponding
signal in Ecliptic coordinates at one single point in time after
integrating the density field along each line-of-sight. The Sun's
position is again marked by an orange dot, but in this case there is
obviously no observer position, since this figure shows the sky as
seen outward from the observer. In this projected 2D space, the
observed structures appear significantly more difficult to visualize
than in 3D space. For instance, while the density of the dust bands
appear symmetric in 3D space, their apparent separation and width as
seen from Earth vary significantly with Ecliptic longitude, as seen in
the second panel; they appear narrower when looking toward the Sun,
where the bands are physically further away from the Earth, and
broader closeby.

The third panel shows the same feature, but now averaged over nearly
a whole year of observations, corresponding to the DIRBE scanning 
strategy, and plotted in Galactic coordinates. This
represents the signal seen in full-mission maps derived from
DIRBE. Since the underlying IPD structure is azimuthally symmetric
about the Sun, the Earth's movement throughout the year also
symmetrizes the total co-added signal, and the dust bands once again
appear symmetric about the Ecliptic plane. However, some small-scale
structures also appear because of small variations in the effective
scanning path of the instrument from day to day; if an entire day's
worth of observations were missing, for instance due to a period of
excessive cosmic ray radiation, strong stripes would appear in this
map.

With the infrastructure for computing such full-mission maps ready at
hand, we can study the impact of each free parameter in greater
detail. As a specific example of this, the bottom panel in
Fig.~\ref{fig:band1} shows the difference between the total signal
obtained when changing the ascending node $\Omega$ for Band~1 by 5\,\%
relative to a suitable set of parameters for our ZL model. 
Intuitively, this corresponds to rotating
the signal in the top figure slightly about the origin. Some parts of
the bands will then appear closer to the Earth, while others will
appear further away. Those regions then in turn appear either red or
blue in the bottom figure. The resulting pattern is a unique
signature for $\Omega$, and if similar structures are observed in the
final ZL cleaned maps, then one should consider modifying this
particular parameter in a future analysis.

Similar figures are provided for all components and all parameters in
Appendices~\ref{sec:zodi-comps} and \ref{sec:param-atlas}, and these
are very useful for building up visual intuition regarding the ZL 
model in use. Quickly scanning through the individual panels in
Figs.~\ref{fig:atlas1} and \ref{fig:atlas2}, we can already now
identify strong degeneracies that are likely to turn out to be problematic
later. For instance, we see that $n_{0,\mathrm{C}}$,
$\alpha_\mathrm{C}$, $n_{0,\mathrm{SR}}$, $T_0$, and $\delta$ are all
dominated by a ring centered along the Ecliptic plane, and these are
likely to interplay significantly. Furthermore, many of these
parameters, such as $n_{0,\mathrm{C}}$, $\sigma_{z,\mathrm{SR}}$, and
$\sigma_{\theta,\mathrm{TF}}$, will obviously also couple
significantly to a wide range of non-ZL type parameters when
integrated into a global analysis framework, including the
all-important CIB monopoles.

\subsection{Numerical optimizations}\label{sect:optimization}
Performing the line-of-sight integrals defined by Eq.~\eqref{eq:los}
is an expensive part in the \cosmoglobe\ DR2 analysis pipeline already
for the DIRBE data, which only comprise 18\,GB after compression. In
principle, this could be done by brute-force for this particular
experiment on modern computer clusters, but such a direct approach
will clearly not be an option for similar analyses of \Planck\ HFI,
\AKARI, and SPHEREx.

When sampling parameters for the ZL model it is possible to include only a
small fraction of the full dataset in each likelihood evaluation,
simply because the signal-to-noise ratio of each sample is so high,
and because of the smooth ZL gradient of the ZL
structure. Intuitively speaking, white instrumental noise is
irrelevant compared to overall systematic model uncertainties, and
some number of consecutive time-domain samples therefore provide
essentially the precisely same information. In our current analysis,
we adopt a thinning factor of eight, meaning that we effectively fit
the data to a time-stream sampled at 1\,Hz rather than the original
8\,Hz DIRBE CIO. In principle, we could have averaged over this time
segment, rather than simply omitting the relevant samples, in order to
suppress instrumental noise; and, in fact, the first implementation of
our computer code did exactly this. However, averaging over 1\,sec
time scales implies that the true underlying model is also smoothed
out to the same time scales, and this increases the overall modeling
errors. Although the differences were generally small, we obtained
slightly better fits by thinning rather than averaging.

A second optimization step is introduced by dividing the parameters in
$\zeta_{\mathrm{z}}$ into a set of so-called sampling groups, and
estimating the free parameters in each group separately. Specifically,
it is worth noting that the overall signal-to-noise ratio for the ZL
component shape parameters, such as $\Omega_{\mathrm{C}}$ or $\delta$,
is vastly dominated by the 12, 25, and 60\,$\mu$m channels. At the same
time, the emissivity and albedo for a given channel depend only on that
same channel, and it is therefore not necessary to process, say, the
1.25\,$\mu$m TOD when estimating the 4.9\,$\mu$m emissivity. In practice,
we therefore first estimate all ZL shape parameters (and the
corresponding emissivities and monopoles) using only the 
12, 25, and 60\,$\mu$m channels. We then estimate the emissivity
and monopole for each of the higher-wavelength channels separately --- now
conditionally on the shape parameters derived before. Finally, 
we estimate the phase function parameters, the albedoes, the emissivities, and 
the monopoles for the 1.25--3.5\,$\mu$m channels jointly, always using the 
previously derived shape parameters. 
The cost of this approach is slightly higher statistical
uncertainties on the shape parameters, since the other channels could
have contributed with some information for these parameters as well,
but the advantage is a computational speed-up of roughly one
order-of-magnitude, and we consider this an excellent trade-off.

\section{Methods}\label{sect:param-estimation}

The main operational goal of this paper is to measure the free ZL
parameters, $\zeta_{\mathrm{z}}$, using time-ordered data from the
DIRBE instrument. However, these data contain many other physical
effects in the form of both instrumental and astrophysical confusion
\citep[e.g.,][]{hauser1998,arendt1998}. In order to estimate
$\zeta_{\mathrm{z}}$ robustly, it is essential to account for all
those other degrees of freedom at the same time. On the other hand,
many of those parameters have only a limited signal-to-noise ratio
with DIRBE data alone, and far stronger constraints will typically
result from combining the DIRBE measurements with external
data. Enabling such global multi-experiment analysis is a main goal of
the \cosmoglobe\ framework \citep{Watts2023}. In this section, we briefly review the key
ideas behind this approach, and we describe the generalizations that
are required for ZL parameter estimation. For full details, we refer
the interested reader to \citet{CG02_01}. However, we
emphasize that the approach presented here is only a first step, and
future work should aim at implementing faster and more robust
algorithms.

\subsection{Data model, posterior distribution, and Gibbs sampling}
\label{sec:gibbs}

The first step in many Bayesian parameter estimation methods is to
write down an explicit parametric data model. For \cosmoglobe\ DR2, we
adopt the following model,
\begin{align}
	\label{eq:model}
	\dv &=\G\left[\B\P\sum_{c=1}^{n_{\mathrm{comp}}}\M_c\a_c+\s_{\mathrm{zodi}} +
          \s_{\mathrm{static}}\right] + \n_\mathrm{corr} + \n_\mathrm{w} \nonumber\\
        &\equiv \s^{\mathrm{tot}} + \n_\mathrm{w},
\end{align}
where $\dv$ denotes observed data; $\G$ denotes an overall calibration
factor; $\P$ and $\B$ represent the instrumental pointing and beam,
respectively; the sum over components $c$ represents the contribution
from astrophysical components (thermal dust, free-free, starlight
emission etc.), each described by an overall amplitude $\a$ (which may be a
pixelized map) and a mixing matrix, $\M$, which depends on some set of
unknown SED parameters, $\beta$; and $\n=\n_\mathrm{corr} + \n_\mathrm{w}$ denotes instrumental
noise. The latter is a sum of correlated noise ($\n_\mathrm{corr}$) and white noise ($\n_\mathrm{w}$). 
We further define $\a_{\mathrm{sky}}$ and
$\beta_{\mathrm{sky}}$ to be the set of all astrophysical component
amplitudes and spectral parameters, and $\xi_{\mathrm{n}}$ to be the
set of all free instrumental noise parameters. We also define
$\a_{\mathrm{static}}$ by
$\s_{\mathrm{static}}=\P_{\mathrm{sol}}\a_{\mathrm{static}}$, where
$\P_{\mathrm{sol}}$ is the pointing in solar-centric
coordinates. Finally, we denote the set of all free parameters in
Eq.~\eqref{eq:model} by $\omega$, and for a full explicit definition
of this parameter set, we refer the interested reader to
\citet{CG02_01}. The foreground model is treated in more 
detail in four companion papers. In particular, we refer the interested reader 
to \citet{CG02_04} for specifics about stars and to \citet{CG02_05,CG02_06,CG02_07} 
for our analysis of thermal dust.

As far as this paper is concerned, the key term is
$\s_{\mathrm{zodi}}$, which is nothing but Eq.~\eqref{eq:los}
evaluated along the line-of-sight defined by the pointing $\P$. This
term depends on $\zeta_{\mathrm{z}}$, and our task in this paper is to
establish an approximation to the marginal posterior distribution,
$P(\zeta_{\mathrm{z}}|\dv)$. One straightforward way of computing this
marginal distribution is, perhaps somewhat surprisingly, to first
consider the much bigger task of estimating the full joint posterior
distribution, $P(\omega|\dv)$ --- which now includes billions of
correlated parameters rather than just a hundred. The reason this is a 
more straightforward problem, computationally speaking, is that 
the joint distribution has a well-defined and simple analytic expression that it
is possible to sample from, while the marginal distribution does not;
for early CMB applications of this two-stage approach, see
\citet{jewell2004,wandelt2004,eriksen:2004}.

In order for this to work, we have to define $\s_{\mathrm{tot}}$ as the sum 
of all components but $\n_\mathrm{w}$, so that $\n_\mathrm{w}=\dv-\s_{\mathrm{tot}}$ 
is a Gaussian distribution with 
zero mean and covariance matrix $\N_{\mathrm{w}}$ that also has free parameters. 
For most instruments this is an excellent approximation. With this assumption, we can
write the likelihood, $\mathcal{L}(\omega) \equiv P(\dv|\omega)$, as 
\begin{align}
-2\ln\mathcal{L}(\omega) &= (\dv-\s^{\mathrm{tot}}(\omega))^t
  \N_{\mathrm{w}}^{-1}(\dv-\s^{\mathrm{tot}}(\omega)) +\ln|\N_{\mathrm w}| \nonumber \\
  &\equiv \chi^2(\omega) +\ln|\N_{\mathrm w}|,
\end{align}
and the posterior distribution is then defined by Bayes' theorem,
\begin{equation}
P(\omega\mid\dv) = \frac{P(\dv\mid\omega) P(\omega)}{P(\dv)} \propto
\mathcal{L}(\omega) P(\omega).
\end{equation}
Here $P(\omega)$ is called the prior, which may be used to inject
prior knowledge about given parameters, while $P(\dv)$ is called the
evidence, which for our purposes is just a normalization constant.

In order to map out this full joint posterior, we use a statistical
method called Gibbs sampling \citep[e.g.,][]{geman:1984}, which allows
us to draw samples iteratively by scanning through all conditional
distributions, as opposed to drawing samples directly from the joint
distribution. Sampling from $N$ conditional distributions, each
defined by a simple analytical distribution, is typically much simpler
than drawing from a single joint $N$-dimensional distribution that
does not have a closed form analytical expression. In practice, for
the \cosmoglobe\ DR2 analysis this translates into the following
so-called Gibbs chain:
\begin{alignat}{11}
    \tens{G} &\,\leftarrow P(\tens{G}&\,\mid &\,\dv,&\, &\,\phantom{\tens{G}} &\,\xi_n, &
    \,\beta_{\mathrm{sky}},& \,\a_{\mathrm{sky}}, &\,\zeta_{\mathrm{z}},
    &\,\a_{\mathrm{static}})\label{eq:gibbs_G}\\
    \xi_{\mathrm{n}} &\,\leftarrow P(\xi_{\mathrm{n}}&\,\mid &\,\dv,&\, &\,\tens{G}, &\,\phantom{\xi_n} &
    \,\beta_{\mathrm{sky}},& \,\a_{\mathrm{sky}}, &\,\zeta_{\mathrm{z}},
    &\,\a_{\mathrm{static}})\\
    \beta_{\mathrm{sky}} &\,\leftarrow P(\beta_{\mathrm{sky}}&\,\mid &\,\dv,&\, &\,\tens{G}, &\,\xi_n, &
    \,\phantom{\beta_{\mathrm{sky}}}& \,\a_{\mathrm{sky}}, &\,\zeta_{\mathrm{z}}, &\,\a_{\mathrm{static}})\\
    \a_{\mathrm{sky}} &\,\leftarrow P(\a_{\mathrm{sky}}&\,\mid &\,\dv,&\, &\,\tens{G}, &\,\xi_n, &
    \,\beta_{\mathrm{sky}},& \,\phantom{\a_{\mathrm{sky}},}
    &\,\zeta_{\mathrm{z}}, &\,\a_{\mathrm{static}})\\
    \zeta_{\mathrm{z}} &\,\leftarrow P(\zeta_{\mathrm{z}}&\,\mid &\,\dv,&\, &\,\tens{G}, &\,\xi_n, &
    \,\beta_{\mathrm{sky}},& \,\a_{\mathrm{sky}},
    &\,\phantom{\zeta_{\mathrm{z}},} &\,\a_{\mathrm{static}})\label{eq:gibbs_zodi}\\
    \a_{\mathrm{static}} &\,\leftarrow P(\a_{\mathrm{static}}&\,\mid &\,\dv,&\, &\,\tens{G}, &\,\xi_n, &
    \,\beta_{\mathrm{sky}},& \,\a_{\mathrm{sky}}, &\,\zeta_{\mathrm{z}} &\,\phantom{\a_{\mathrm{static}}})\label{eq:gibbs_static},
\end{alignat}
where $\leftarrow$ indicates the process of drawing a sample from the
distribution on the right-hand side. Each sampling step in this chain
is described by \citet{CG02_01} and references therein --- except for
Eq.~\eqref{eq:gibbs_zodi}, which is the main topic of this paper.

Based on the data model in Eq.~\eqref{eq:model}, we can define the
following residual
\begin{equation}
\r = \dv - (\G\P\B\sum_{c=1}^{n_{\mathrm{comp}}}\M_c\a_c +
          \s_{\mathrm{static}} + \n_\mathrm{corr}),
\end{equation}
and this should ideally only contain ZL and white instrumental noise.
As such, the assumption that this noise is also Gaussian defines the
conditional distribution in Eq.~\eqref{eq:gibbs_zodi}, and we may
write
\begin{equation}
  -2\ln P(\zeta_{\mathrm{z}}|\dv, \ldots) = \sum_{\nu}
  \left(\frac{\r_{\nu} -
    \s_{\nu,\mathrm{zodi}}(\zeta_\mathrm{s})}{\sigma_{\nu}}\right)^2 \equiv
  \chi^2 (\zeta_{\mathrm{z}}),
  \label{eq:gibbs_chisq}
\end{equation}
where we have introduced multiple data frequency channels, denoted by
$\nu$, and also for simplicity neglected the prior,
$P(\zeta_{\mathrm{z}})$. We also define the reduced chi-squared
$\chi^2_{\mathrm{red}}=\chi^2/n_{\mathrm{TOD}}$, where
$n_{\mathrm{TOD}}$ is the number of TOD samples included in the
likelihood evaluation. In this framework, ZL parameter estimation is
thus nothing but a traditional Gaussian $\chi^2$ optimization (or
sampling) problem after all non-ZL contributions have been subtracted
from $\dv$. Precisely how we implement this operation in the
current pipeline is described in Sect.~\ref{sec:nonlin}.

\begin{figure*}
    \centering
    \resizebox{\textwidth}{!}{%
    \includegraphics[height=1cm]{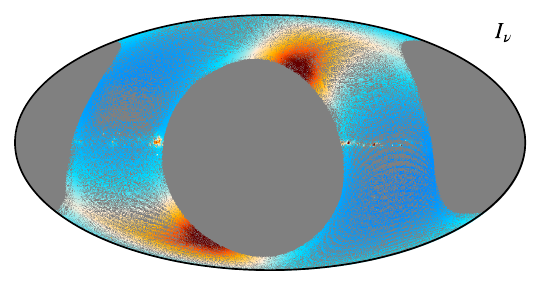}%
    \includegraphics[height=1cm]{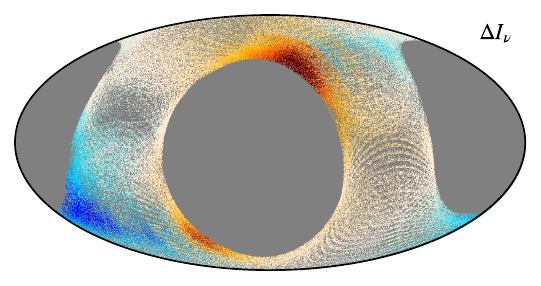}%
    }\\
    \resizebox{\textwidth}{!}{%
    \includegraphics[height=1cm]{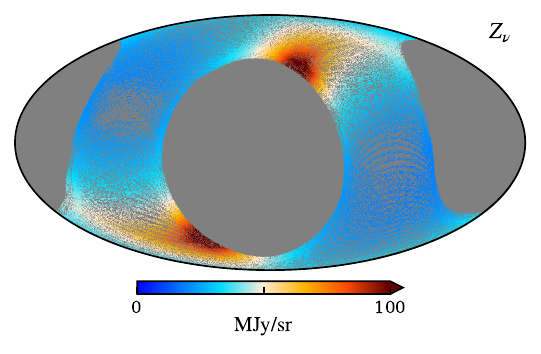}%
    \includegraphics[height=1cm]{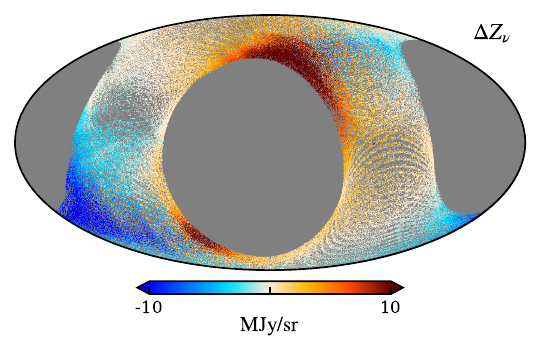}%
    }\\
    \resizebox{\textwidth}{!}{%
    \includegraphics[height=1cm]{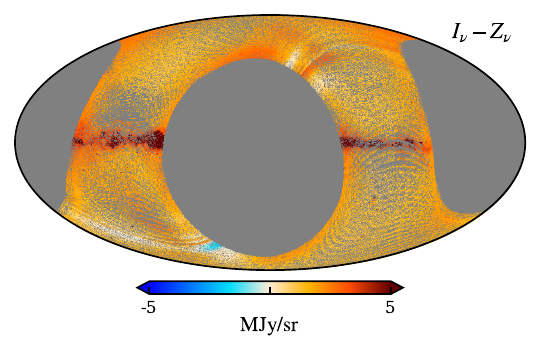}%
    \includegraphics[height=1cm]{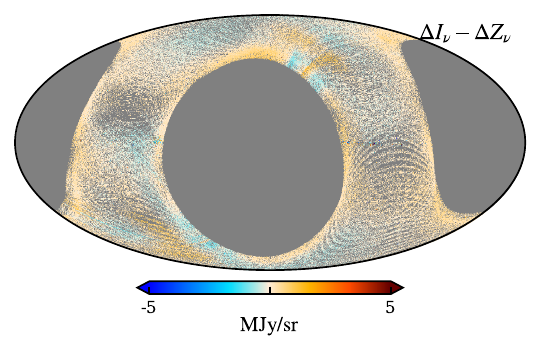}%
    }\\
    \caption{Illustration of the basic sky maps involved in the ZL 
    fitting algorithms adopted by the \Cosmoglobe\ (\emph{left column}) 
    and K98 (\emph{right column}) pipelines for one week of 
    $25\,\mu\mathrm{m}$ observations, both adopting the K98 model. The basic data element in 
    \Cosmoglobe\ is the full sky signal, $I_{\nu}$ (\emph{top left}), 
    which is fitted with the full ZL model (K98 in this case), $Z_{\nu}$ 
    (\emph{middle left}), both modeled in time-domain. The $\chi^2$ 
    used in the \Cosmoglobe\ analysis minimizes the total signal-minus-model residual, 
    $I_{\nu}-Z_{\nu}$ (\emph{bottom left}). In contrast, the 
    K98 pipeline used exclusively differences between weekly and 
    full-season maps, both for the observed signal, 
    $\Delta I_{\nu} \equiv I_{\nu}-\left<I_{\nu}\right>$ (\emph{top right}), 
    and the ZL model, $\Delta Z_{\nu} = Z_{\nu}-\left<Z_{\nu}\right>$ (\emph{middle right}), 
    where brackets indicate full-survey averages. Correspondingly, the 
    final $\chi^2$ is defined through $\Delta I_{\nu} - \Delta Z_{\nu}$ 
    (\emph{bottom right}), and is by construction only sensitive to 
    time-variable signals. The main advantage of the 
    K98 approach is insensitivity to stationary sky signals, in 
    particular thermal dust and CIB, while the main advantage of the 
    \Cosmoglobe\ approach is a much higher effective signal-to-noise 
    ratio, both to ZL parameters and zero-levels, as seen by comparing
    the two bottom panels.}
    \label{fig:week_vs_full}
\end{figure*}

So far we have silently skipped past one important term in
Eq.~\eqref{eq:model}, namely $\s_{\mathrm{static}}$, which is
discussed in detail by \citet{CG02_01}.  As noted by
\citet{hauser1998} and \citet{kelsall1998}, the DIRBE TOD contain
significant excess radiation that is not well described by the
low-dimensional parametric K98 model. Shortly after those
observations, \citet{leinert:1998} showed that some of this radiation
appeared to be stationary in solar-centric coordinates; see their
Fig.~54. Such radiation can in principle be created through several
different physical mechanisms. For instance, a yet unknown zodiacal
component could create a signal that appears stationary in
solar-centric coordinates, just like the circumsolar ring described
in Sect.~\ref{sec:ring}, or it could also be caused by stray-light
contamination in the DIRBE optics. However, even though this radiation
was first noted more than two decades ago, it was never mapped out
systematically until now, as part of the current \cosmoglobe\ DR2
analysis \citep{CG02_01}. For the time being, we choose to remain
agnostic regarding the physical origin of this signal, and therefore,
strictly speaking, the ZL model presented in the current paper only
describes the parts of the total observed ZL that is attributable to
the K98 parametrization. In the future, it is possible that the static
component presented by \citet{CG02_01} should also be added to this
model. However, before that is done, it is imperative to rule out the
stray-light hypothesis, and that will require both detailed modeling
of the DIRBE instrument and joint analysis with other experiments,
such as \AKARI, \IRAS, \Planck\ HFI, and SPHEREx. Doing that is beyond the scope
of the current \cosmoglobe\ data release, but it will certainly be a
main topic for future work.

Related to this, we also note that the circumsolar ring and trailing
feature discussed in Sect.~\ref{sec:ring} are completely degenerate
with a general pixelized static component in solar-centric
coordinates, and it is therefore not possible to fit these and the
excess radiation component simultaneously. For this reason, we fix the
circumsolar ring and trailing feature parameters at their K98 values,
and note that these will have to be revisited once a physical model
for the excess radiation has been established.

\subsection{Comparison with K98 fitting algorithm}
\label{sec:algorithm_comparison}

Before describing the practical numerical implementation used for
sampling from Eq.~\eqref{eq:gibbs_zodi} in this paper, it is worth
considering the more important fundamental differences between
our approach and that adopted by the DIRBE team as described by
\citet{kelsall1998}. The first difference worth noting in this respect
is that while our $\chi^2$ statistic is defined directly in terms of
TOD, the K98 parameter estimation method works with weekly maps. That
is, the raw data are co-added week-by-week into pixelized maps, and
these are fed into a corresponding pixel-based $\chi^2$ statistic. One
important motivation for working with weekly maps rather than single
TOD samples is lower computational requirements, which was more
important two decades ago than it is today.

A second important difference between the two algorithms --- and this
is conceptually a far more important one --- is the fact that while our
method makes active use of an explicit parametric data model for all
non-ZL components, the K98 algorithm eliminates any contributions from
non-ZL components by only considering differences between weekly maps
and the corresponding full-mission average map in their $\chi^2$
statistic. That is, rather than optimizing the full $\chi^2$ as
defined in Eq.~\eqref{eq:gibbs_chisq}, their algorithm optimizes
\begin{align}
  -2\ln P_{\mathrm{K98}}(\zeta_{\mathrm{z}}|\dv, \ldots) &= \sum_{i,\nu}
  \left(\frac{\Delta\dv_{i,\nu} -
    \Delta\s_{i,\nu,\mathrm{zodi}}(\zeta_\mathrm{s})}{\sigma_{\nu}}\right)^2 \nonumber\\
    &\equiv \chi^2_{\mathrm{K98}} (\zeta_{\mathrm{z}}),
  \label{eq:k98_chisq}
\end{align}
where $i$ indicate week number, $\Delta \dv_{i,\nu} = \dv_{i,\nu} -
\left<\dv_{i,\nu}\right>$, $\Delta \s_{i,\nu,\mathrm{zodi}} =
\s_{i,\nu,\mathrm{zodi}} - \left<\s_{i,\nu,\mathrm{zodi}}\right>$, and brackets
denote averaging over the full mission.

Clearly, this statistic has a key philosophical advantage as compared
to the full-signal statistic in Eq.~\eqref{eq:gibbs_chisq}: it does
not require any assumptions regarding the astrophysical nature of a 
complicated infrared sky. At least to first order, it is by
construction safe against biases from foreground modeling
errors. However, this bias immunity also comes at a massive cost in
terms of statistical uncertainties, because it is not only immune to
astrophysical bias, but it is also by construction blind to the monopole 
created when producing mission averaged maps, which can be as bright as 
~20 MJy/sr at the 25$\mu$m channel ZL signal, meaning that a lot of 
signal-to-noise is lost with this approach.

Figure~\ref{fig:week_vs_full} illustrates this difference. The top
left panel shows the full intensity signal as analyzed in the
\cosmoglobe\ algorithm for one single week of 25\,$\mu$m
observations. The visual imprint is strongly dominated by the ZL
features as seen in the total instantaneous ZL view in Appendix~\ref{sec:zodi-comps}, aligned with the Ecliptic
plane. The right panel shows the same after subtracting the
full-mission mean; the characteristic cloud pattern has now turned
into differential structures that are difficult to interpret
visually. The middle row shows the same for the ZL signal as predicted by the K98 model,
and the bottom row shows the difference between the top and middle
rows, which serves as the input to the $\chi^2$ evaluations.

Several points are worth noting in these figures. First of all, we
immediately note that the color scale is one order of magnitude narrower
in the right column than in the left column; this will translate
directly into lower constraining power for the differential
approach.

Second, we see that the Galactic plane signal represents a strongly
sub-dominant component in the total signal amplitude. Even relatively
large errors made in the model assumptions of these will have a very
small impact on the overall ZL estimates, and it is also
straightforward to mitigate this effect further by masking out any
samples that are close to known bright Galactic sources; this is fully
equivalent to what is done in the CMB field when estimating the CMB
power spectrum.

Third, as seen in the bottom panel, the $\chi^2_{\mathrm{K98}}
(\zeta_{\mathrm{z}})$ is also by construction entirely blind to the
zero-level of the ZL model: the large relative monopole error between
the data and the model seen in the bottom right panel is entirely
suppressed in the bottom left panel, and there is by construction no
way for the differential method to constrain the absolute level ZL
monopole. On the one hand, such blindness may certainly be considered
to be an algorithmic strength, as indeed argued by
\citet{kelsall1998}, given that one of the main goals of the entire
DIRBE mission was to precisely measure the CIB monopole spectrum
\citep{hauser1998}. Nevertheless, the final derived CIB constraints do
of course still depend directly on reconstructed ZL monopole, whatever
it may be. Intuitively speaking, the differential method aims to
measure the ZL monopole using derivatives alone. Whether that task is
easier or harder than to establish a sufficiently accurate model of
the Milky Way can only be determined by trying both methods, and
comparing the results. In addition, it is also worth noting that
derivative measurements in general are far more susceptible to
systematic biases from any non-Galactic source than absolute intensity
measurements. One important example in this respect is optical
non-idealities.

\subsection{Posterior sampling by non-linear optimization}
\label{sec:nonlin}

To complete the algorithm in Sect.~\ref{sec:gibbs}, we still need to
specify the details of the algorithm used to draw samples from
Eq.~\eqref{eq:gibbs_zodi}. A broad range of Bayesian sampling methods
can be envisioned for this purpose, from simple
Metropolis-Hastings (MH) accept-reject samplers to various
incarnations of Hamiltonian samplers that exploit derivative
information. Indeed, our very first implementation employed a simple
MH sampler with manually tuned step lengths, and this was used for
early model exploration and code testing. However, this approach was
quickly abandoned because it, after a short burn-in period, very
quickly got stuck in a local minimum, and all subsequent proposed
samples were rejected.

As noted already in Sect.~\ref{sect:zodi-model}, the main challenge
with the ZL posterior distribution is a large number of
degeneracies. These translate into a highly structured posterior
distribution with many local maxima, and it is generally difficult for
most iterative non-linear optimization or MCMC methods to map out such
parameter spaces efficiently.

The original K98 analysis used a standard Levenberg-Marquardt
algorithm to compute the best-fit model. This is a non-linear
optimization algorithm that essentially interpolates between a
Gauss-Newton and a gradient descent algorithm. As such, that analysis
was also susceptible to getting trapped in a local minimum.

In the current paper, we adopt a pragmatic approach to this problem
that uses ideas from both the MH and non-linear optimization
approaches. The non-linear optimization step is for now performed with
a simple Powell search, which uses a series of bi-directional line
searches along a set of conjugate directions. The only reason for
choosing this method over, say, Levenberg-Marquardt was one of
implementational ease; it does not require derivative information.

The algorithm is defined as follows: we start by initializing all free
ZL parameters at the values listed in \citet{kelsall1998} and \citet{Hong}, and run one Powell optimization to
reach the nearest local posterior maximum. We then iterate through all
non-ZL Gibbs steps, to obtain a new foreground and instrument model
given the new ZL model. When returning to the ZL sampler, we add a
Gaussian random fluctuation to each $\zeta_{\mathrm{z}}$ parameter,
which for now is defined by a root-mean square (RMS) of 1\,\% of its original value. We
then run a new Powell optimization to identify a new local ZL
parameter maximum. This new model is then accepted or rejected based
on the following Metropolis-like accept probability,
\begin{equation}
a = \mathrm{min}\left(e^{-\frac{1}{2}\frac{\chi^2_{\mathrm{red,new}}-\chi^2_\mathrm{red,old}}{\delta_{a}}},1\right),
\label{eq:accept}
\end{equation}
where $\delta_a$ is a tunable parameter that determines the strictness
of the rejection criterion. Explicitly, if a proposed
$\zeta_{\mathrm{z}}$ has a lower reduced chi-squared $\chi_{\mathrm{red}}^2$ than the
previous sample, it is always accepted; if it has a higher
$\chi_{\mathrm{red}}^2$, it is accepted with probability $a$. We
comment below on the reason this definition uses $\chi^2_{\mathrm{red}}$
rather than the absolute $\chi^2$, as well as how we set $a$.

In practice, many adjustments have been made gradually, over the course
of many months and hundreds of short individual test runs to the
\cosmoglobe\ DR2 data selection, analysis masks, and parametric data
model. In each case, the new job has been restarted on the previous
best-fit solution. This process of slow, gradual and continuously
supervised improvements also serves as a safe-guard against nonphysical
local posterior maxima, and effectively adds a meta-layer of simulated
annealing to the overall algorithm. Typically, the test runs were only
run for a few full Gibbs iterations, as in one to ten full
samples. Then, once the data and model configuration was considered
sufficiently mature for production, a longer run with hundreds of
samples distributed over multiple chains were produced.

\begin{figure}[t]
    \centering
    \includegraphics[width=\linewidth]{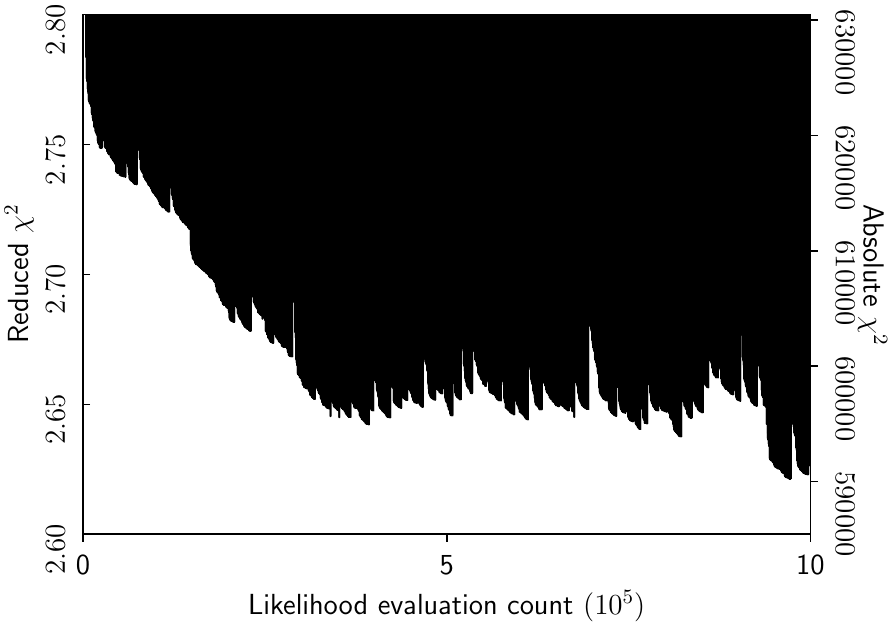}
    \caption{Reduced $\chi^2$ as a function of Powell likelihood evaluation count for one 
    single pre-production Gibbs chain, showing the burn-in phase. Each discrete jump indicates the 
    start of a new Gibbs sample, which is initialized on a new random point that is close to the previous iteration. 
    The following systematic decline within each main Gibbs iteration indicates the non-linear optimization 
    performed by the Powell algorithm. The solid dark region corresponds to a large number of highly 
    sub-optimal parameter trials. }
    \label{fig:powell_chisq_iter}
\end{figure}

\begin{figure}[t]
    \centering
    \includegraphics[width=\linewidth]{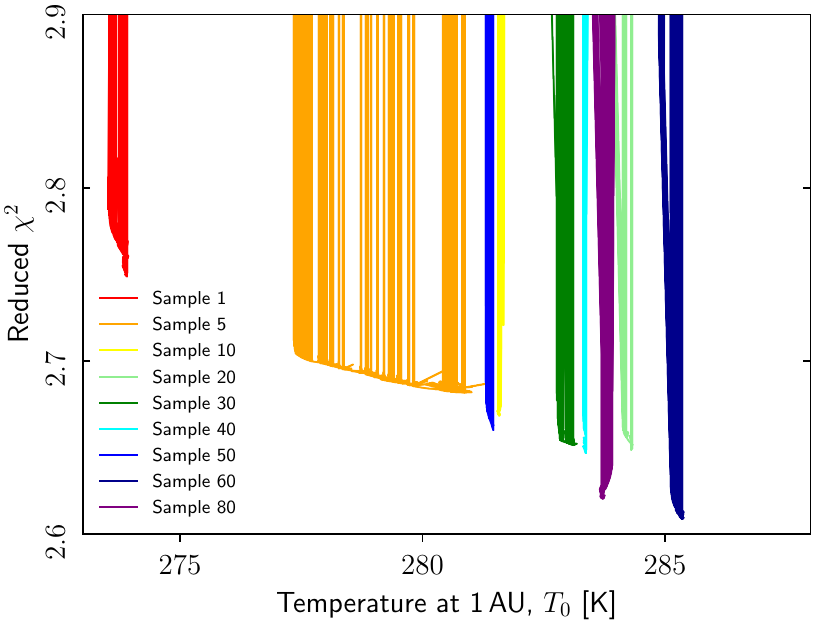}
    \caption{Reduced $\chi^2$ as a function of the temperature at 1\,AU, $T_0$, for the same run shown 
    in Fig.~\ref{fig:powell_chisq_iter}. Each curve shows the full set of parameter trials within one 
    single main Gibbs iteration (or Powell call), and different colors indicate different Gibbs iterations. 
    Redder colors are earlier in the chain.}
    \label{fig:powell_T0}
\end{figure}

The overall behavior of this hybrid MCMC+optimization algorithm is
illustrated in Fig.~\ref{fig:powell_chisq_iter}, which shows the
$\chi^2$ obtained from the first $10^6$ Powell likelihood evaluations
in a preliminary analysis run for a 12+25\,$\mu$m ZL sampling
group; we show the preliminary chain here rather than the final
production run to illustrate the burn-in phase. Each discrete jump
corresponds to one new main Gibbs iteration, while the smooth descent
between two jumps correspond to the Powell search. The dense black
``roof'' corresponds to a large number of very poor parameter
estimates made within each Powell search. Only the very last point in
each block is propagated forward to the rest of the Gibbs chain. Each
likelihood evaluation takes about 0.3\,sec wall-time on 128 cores, and
the evaluations shown in this figure therefore correspond to about
10\,000 CPU-hrs.

We see that the burn-in period appears to be roughly 20 main Gibbs
iterations, or about 300\,000 likelihood evaluations, after which the
chain appears reasonably stationary as measured in terms of
$\chi^2$. As for any MCMC sampler, the chain scatters up and down in
$\chi^2$ as it explores the overall parameter space, and sometimes
worse models than the previous best-fit are accepted.

Further intuition regarding this pseudo-MH algorithm and the
underlying multi-peaked likelihood surface is provided in
Fig.~\ref{fig:powell_T0}, which again shows $\chi^2_{\mathrm{red}}$
for a large number of likelihood evaluations. However, this time this
is plotted as a function of the overall reference temperature of the
IPD cloud, $T_0$, and also plotted separately for nine individual main
Gibbs samples; the sample numbers are color coded according to a
rainbow scheme, such that redder colors are closer to the beginning of
the Markov chain. Intuitively, the lower edge of each curve represents
a snapshot of the local neighborhood around each likelihood
minimum. We note that each such likelihood neighborhood is quite
sharp, indicating that the optimization algorithm is traversing a
rather narrow ``likelihood valley''. At the same time, different
samples reach a different absolute minimum level that is not a smooth
function of $T_0$; this is because of all the other parameters in
$\zeta_{\mathrm{z}}$ that are effectively marginalized over in this
figure. As far as burn-in and convergence is concerned, we see that
the Markov chain moves systematically from its initial value at
$T_0\sim273\,$K to $\sim$\,284\,K during the first 20 samples, and
after that it moves randomly between 282 and 286\,K. In sum, this
figure illustrates well the highly complex and multi-peaked likelihood
surface that must be explored when sampling ZL parameters.

\begin{figure}
    \centering
    \includegraphics[width=\columnwidth]{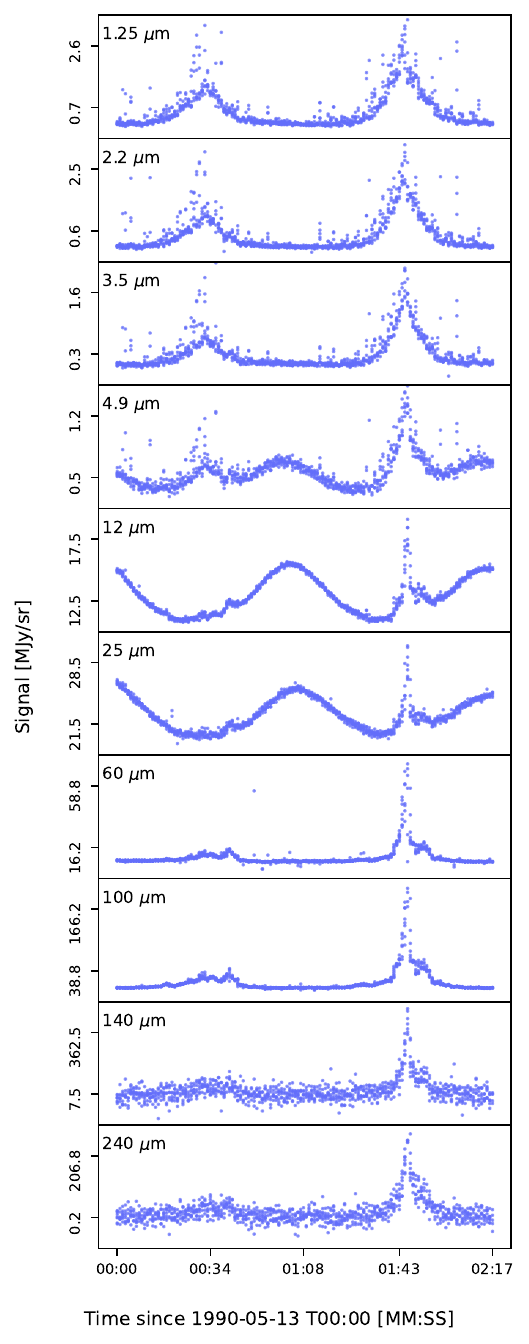}
    \caption{Subsample of the pre-processed TOD used in this analysis for all ten DIRBE bands. The time-streams
	show approximately one rotation of the \textit{COBE} satellite which includes two crossing of the Galactic and Ecliptic planes.}
    \label{fig:tod_zodi}
\end{figure}

\begin{figure}
    \centering
    \includegraphics[width=\columnwidth]{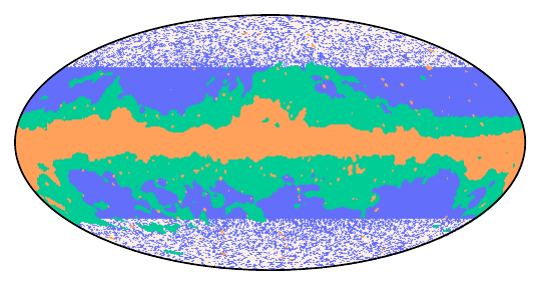}
    \caption{Three of the processing masks used when estimating ZL parameters. The blue mask is the 
    mask used in the stellar emission dominated 1.25 $\mathrm{\mu m}$ band, the orange mask is used 
    in the ZL dominated 25 $\mathrm{\mu m}$ band, and the green mask is used in the thermal dust dominated 
    240 $\mathrm{\mu m}$ band.}
    \label{fig:zodi-procmask}
\end{figure}

\begin{figure*}[t]
    \centering
    \includegraphics[scale=0.8]{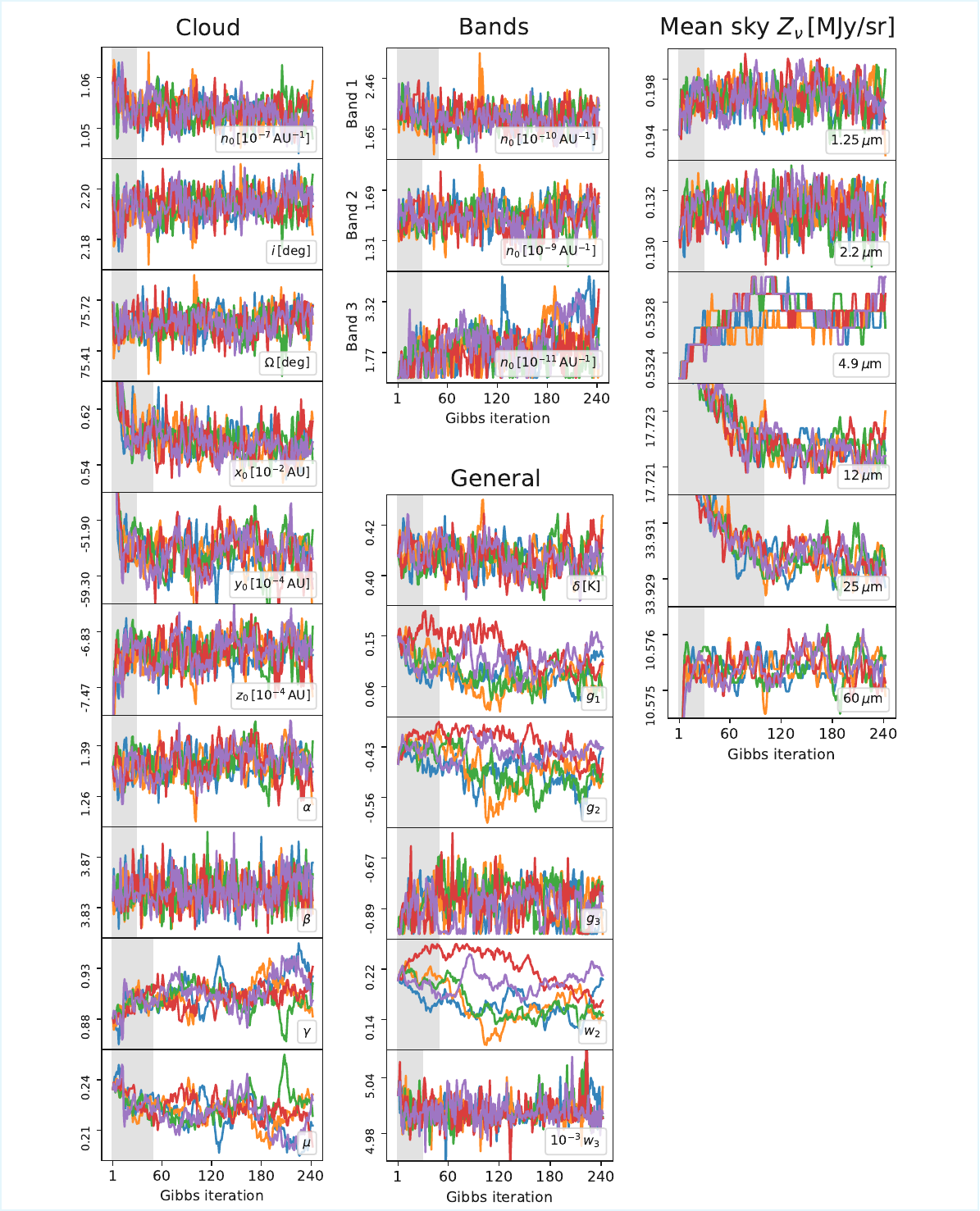}
    \caption{Traceplots of all the geometrical and shape IPD parameters kept free in 
    our model (\textit{Cloud} and \textit{Bands}), the power-law exponent of the IPD
    temperature and the phase function parameters (\textit{General}), along with the full-sky averaged 
    ZL intensity traces for six selected DIRBE channels (\textit{Mean sky $Z_\nu$}). 
    Different colors indicate five independent Markov chains, the gray regions indicate the burn-in set for each 
    parameter, and vertical axes were cropped to emphasize convergence.}
    \label{fig:trace-ipd}
\end{figure*}

\begin{figure*}
    \centering
    \includegraphics[width=0.985\textwidth]{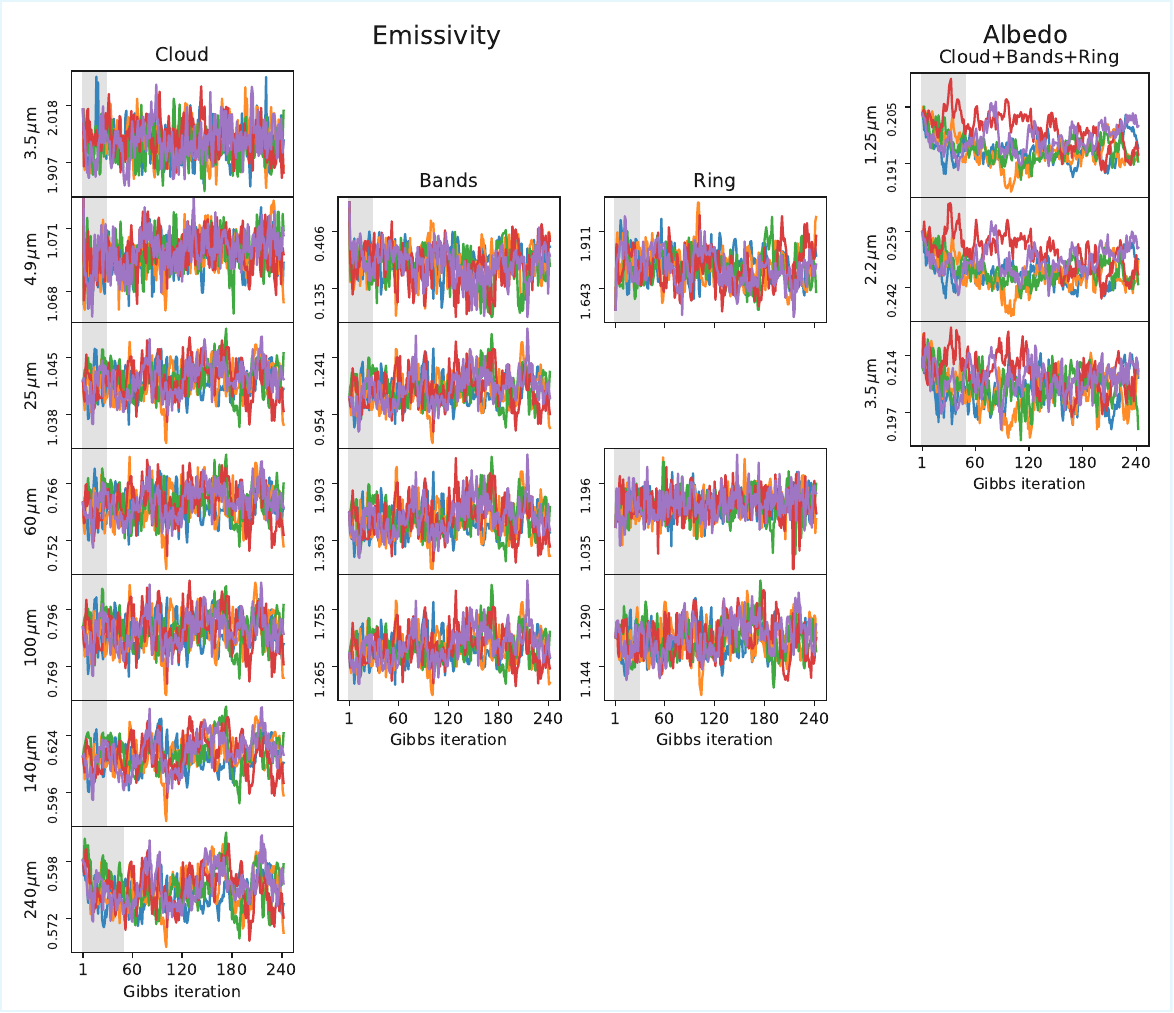}
    \caption{Traceplots of the wavelength dependent emissivity and albedo parameters. 
    Different colors indicate five independent Markov chains, the gray regions indicate the burn-in set for each 
    parameter, and vertical axes were cropped to emphasize convergence.}
    \label{fig:trace-emissivity-albedo}
\end{figure*}

\begin{figure}
    \centering
    \includegraphics[width=\linewidth]{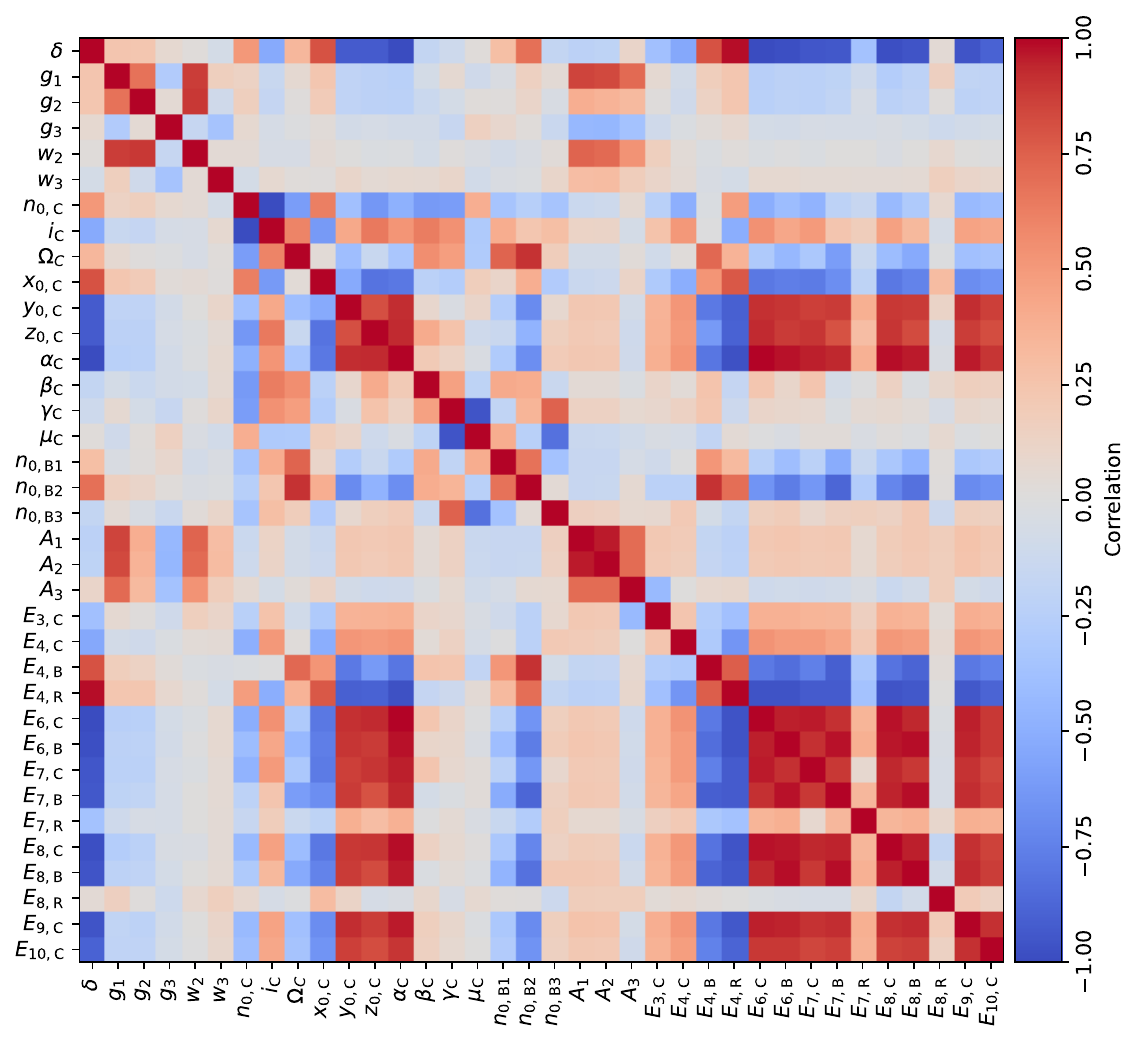}
    \caption{Correlations between the fitted ZL parameters $\zeta_{\mathrm{z}}$.}
    \label{fig:correlations}
\end{figure}

As noted above, we define the accept/reject criterion in terms of a
reduced $\chi^2_{\mathrm{red}}\equiv\chi^2/n_{\mathrm{TOD}}$, rather
than the absolute $\chi^2$, as dictated by the traditional
Metropolis-Hastings rule. The reason for this is illustrated in
Fig.~\ref{fig:powell_chisq_iter}, as both the absolute and reduced
$\chi^2$ are shown in the right and left $y$-axes. The point is simply
that the main uncertainties in these evaluations are not given by white
instrumental noise, but rather model errors and residual
systematics. As far as overall goodness of fit is concerned, all
models that are derived in the post-burn-in phase appear visually to be
equally good, despite the fact that the absolute $\chi^2$ varies with
$\mathcal{O}(10^5)$ from model to model. This quantity is therefore a
very poor measure for overall acceptability. In contrast, the reduced
$\chi^2$ varies with $\mathcal{O}(10^{-3})$ from sample to sample, and
that is a much more meaningful measure for overall
goodness of fit. For now, we conservatively set $\delta_a = 0.1$, and
the accept probability in Eq.~\eqref{eq:accept} is then primarily a
safe-guard against accepting pathologically poor models that lie
outside the range seen in Fig.~\ref{fig:powell_chisq_iter}. 

It is important to stress that this algorithm --- because of the
random jump proposal rule, the subsequent non-linear optimization
phase, and the accept rule based on a reduced $\chi^2$--- does not
formally satisfy the mathematical demands to a Monte Carlo Markov
Chain sampler in terms of ergodicity or reversibility. It is therefore
not guaranteed to converge to the true posterior distribution, even in
the limit of an infinite number of samples. Rather, this algorithm is
essentially simply a pragmatic solution that allows the previous strict
$\chi^2$ optimization algorithm used by \citet{kelsall1998} to evade
local posterior maxima, and explore larger regions of parameter space
without getting trapped. Significant additional algorithm development
efforts should be invested in establishing efficient methods for this
particular posterior distribution.

\section{Data}\label{sect:data}

\subsection{DIRBE Calibrated Individual Observations}
The main dataset used in the \cosmoglobe\ DR2 analysis is the
publicly available DIRBE Calibrated Individual Observations
(CIO). These are a user-friendly pre-calibrated version of the raw TOD
observed by DIRBE. The CIO are pixelized according to the \textit{COBE}
Quadrilateral cube projection. As part of the data prepossessing, we
convert these to corresponding HEALPix\footnote{\url{http://healpix.sourceforge.net/}} 
\citep{healpix} pixel indices and re-order
them into a time-ordered format. We note that the \cosmoglobe\ DR2
maps are binned with $7\arcm \times 7\arcm$ pixels, corresponding to a
HEALPix resolution of $N_{\mathrm{side}}=512$, which is substantially
higher resolution than the pixel size of $19\arcm \times 19\arcm$ used
in the original DIRBE analysis. For full details on the
preprocessing of the DIRBE CIO we refer to \cite{CG02_01}.

As already noted by \citet{kelsall1998}, a sharp edge may be seen in
the K98 ZSMA 25\,$\mu$m map at Galactic coordinates
$(l,b)\sim(190^{\circ},15^{\circ})$. This corresponds to the very end
of the DIRBE observing period. \citet{CG02_01} study this effect in
greater detail, and we find that this artifact may be mitigated by
omitting the last two weeks for the 1.25--3.5\,$\mu$m channels, and
the last month for the 4.9--100$\,\mu$m channels. A similar sharp
feature may be detected near the start of the survey as well, and we
therefore also remove the first week of observations in all ten frequency
channels. Other features that could be handled in a future release
include comet trails, which have been found at the 1\% level in the 12
and 25\,$\mu$m bands \citep{arendt}.

In addition, to minimize the impact of the excess radiation component,
\citet{CG02_01} define a set of masks in solar-centric coordinates for
each frequency channel between 1.25 and 100\,$\mu$m, and any TOD
sample that is excluded by these masks is removed from further
analysis. This approach plays exactly the same role as the solar
elongation cut used by the DIRBE team to produce their final ZSMA
maps, but provides much better precision in terms of removing specific
systematic features.

In the following, we will refer to the pre-processed CIO as TOD.  A
subsample of the TOD for each DIRBE band as used in our analysis can
be seen in Fig.~\ref{fig:tod_zodi}, where we show data for one
rotation of the spacecraft about its boresight. Within this rotation,
both the Galactic and Ecliptic planes are crossed twice. The two peaks
at around 30\,sec and 1\,min 45\,sec correspond to the Galactic plane
crossings, while the sine-like waves in bands 4.9, 12, and 25\,$\mu$m
are due to the ZL emission which peaks in the Ecliptic plane.

In order to trace seasonal ZL modeling errors, we divide the TOD for
each DIRBE channel into two, corresponding to the first and second
half of the full survey. All ZL parameters are fitted jointly and
simultaneously using all data, but the resulting ZL cleaned TOD are
binned separately into two independent half-mission maps, and the
corresponding half-mission difference maps therefore provide
information about seasonal ZL variations not captured by the model.

As summarized by Eq.~\eqref{eq:model}, the \cosmoglobe\ DR2 data model
includes a wide range of both astrophysical and instrumental
parameters. The DIRBE CIO are clearly not able to constrain this
model very well on their own, and we therefore include 
ancillary datasets. For full details, see
\citet{CG02_01,CG02_03,CG02_04,CG02_05,CG02_06,CG02_07}.

\subsection{Masks}
\label{sec:masks}

As discussed in Sect.~\ref{sec:algorithm_comparison}, the
\cosmoglobe\ DR2 parameter estimation algorithm considers the full
intensity measured by DIRBE when fitting $\zeta_{\mathrm{z}}$, as
opposed to only considering differences between weekly and
full-mission maps as \citet{kelsall1998} did. In order to minimize the
risk of confusion from Galactic thermal dust and starlight emission,
it is useful to define a confidence mask for each channel that
identifies regions with a high signal-to-noise ratio (S/N) for ZL, 
and low S/N for Galactic signals.

For the three channels between 1.25 and 3.5\,$\mu$m, which are
strongly dominated by starlight emission, we generate the ZL
confidence mask by thresholding the bright compact source model
evaluated at 1.25\,$\mu$m at 20\,kJy/sr. In addition we remove all
observations with an absolute Galactic latitude $|b|<45^{\circ}$. The
resulting mask is shown in blue in Fig.~\ref{fig:zodi-procmask}, and
leaves 18\,\% of the sky available for analysis.

For wavelengths longer than and equal to 60\,$\mu$m, the starlight
emission is negligible, and the Galactic signal is instead dominated
by thermal dust emission. At these channels, we therefore instead use
the sum of the three thermal dust component maps as the main Galactic
tracer. For instance, at 100$\,\mu$m we threshold this map at
3\,MJy/sr. In addition, we also remove any pixels for which the
absolute data-minus-model residual exceeds a given threshold, which
for the 100$\,\mu$m channel was set to 0.8\,MJy/sr. The orange and
green pixels in Fig.~\ref{fig:zodi-procmask} shows the final masks for
the 25 and 240\,$\mu$m channels, which leaves 81 and 52\,\% of the sky
available for analysis, respectively.

For the intermediate channels between 4.9 and 25\,$\,\mu$m, we
threshold on both diffuse and starlight emission. In all cases, we
overlay the final masks on the data-minus-model residual map, and
verify by eye that no obvious bright Galactic residuals remain after
masking. 

\section{Results}\label{sect:improved-model}

We are now finally ready to present our new ZL model obtained
by analyzing the data summarized in Sect.~\ref{sect:data} with the
algorithm outlined in Sect.~\ref{sect:param-estimation}. We have
produced a total of 1210 full Gibbs samples, distributed over five
Markov chains. A number of samples between 30 and 50, tuned on each parameter 
individually, was removed from each chain as burn-in, 
leaving a total number of samples between 1060 and 960 for final analysis. In total,
1920 computing cores were used to produce this sample set, for a
total cost of about 289k CPU-hrs.

\subsection{Markov chains}

We start our presentation with a visual inspection of the individual
Markov chains for each one of the 36 free ZL parameters in $\zeta_{\mathrm{z}}$. 
These are shown for the general ZL shape parameters in
Fig.~\ref{fig:trace-ipd} and for the amplitude (emissivity and albedo)
parameters in Fig.~\ref{fig:trace-emissivity-albedo}. The gray regions indicate the discarded burn-in, 
which, at this stage, was tuned for each parameter individually. 

Note that we are not attempting to fit the geometrical parameters of the three dust bands, 
but only their number densities. In particular, throughout our many attempts, 
we found that we were unable to constrain the signal in dust band 3 with the DIRBE data alone, 
as the band would effectively disperse. This phenomenon was previously noted by \citet{Spiesman}. 
Moreover, since the degeneracy between $T_0$ and emissivity created long 
correlations in previous test runs, here the temperature of IPD at 1\,AU was 
kept fixed at 286\,K. The same approach was already adopted by \citet{kelsall1998}. 
Finally, the 25\,$\mu$m emissivities for the ring and feature were fixed at the K98 value, 
as these are strongly degenerate with the solar-centric component discussed by \citet{CG02_01}.

Examining the traceplots helps to evaluate the degree of convergence for individual parameters. Starting
with the shape parameters for the dominant cloud component shown in
the leftmost column in Fig.~\ref{fig:trace-ipd}, we see that many
parameters appear to drift slightly during the first 30--50 samples or so, 
but after that, they become more stationary. We also
see that the mixing between different chains for $\gamma$ and $\mu$ is notably slower than
what is common for most Monte Carlo methods discussed in the
literature, and this indicates rather long correlation lengths. A similar behavior is observed for the 
phase function parameters, as shown in the bottom plots in the central column of Fig.~\ref{fig:trace-ipd}. Future
algorithm development work will aim at decreasing
these correlation lengths through better sampling algorithms, as well
as reducing the cost per Monte Carlo sample, such that longer Markov
chains can be produced. Still, even with the currently available
limited sample set, it does appear possible to derive sensible
estimates of both the posterior mean and standard deviation for most
parameters. It should be noted that the intial values for these parameters are the best-fit 
parameters of previous trial runs and not the final K98 parameters.

Looking at the dust bands' parameters (uppermost plots in the central column of Fig.~\ref{fig:trace-ipd}), 
it is worth noting that the number density of dust band 3 
seems to be reaching a prior bound. As already pointed out, with the DIRBE data this band is very faint 
and cannot be robustly measured. Therefore, we aim here to constrain an upper limit for it, 
rather than attempting to provide a point estimate. A similar effect characterizes the $g_3$ phase function 
parameter, which consequently appears to be prior-determinated.

Regarding the emissivity and albedo parameters shown in
Fig.~\ref{fig:trace-emissivity-albedo}, we see much of the same
quantitative behavior. A slight drift is always visible during burn-in, 
which was set to 30 samples for all parameters apart from the emissivity 
at 240\,$\mu$m and the albedoes that required more time to converge, 
but then all the traceplots seem quite stable. Long correlation lenghts are noticeable 
in the albedoes plots, which could benefit from the algorithmic improvements already noted.

Because of the combination of a very low absolute ZL signal-to-noise ratio of the
far-infrared DIRBE channels with strong diffuse Galactic thermal dust
emission, there are strong degeneracies between the
$\zeta_{\mathrm{z}}$ and the Galactic model, which are explored
rather inefficiently with the current Gibbs sampler. A future
implementation of this framework could consider fitting an overall
parametric function for the ZL SED that is smooth in wavelength,
rather than free amplitudes at each channel such that the low S/N
channels are supported by the stronger ZL channels in the mid-infrared
regime. 
In addition, we note that only the sum over all ZL
components matters for cosmological analysis, not each individual
component, and internal degeneracies are therefore of lower overall
concern.

Figure~\ref{fig:correlations} shows the Pearson correlation
coefficient between pairs of parameters in $\zeta_{\mathrm{z}}$,
evaluated directly from the post-burn-in Gibbs sample set. Here, the burn-in was set to 
50 samples to match the longest convergence period among all parameters, ensuring that an 
equal number of post-burn-in samples was used for all correlations.
This plot is highly structured, reflecting the complexity of the ZL likelihood. 
Particularly strong are the correlations between the emissivity parameters in the 60--240\,$\mu$m channels. 
As already noted above, fitting a smooth parametric ZL SED function rather than individual amplitude
factors at each channel should be beneficial in many ways; both in
terms of physical interpretation, but also in terms of overall
sampling efficiency in a Gibbs sampler. 

\begin{figure}[t]
    \centering
    \includegraphics[width=\columnwidth]{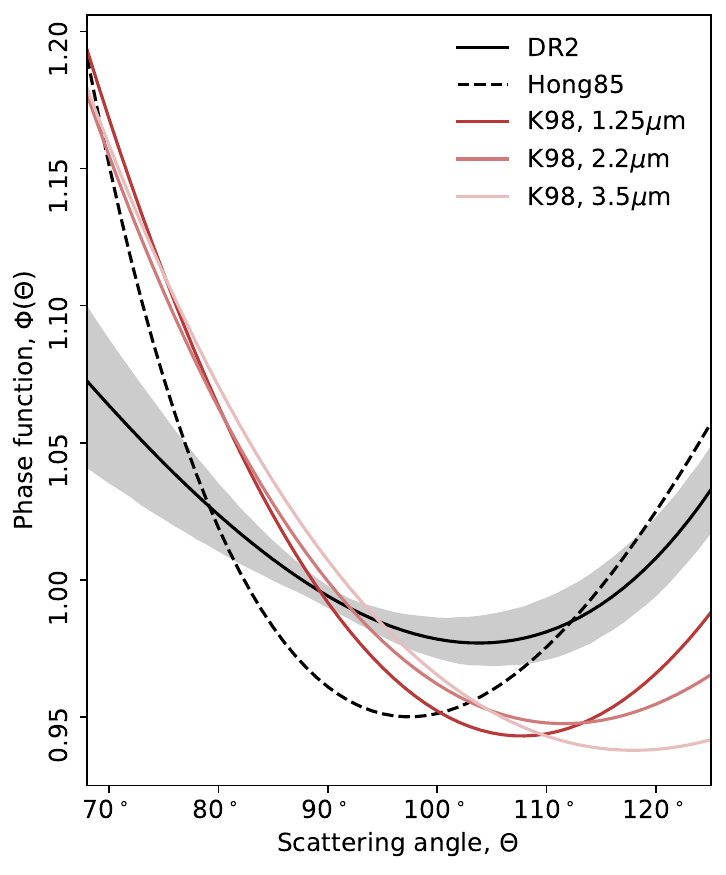}
    \caption{Comparison of different parametrizations of the phase function $\Phi(\Theta)$, normalized 
    over the solar elongation angle interval observed by DIRBE ($\Theta \in [65^\circ,\,125^\circ]$). 
    The red lines represent $\Phi(\Theta)$ as defined by the original K98 model 
    in the frequency channels 1.25 -- 3.5\,$\mu$m; the black dashed line is the phase function 
    defined in \citet{Hong}; while the solid black line is the DR2 function. This final one 
    follows the definition of Eq.~\ref{eq:phase_function} and uses all the post-burn-in parameters 
    to obtain the mean value and the standard deviation of $\Phi$, 
    shown as a line and a grey area respectively in the figure.
    }
    \label{fig:phase_function}
\end{figure}

\begin{figure}[t]
    \centering
    \vspace{1pt}
    \includegraphics[width=\columnwidth]{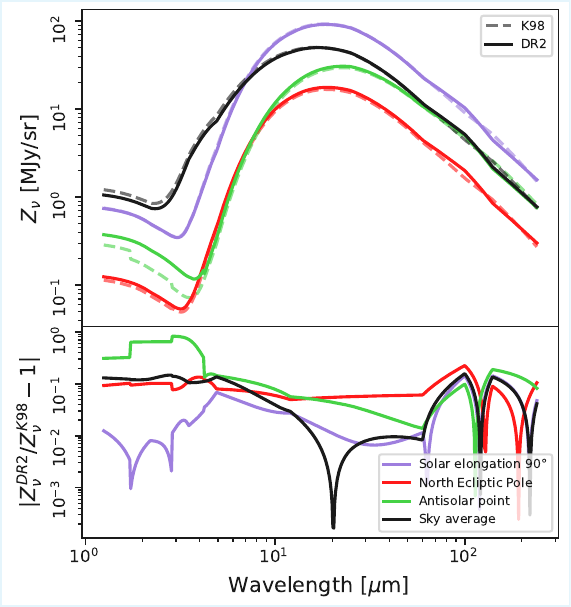}
    \caption{(\textit{Top panel}:) ZL intensity as a function of wavelength on the 2024-01-01
      as predicted by the DR2 (solid lines) and K98 (dashed lines)
      models. Colors show three different positions on the sky with Galactic coordinates 
      ($l=39.91^\circ,\,b=-58.70^\circ$), ($l=96.38^\circ,\,b=29.81^\circ$), and ($l=191.41^\circ,\,b=8.68^\circ$) 
      corresponding to the solar elongation at $90^\circ$, the North Ecliptic Pole, 
      and the antisolar point, respectively. The black curves instead show the full-sky average. 
      (\textit{Bottom}:) Fractional difference between the DR2 and 98 models, 
      evaluated from the results shown in the top panel.
    }
    \label{fig:zodi-intensity}
\end{figure}

\begin{figure}[t]
    \centering
    \includegraphics[width=0.9\columnwidth]{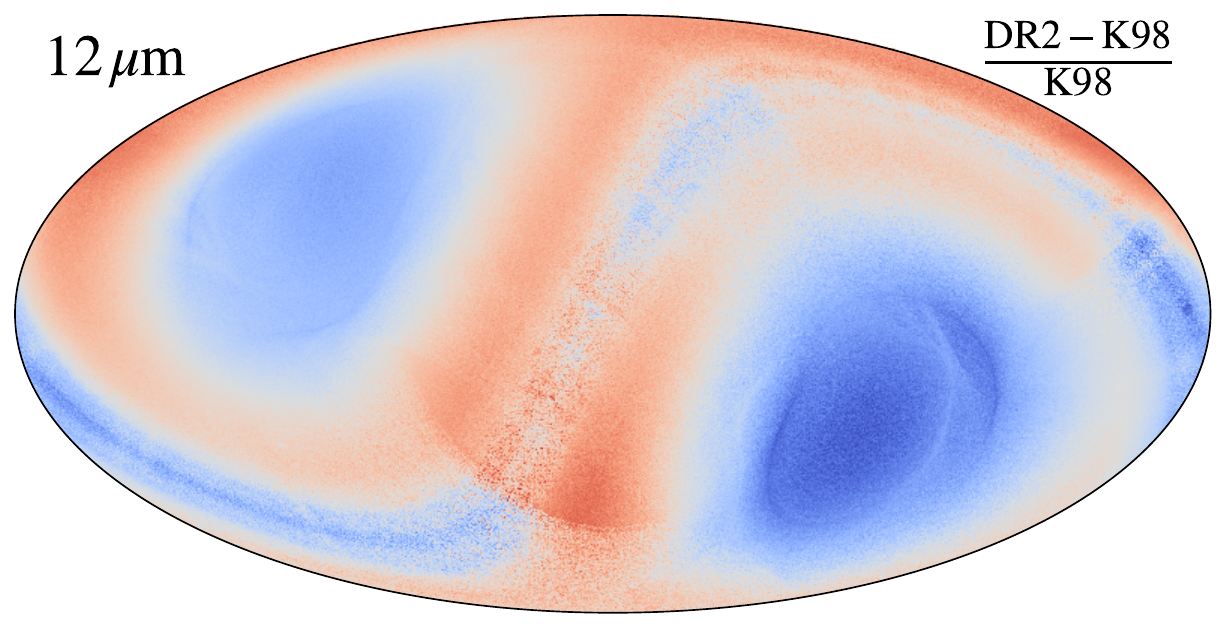}\\
    \includegraphics[width=0.9\columnwidth]{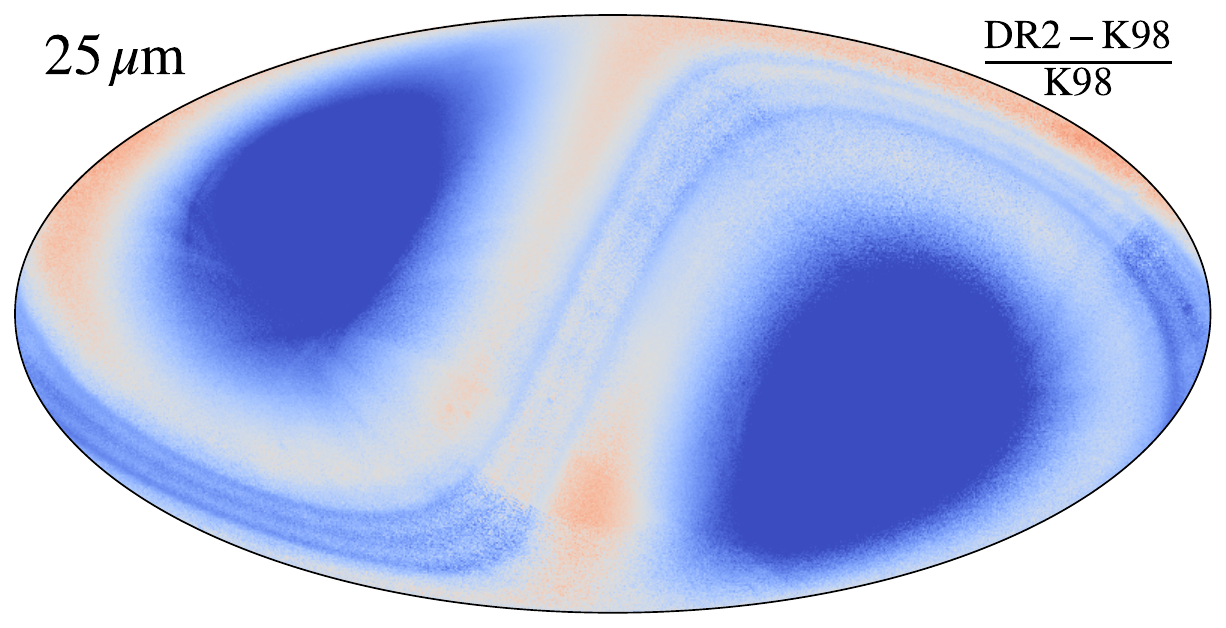}
    \includegraphics[width=0.55\columnwidth]{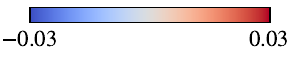}
    \caption{Relative fractional difference between mission averaged ZL simulations 
    of our best-fit model and K98 at 12 and 25\,$\mu$m. To highlight structural zodiacal light 
    differences, an estimated value of the ZL monopole (as evaluated in Fig.~\ref{fig:zsma_mean}) 
    was subtracted from the DR2 model in both channels before the comparison was carried out.}
    \label{fig:reldiff}
\end{figure}

\subsection{Updated ZL model}

To estimate the marginal posterior distribution for each ZL parameter
individually, we compute the mean and standard deviation of all
post-burn-in samples. For each parameter the number of burn-in samples
was tuned individually, as shown in Figs.~\ref{fig:trace-ipd} and
\ref{fig:trace-emissivity-albedo}. The resulting values for the shape
parameters are tabulated in the fourth column of
Table~\ref{table:zodi-params-geo}, the amplitude parameters are
provided in Table~\ref{table:zodi-params-source}. For comparison, the
corresponding values from \citet{kelsall1998} are listed in the third
column in each table. The fifth and final column lists the uniform
priors imposed on each parameter in the DR2 analysis. Corresponding
phase function parameters can be found in
Table~\ref{table:zodi-params-phase}, and are in this case compared with \citet{Hong}.

\renewcommand{\arraystretch}{1.5} %
\begin{table*}[!h]
    \small
    \centering
    \caption{Best-fit interplanetary dust parameter estimates and uncertainties in the DR2 analysis,
      comparing values with the K98 model. Parameters that are not listed are fixed at the respective K98 values.}
   \label{table:zodi-params-geo}
    \newcolumntype{C}{ @{}>{${}}r<{{}$}@{} }
    \begin{tabular}{l l *2{rCl}c}
    \hline
    \hline
     Parameter & Description & \multicolumn{3}{c}{K98} & \multicolumn{3}{c}{DR2} & Uniform prior\\ 
     \hline
     \multicolumn{9}{c}{Smooth Cloud}\\
     \hline
     $n_{0, \mathrm{C}}$ [$10^{-8}$ AU$^{-1}$]\dotfill& Number density at 1 AU & 11.3 &\pm& 0.1 & 10.56 & \pm & 0.03 & [$10^{-3}$, $10^3$]\\
     $\alpha$\dotfill& Radial power-law exponent \quad& 1.34 &\pm& 0.02 & 1.34 & \pm & 0.04 & [1, 2]\\
     $\beta$\dotfill& Vertical shape parameter & 4.14 &\pm& 0.07 & 3.84 & \pm & 0.01 & [3, 5]\\
     $\gamma$\dotfill& Vertical power-law exponent & 0.94 &\pm& 0.03 & 0.91 & \pm & 0.02 & [0.3, 1.1]\\
     $\mu$\dotfill& Widening parameter & 0.189 &\pm& 0.014 & 0.223 & \pm & 0.009 & [0.1, 0.4] \\
     $i$ [deg]\dotfill& Inclination & 2.03 &\pm& 0.02 & 2.195 & \pm & 0.007 & [$-30$, 30]\\
     $\Omega$ [deg]\dotfill& Ascending node & 77.7 &\pm& 0.6 & 75.6 & \pm & 0.1 & [$-720$, 720]\\
     $x_0$ [$10^{-2}$ AU]\dotfill& x-offset from the Sun  & 1.2 &\pm& 0.1 & 0.57 & \pm & 0.02 & [$-4$, 4]\\
     $y_0$ [$10^{-2}$ AU]\dotfill& y-offset from the Sun &  0.55 &\pm& 0.8 & $-0.56$ & \pm & 0.02 & [$-2$, 2]\\
     $z_0$ [$10^{-3}$ AU]\dotfill& z-offset from the Sun & $-2.2$ &\pm& 0.4 & $-0.71$ & \pm & 0.02 & [$-20$, 20]\\
     \hline
     \multicolumn{9}{c}{Dust bands}\\
     \hline
     $n_{0, \mathrm{B}_1}$ [$10^{-10}$ AU$^{-1}$]\dotfill& Number density at 1 AU for dust band 1 & 5.6 &\pm& 0.7 & 1.8 & \pm & 0.2 & [0.1, $10^5$]\\
     $n_{0, \mathrm{B}_2}$ [$10^{-9}$ AU$^{-1}$]\dotfill& Number density at 1 AU for dust band 2 & 1.99 &\pm& 0.13 & 1.49 & \pm & 0.10 & [0.1, $10^5$]\\
     $n_{0, \mathrm{B}_3}$ [$10^{-10}$ AU$^{-1}$]\dotfill& Number density at 1 AU for dust band 3 & 1.4 & \pm & 0.2 & \multicolumn{3}{c}{< 0.32} & [0.1, $10^5$] \\
     \hline
     &&&&&&\\
    \end{tabular}
    \end{table*}

\begin{table*}[!h]
    \small
    \centering
    \caption{Best-fit estimates and uncertainties of the source-function parameters in the DR2 analysis,
      comparing values with the K98 model. Parameters that are not listed are fixed at the respective K98 values.
    }
    \label{table:zodi-params-source}
    \newcolumntype{C}{ @{}>{${}}r<{{}$}@{} }
    \begin{tabular}{l l *2{rCl}c}
    \hline
    \hline
    Parameter & Description & \multicolumn{3}{c}{K98} & \multicolumn{3}{c}{DR2} & Uniform prior\\ 
    \hline
    \multicolumn{9}{c}{All zodiacal components}\\
    \hline
    $\delta$ \dotfill & Temperature power-law exponent  & 0.467 &\pm& 0.004 & 0.406& \pm & 0.009 & [0.3, 0.5]\\
    $A_1$ \dotfill & Albedo at 1.25$\mu $m & 0.204 &\pm& 0.001 & 0.195 &\pm& 0.004 & [0, 1]\\
    $A_2$ \dotfill & Albedo at 2.2$\mu $m & 0.255 &\pm& 0.002 & 0.248 &\pm& 0.005 & [0, 1]\\
    $A_3$ \dotfill & Albedo at 3.5$\mu $m & 0.21 &\pm& 0.02 & 0.205 &\pm& 0.006 & [0, 1]\\
    
    \hline
    \multicolumn{9}{c}{Smooth Cloud}\\
    \hline

    $E_1$\dotfill & Emissivity at 1.25$\mu $m  & 1 && Fixed & 1 && Fixed &\\
    $E_2$\dotfill & Emissivity at 2.2$\mu $m  & 1 && Fixed & 1 && Fixed &\\
    $E_3$\dotfill & Emissivity at 3.5$\mu $m  & 1.66 &\pm& 0.09 & 1.95 &\pm& 0.03 & [0, 10] \\
    $E_4$\dotfill & Emissivity at 4.9$\mu $m  & 0.997 &\pm& 0.004 & 1.070 &\pm& 0.001 & [0, 10]\\
    $E_5$\dotfill & Emissivity at 12$\mu $m  & 0.958 &\pm& 0.003 & 1 &&  Fixed & \\
    $E_6$\dotfill & Emissivity at 25$\mu $m  &  1 && Fixed & 1.042 &\pm& 0.002 & [0, 10]\\
    $E_7$\dotfill & Emissivity at 60$\mu $m  & 0.733 &\pm& 0.006 & 0.761 &\pm& 0.004 & [0, 10]\\
    $E_8$\dotfill & Emissivity at 100$\mu $m  & 0.647 &\pm& 0.012 & 0.786 &\pm& 0.008 & [0, 10]\\
    $E_9$\dotfill & Emissivity at 140$\mu $m  & 0.677 &&  & 0.615 &\pm& 0.008 & [0, 10]\\
    $E_{10}$\dotfill & Emissivity at 240$\mu$m  & 0.519 &&  & 0.586 &\pm& 0.008 & [0, 10]\\
    \hline
    \multicolumn{9}{c}{Dust bands}\\
    \hline
    $E_1$\dotfill & Emissivity at 1.25$\mu $m  & 1 && Fixed & 1 && Fixed &\\
    $E_2$\dotfill & Emissivity at 2.2$\mu $m  & 1 && Fixed & 1 && Fixed &\\
    $E_3$\dotfill & Emissivity at 3.5$\mu $m  & 1.66 && Fixed to cloud & 1.95 &&  Fixed to cloud & \\
    $E_4$\dotfill & Emissivity at 4.9$\mu $m  & 0.36 &\pm& 0.05 & 0.25 &\pm& 0.08 & [0, 10]\\
    $E_5$\dotfill & Emissivity at 12$\mu $m  & 1.0 &\pm& 0.2 & 1 && Fixed &\\
    $E_6$\dotfill & Emissivity at 25$\mu $m  & 1 && Fixed & 1.09 &\pm& 0.08 & [0, 10]\\
    $E_7$\dotfill & Emissivity at 60$\mu $m  & 1.3 &\pm& 0.3 & 1.6  &\pm& 0.2 & [0, 10]\\
    $E_8$\dotfill & Emissivity at 100$\mu $m  & 1.5 &\pm& 0.65 & 1.4 &\pm& 0.1 & [0, 10]\\
    $E_9$\dotfill & Emissivity at 140$\mu $m  & 1.13 && & 0.615 && Fixed to cloud\\
    $E_{10}$\dotfill & Emissivity at 240$\mu $m  & 1.40 && & 0.586 && Fixed to cloud\\
    \hline
    \multicolumn{9}{c}{Ring and Trailing Feature}\\
    \hline
    $E_1$\dotfill & Emissivity at 1.25$\mu $m  & 1 && Fixed & 1 && Fixed &\\
    $E_2$\dotfill & Emissivity at 2.2$\mu $m  & 1 && Fixed & 1 && Fixed &\\
    $E_3$\dotfill & Emissivity at 3.5$\mu $m  & 1.66 && Fixed to cloud & 1.95 &&  Fixed to cloud & \\
    $E_4$\dotfill & Emissivity at 4.9$\mu $m  & 1.06 &\pm& 0.09 & 1.74 &\pm& 0.08 & [0, 10]\\
    $E_5$\dotfill & Emissivity at 12$\mu $m  & 1.06 &\pm& 0.01 & 1 && Fixed &\\
    $E_6$\dotfill & Emissivity at 25$\mu $m  & 1 && Fixed & 1 && Fixed &\\
    $E_7$\dotfill & Emissivity at 60$\mu $m  & 0.873 &\pm& 0.004 & 1.14  &\pm& 0.04 & [0, 10]\\
    $E_8$\dotfill & Emissivity at 100$\mu $m  & 1.10 &\pm& $7 \times 10^{-6}$ & 1.22 &\pm& 0.04 & [0, 10]\\
    $E_9$\dotfill & Emissivity at 140$\mu $m  & 1.13 && & 0.615 && Fixed to cloud\\
    $E_{10}$\dotfill & Emissivity at 240$\mu $m  & 1.40 && & 0.586 && Fixed to cloud\\
    \hline
    &&&&&&\\
    \end{tabular}
\end{table*}

\begin{table}
  \small
  \centering
  \caption{Best-fit phase function parameter estimates and uncertainties in the DR2 analysis,
    comparing values with those listed in \citet{Hong}.}
 \label{table:zodi-params-phase}
  \newcolumntype{C}{ @{}>{${}}r<{{}$}@{} }
  \begin{tabular}{l c rCl c}
  \hline
  \hline
   Parameter & Hong & \multicolumn{3}{c}{DR2} & Uniform prior\\ 
   \hline
   \multicolumn{6}{c}{Asymmetry factors}\\
   \hline
   $g_1$\dotfill& 0.70  & 0.09 & \pm & 0.03 & [$-1$, 1]\\
   $g_2$\dotfill& $-0.20$  & $-0.47$ & \pm & 0.05 & [$-1$, 1]\\
   $g_3$\dotfill& $-0.81$  & $-0.86$ & \pm & 0.09 & [$-1$, 1]\\
   \hline
   \multicolumn{6}{c}{Weights}\\
   \hline
   $w_2$\dotfill& 0.330  & 0.18 & \pm & 0.03 & [0, 1]\\
   $w_3$\dotfill& 0.005  & 0.005 & \pm & $1 \times 10^{-5}$ & [0, 1]\\
   \hline
   &&&&&\\
\end{tabular}
\end{table}

\begin{figure*}[t]
    \centering
    \includegraphics[width=0.22\linewidth]{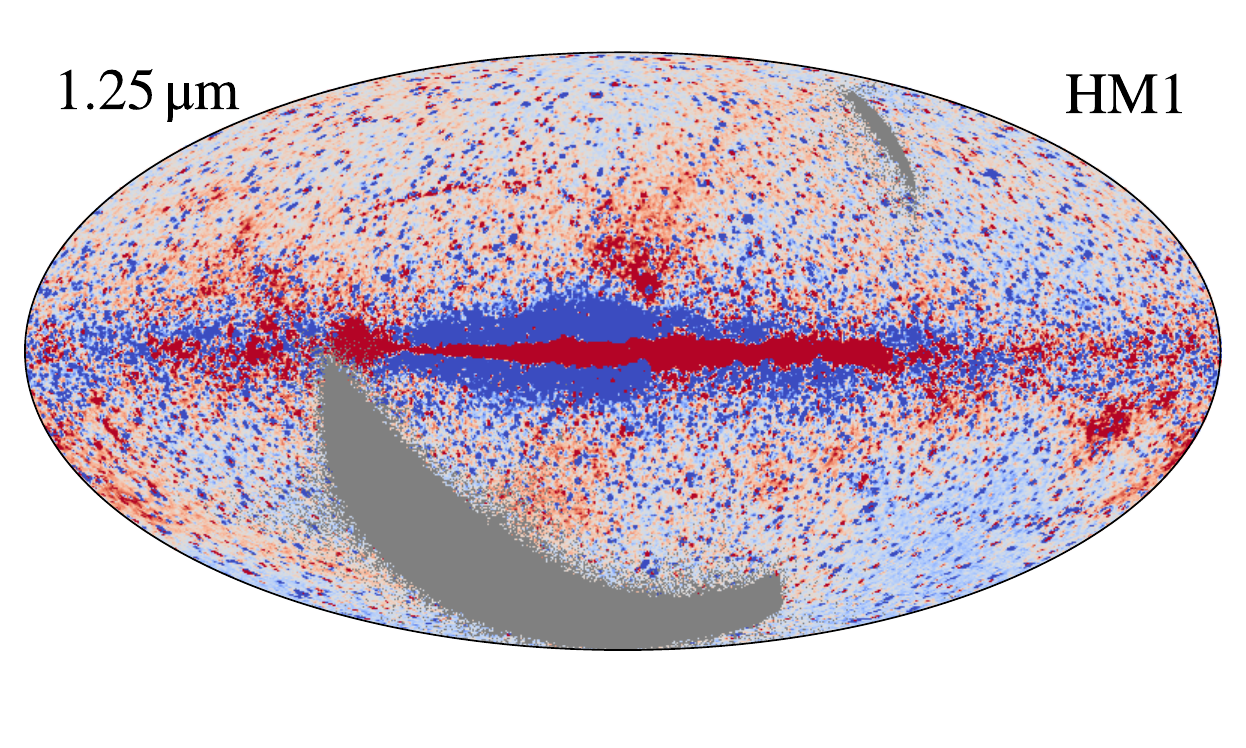}%
    \includegraphics[width=0.22\linewidth]{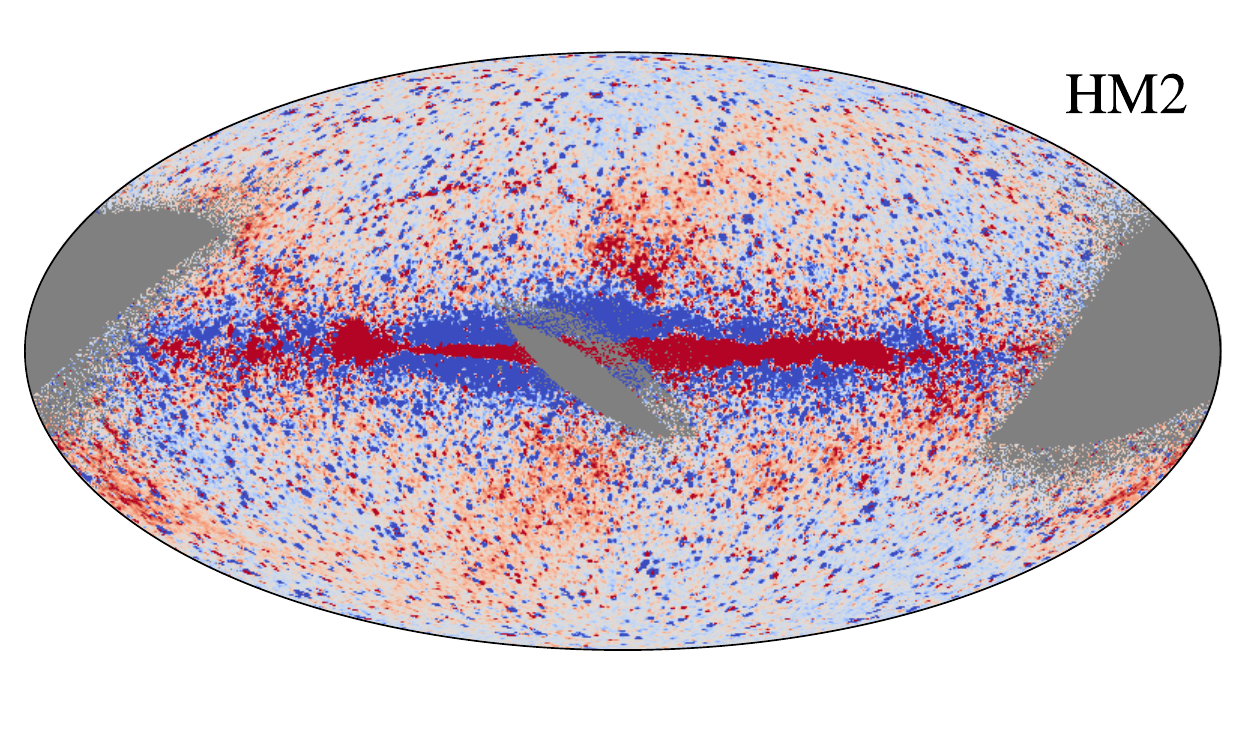}%
    \includegraphics[width=23mm,angle=90]{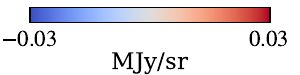}\hspace*{3mm}
    \includegraphics[width=0.22\linewidth]{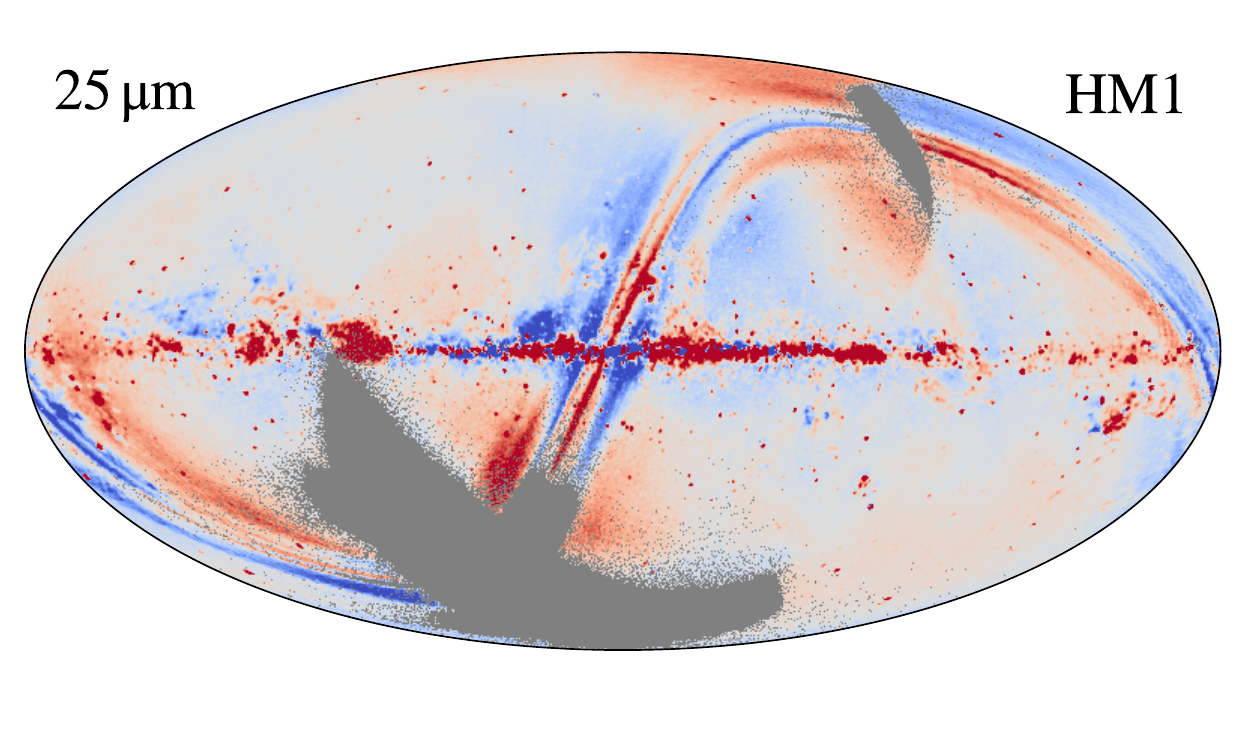}%
    \includegraphics[width=0.22\linewidth]{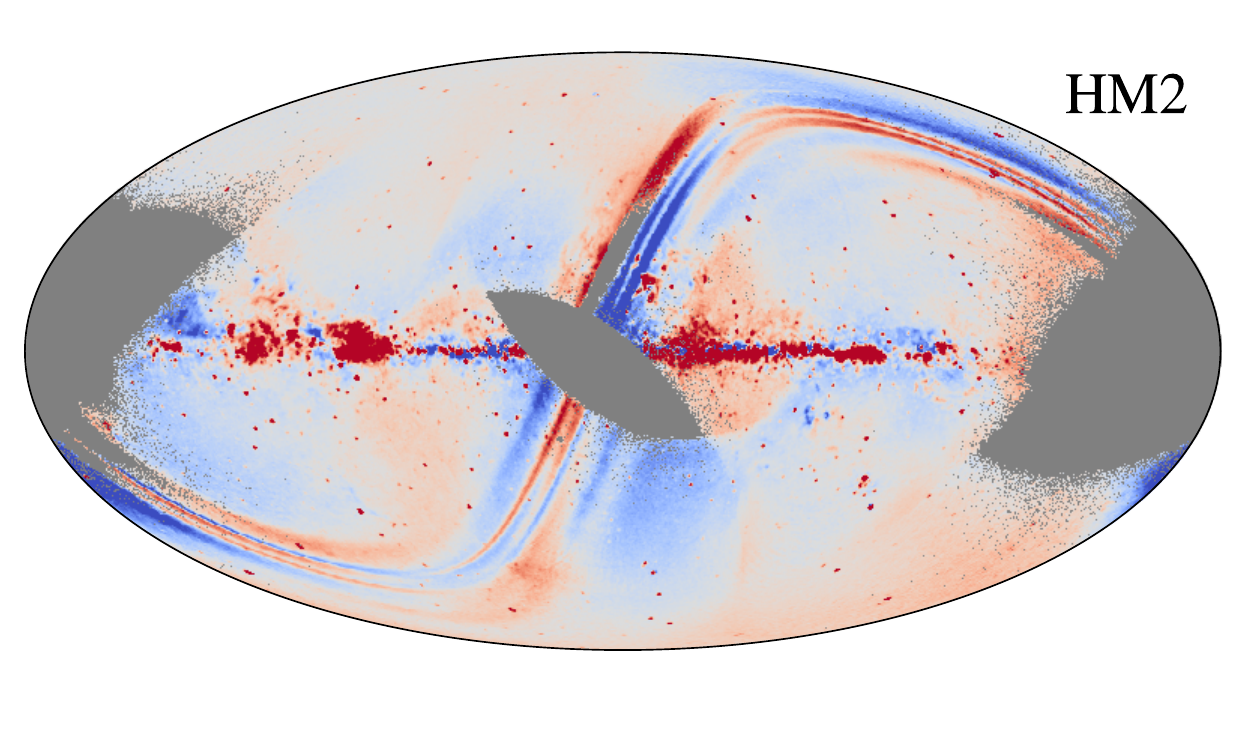}%
    \includegraphics[width=23mm,angle=90]{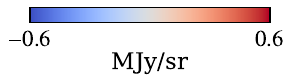}\\
    \includegraphics[width=0.22\linewidth]{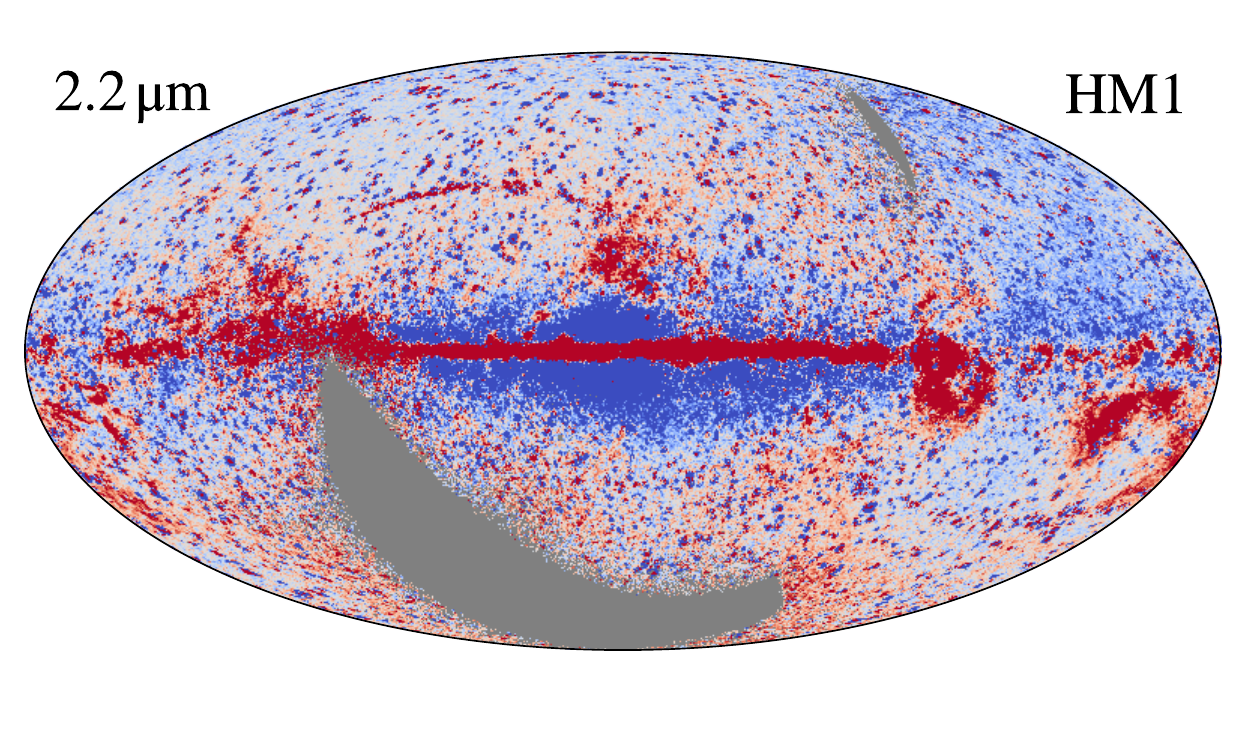}%
    \includegraphics[width=0.22\linewidth]{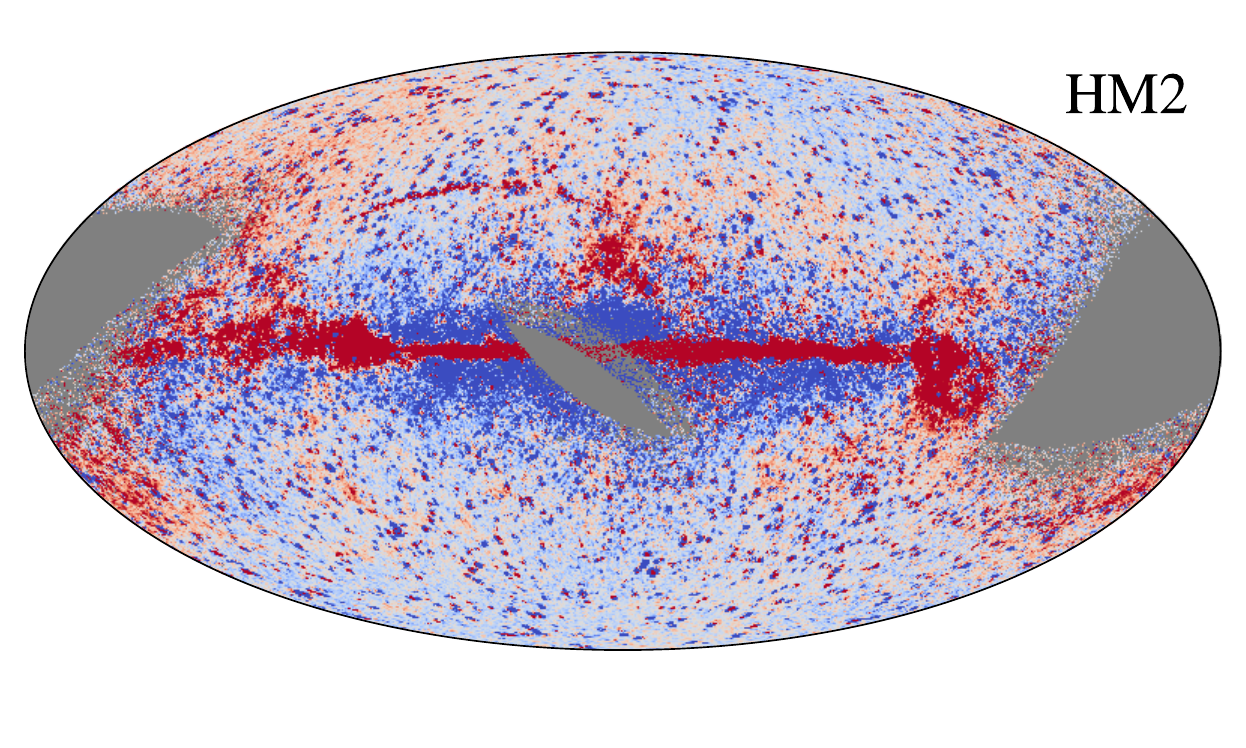}%
    \includegraphics[width=23mm,angle=90]{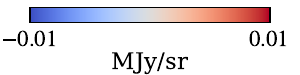}\hspace*{3mm}
    \includegraphics[width=0.22\linewidth]{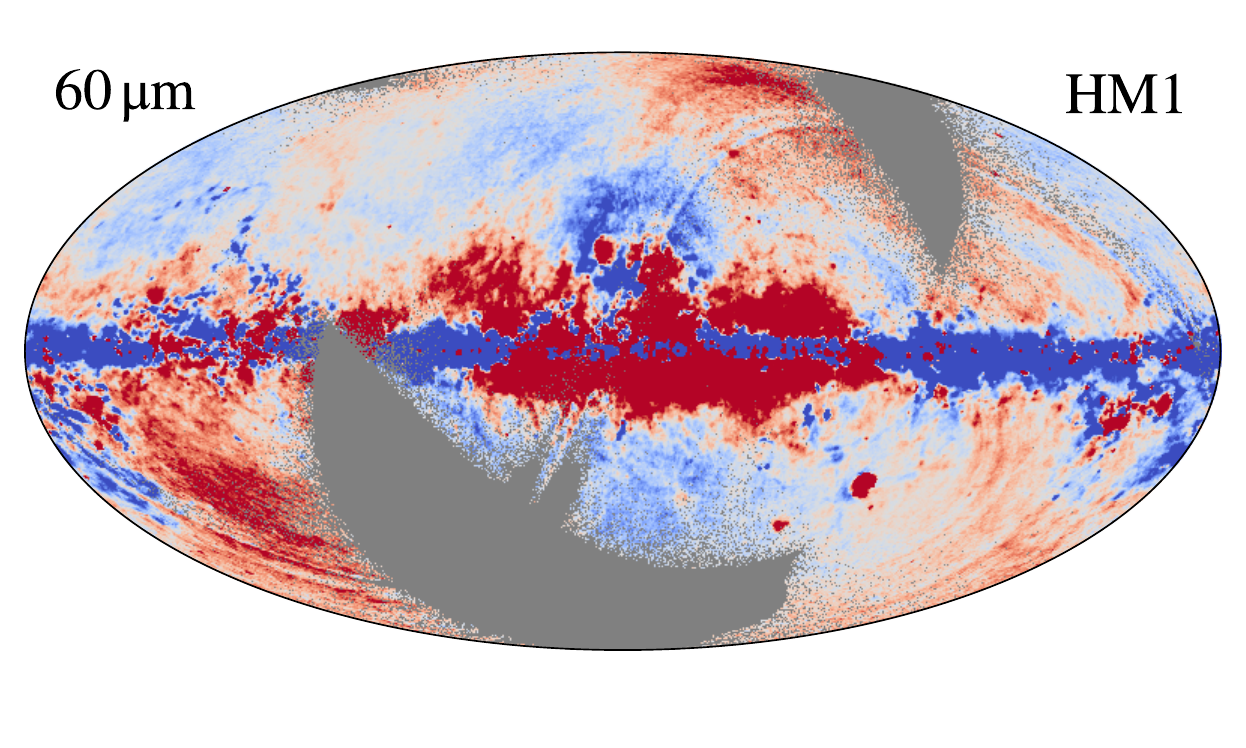}%
    \includegraphics[width=0.22\linewidth]{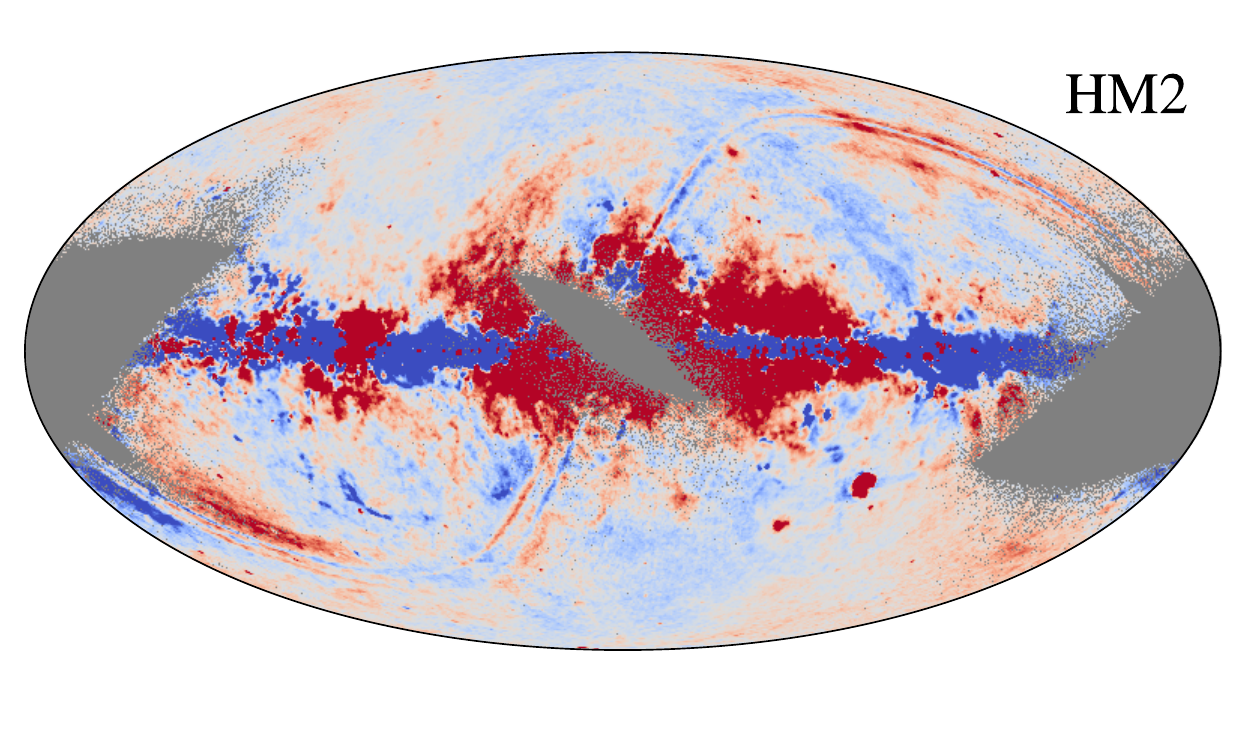}%
    \includegraphics[width=23mm,angle=90]{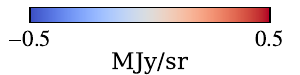}\\
    \includegraphics[width=0.22\linewidth]{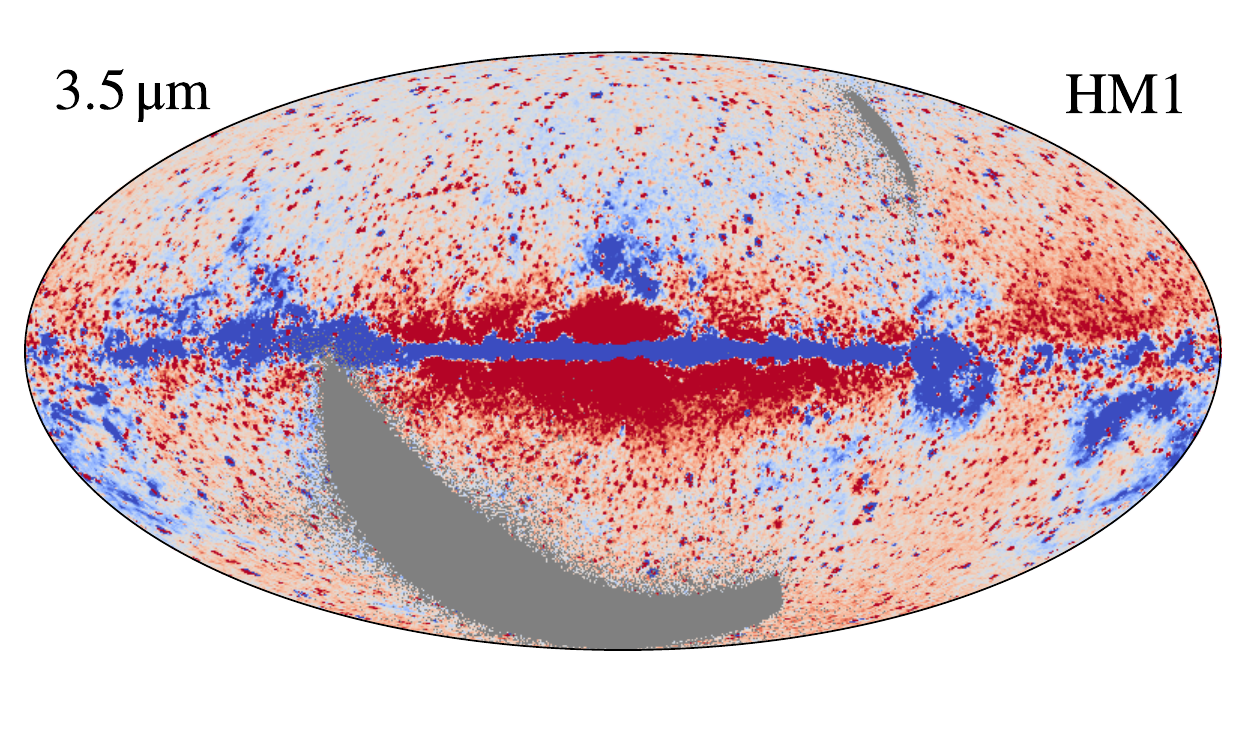}%
    \includegraphics[width=0.22\linewidth]{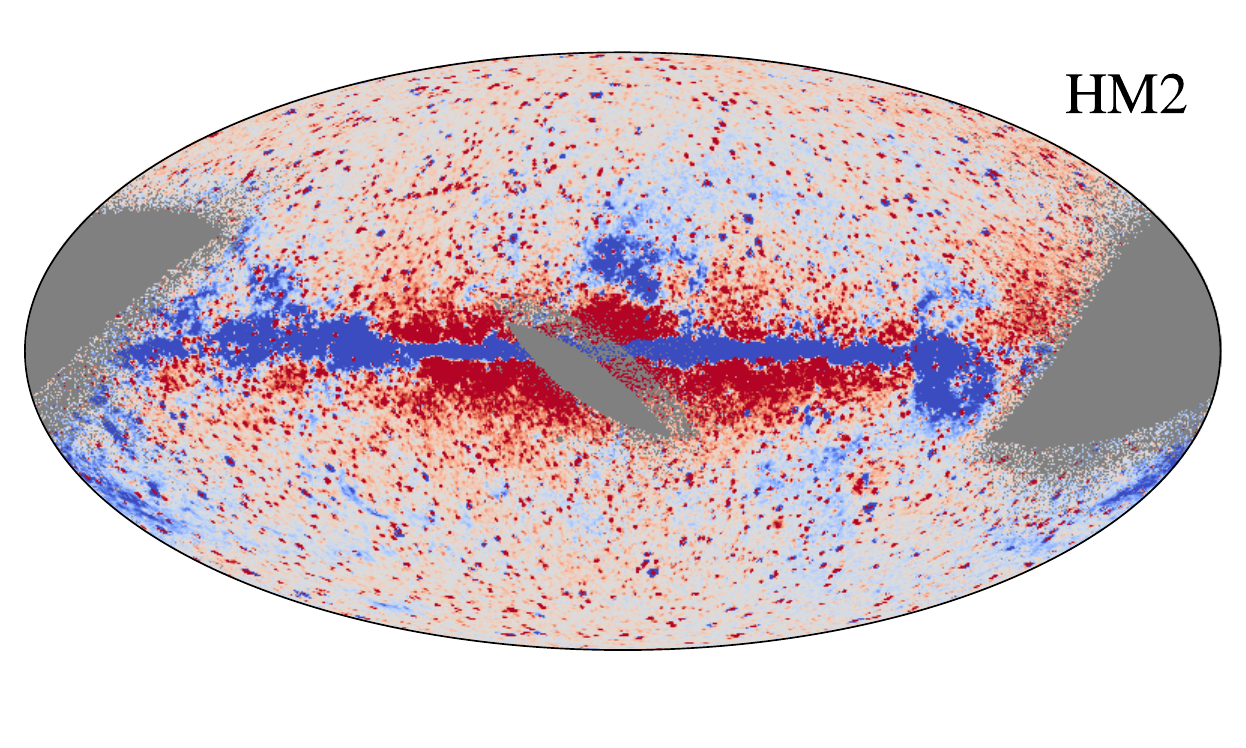}%
    \includegraphics[width=23mm,angle=90]{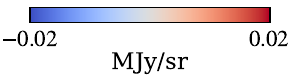}\hspace*{3mm}
    \includegraphics[width=0.22\linewidth]{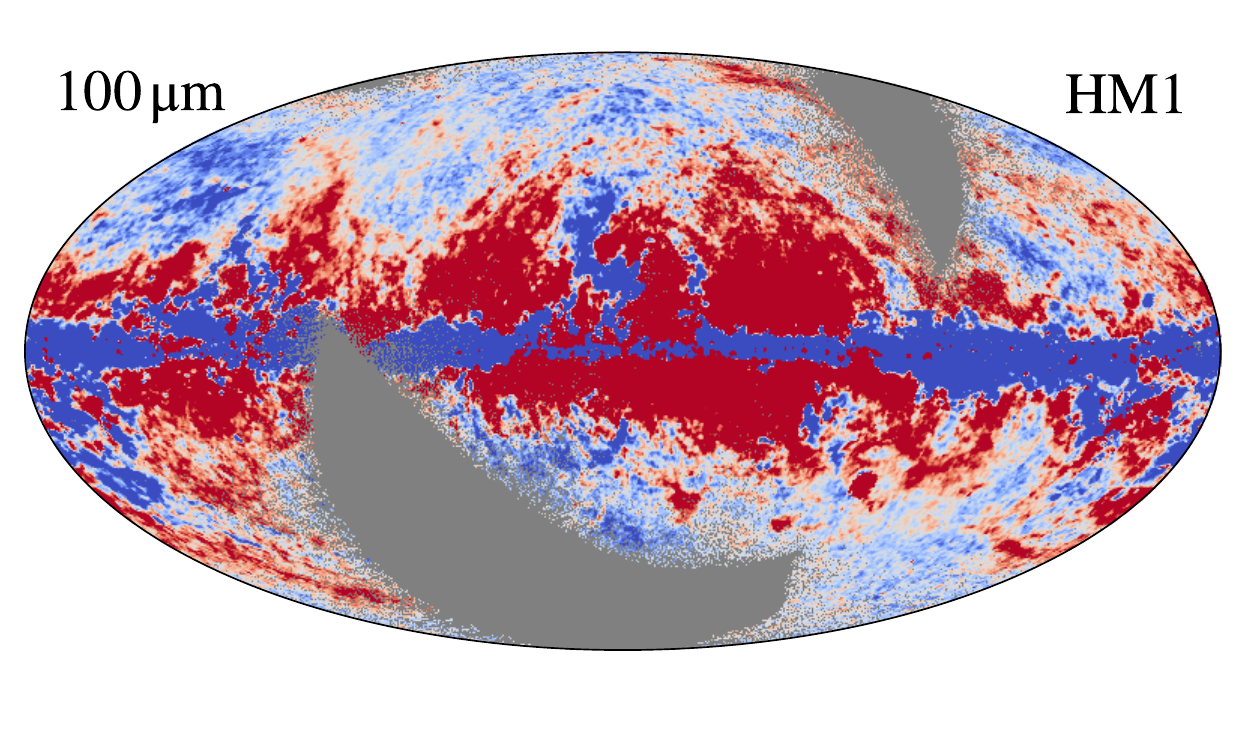}%
    \includegraphics[width=0.22\linewidth]{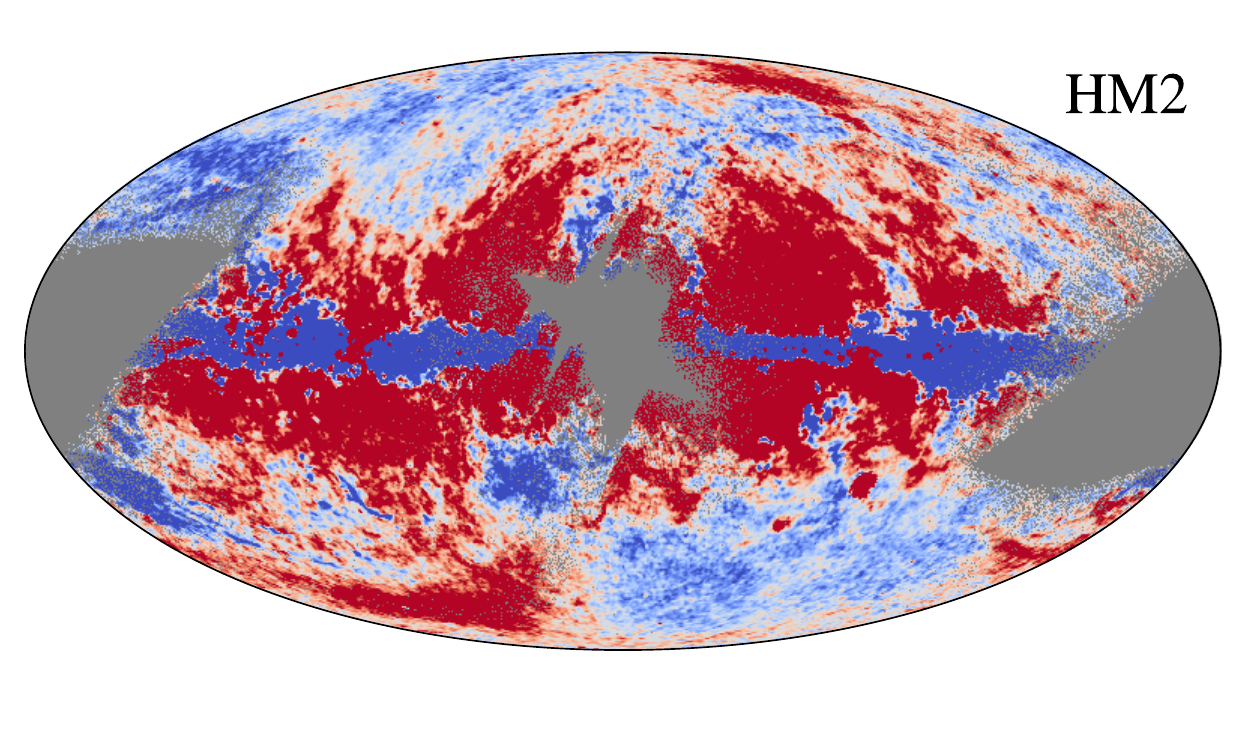}%
    \includegraphics[width=23mm,angle=90]{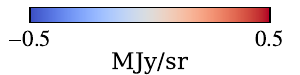}\\
    \includegraphics[width=0.22\linewidth]{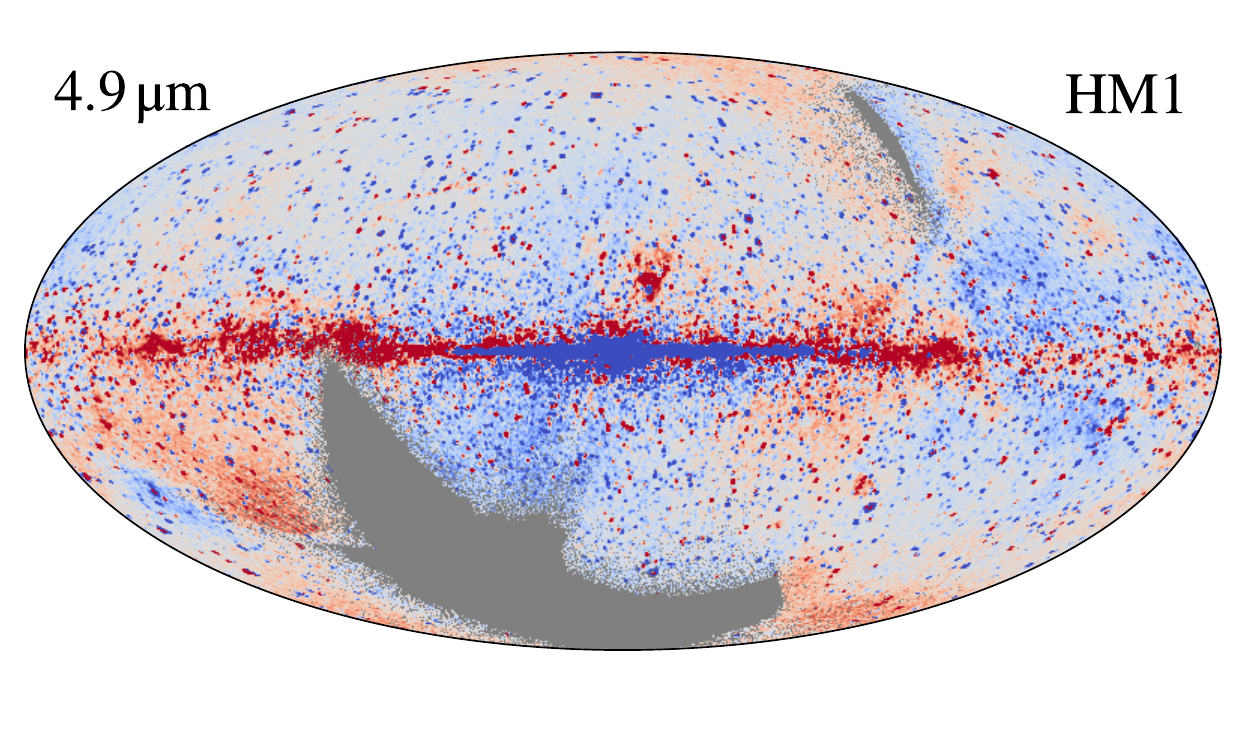}%
    \includegraphics[width=0.22\linewidth]{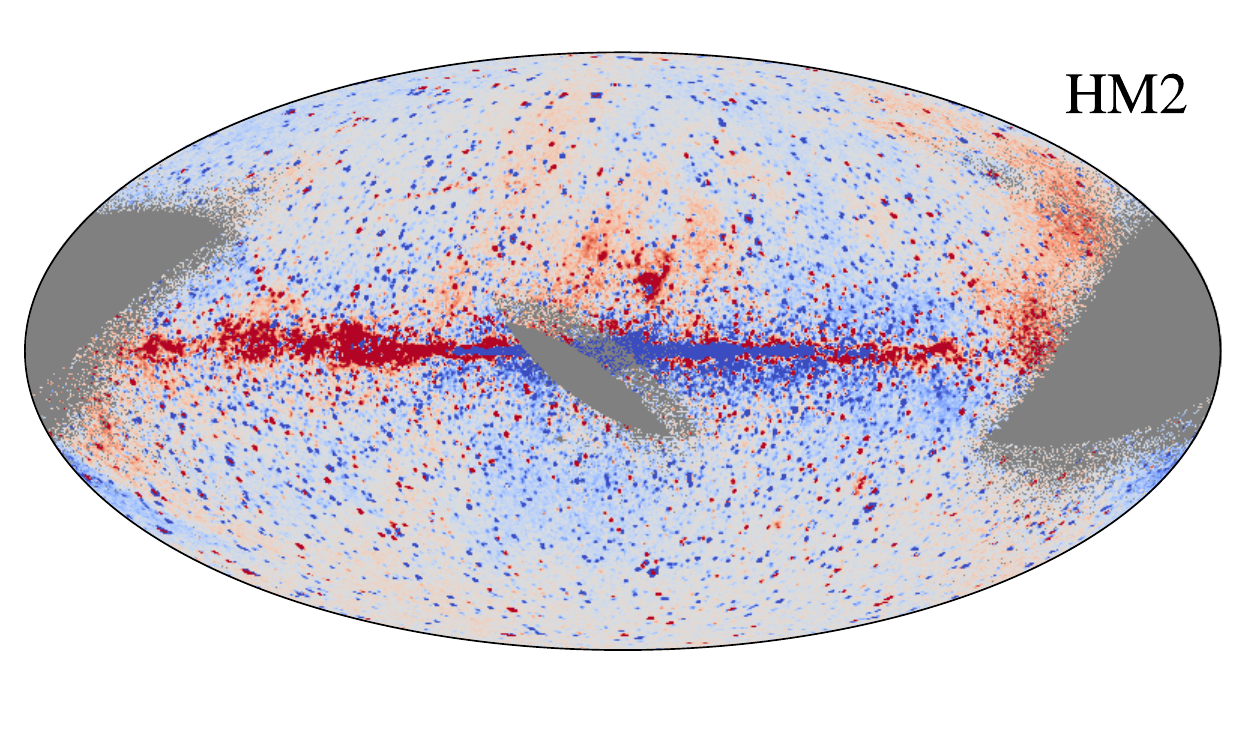}%
    \includegraphics[width=23mm,angle=90]{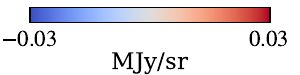}\hspace*{3mm}
    \includegraphics[width=0.22\linewidth]{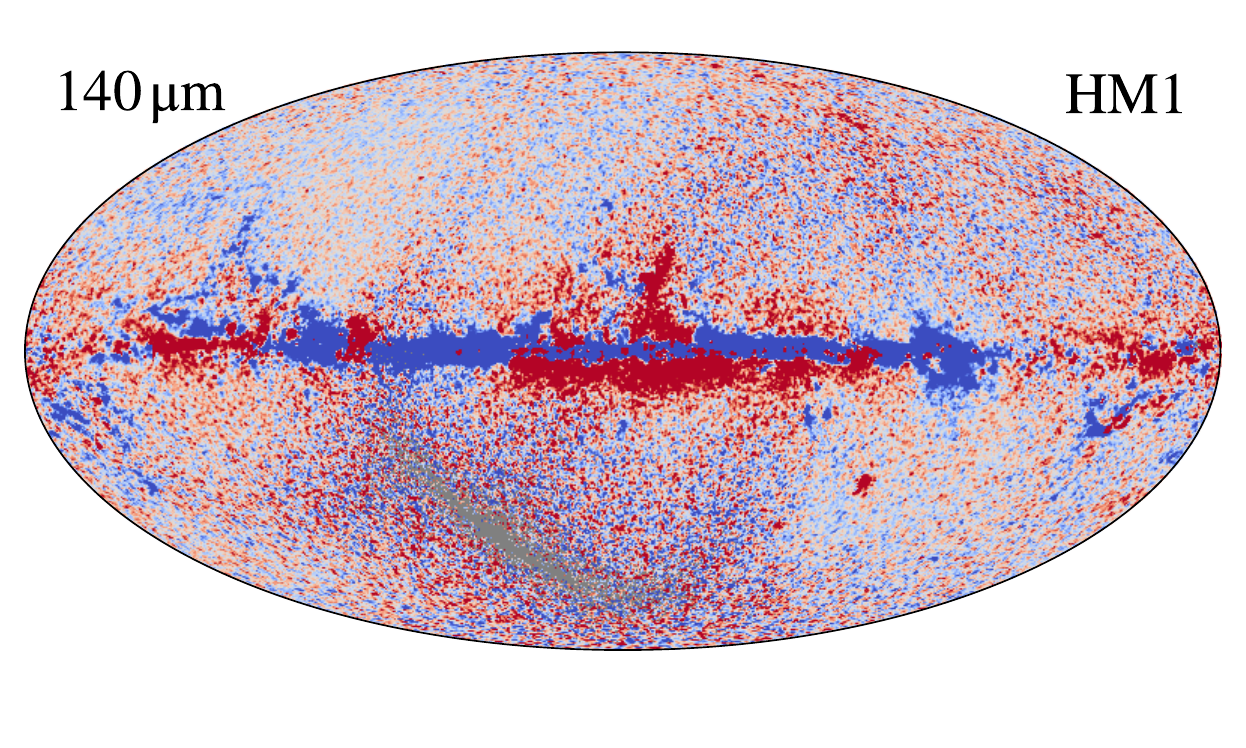}%
    \includegraphics[width=0.22\linewidth]{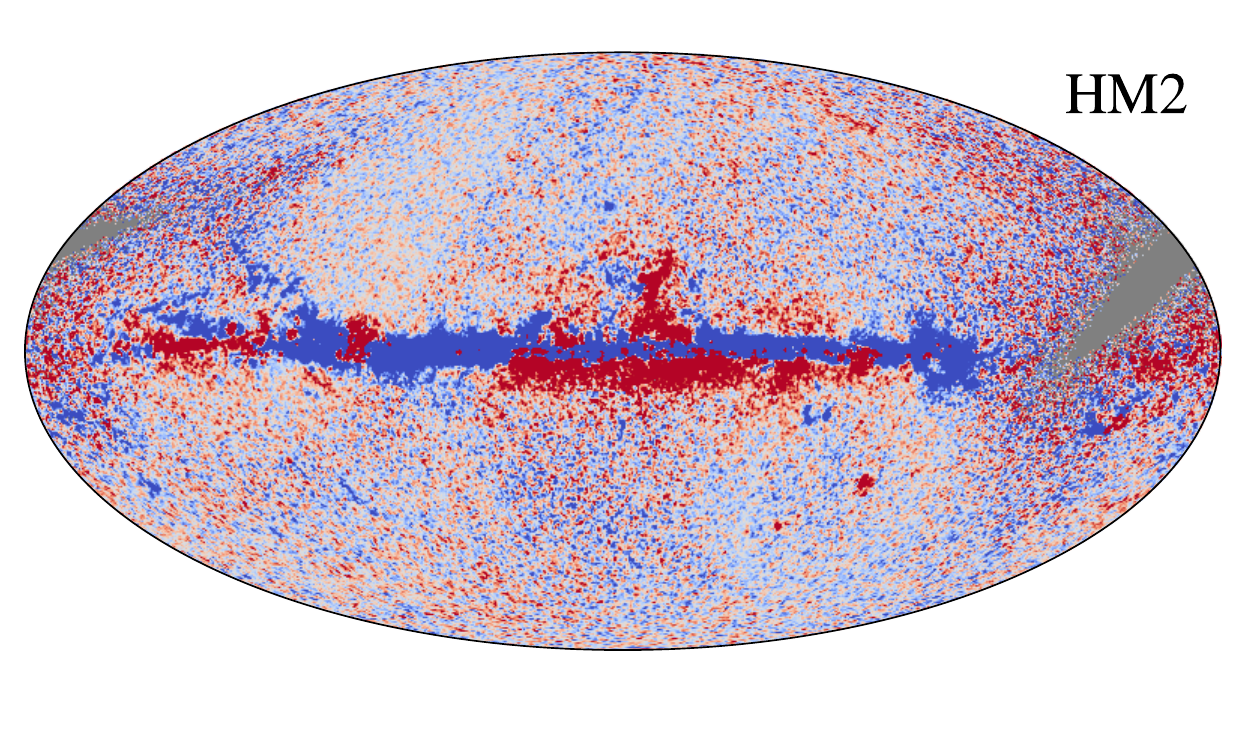}%
    \includegraphics[width=23mm,angle=90]{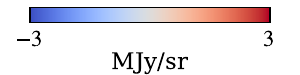}\\
    \includegraphics[width=0.22\linewidth]{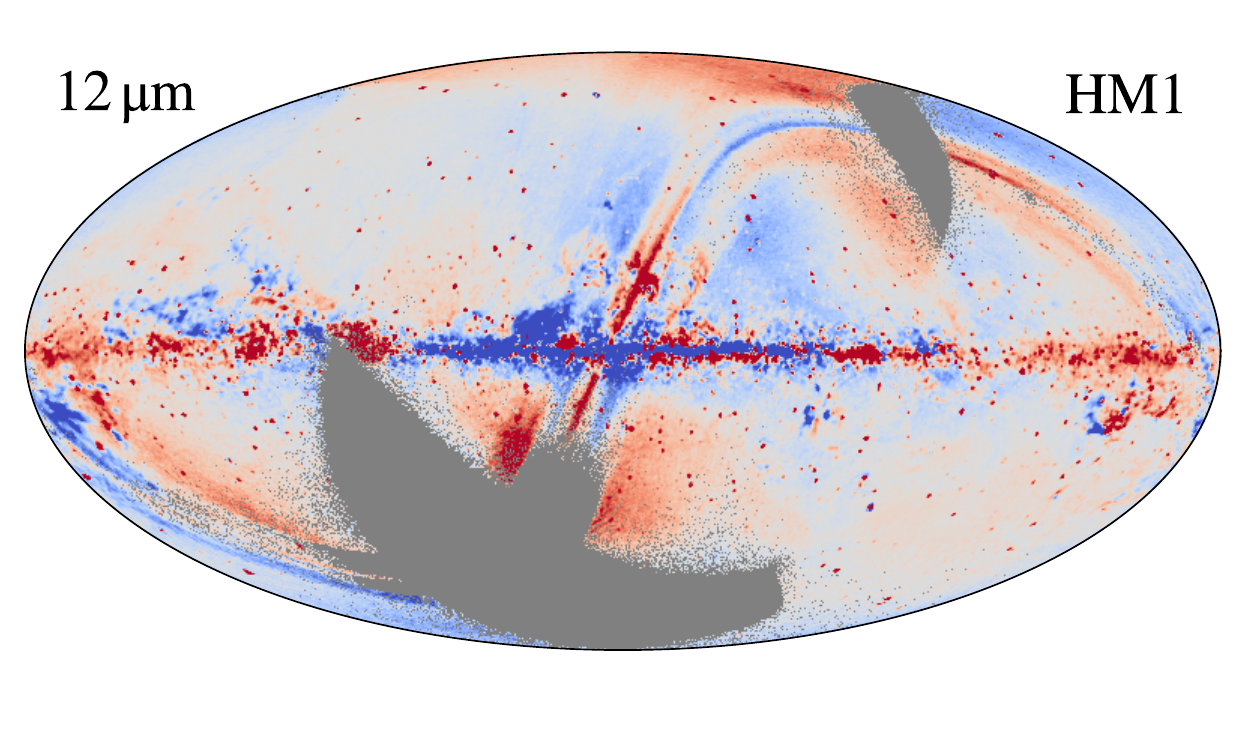}%
    \includegraphics[width=0.22\linewidth]{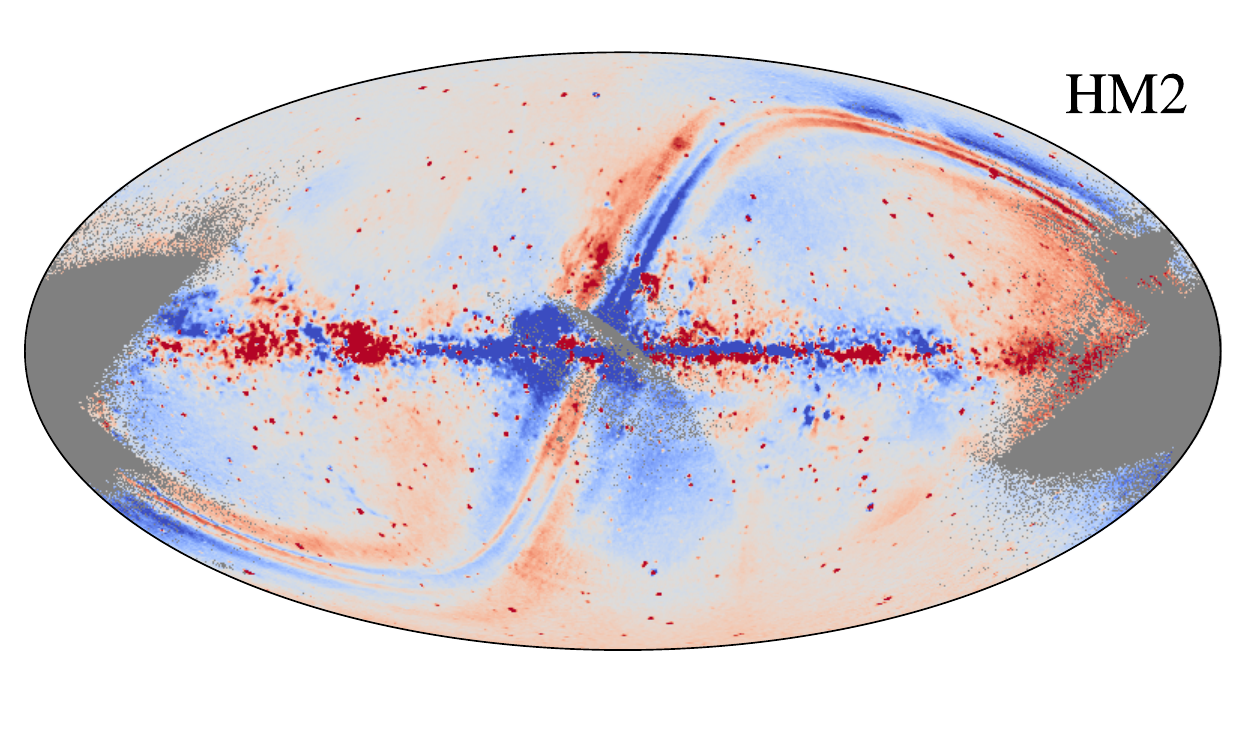}%
    \includegraphics[width=23mm,angle=90]{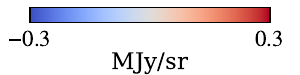}\hspace*{3mm}
    \includegraphics[width=0.22\linewidth]{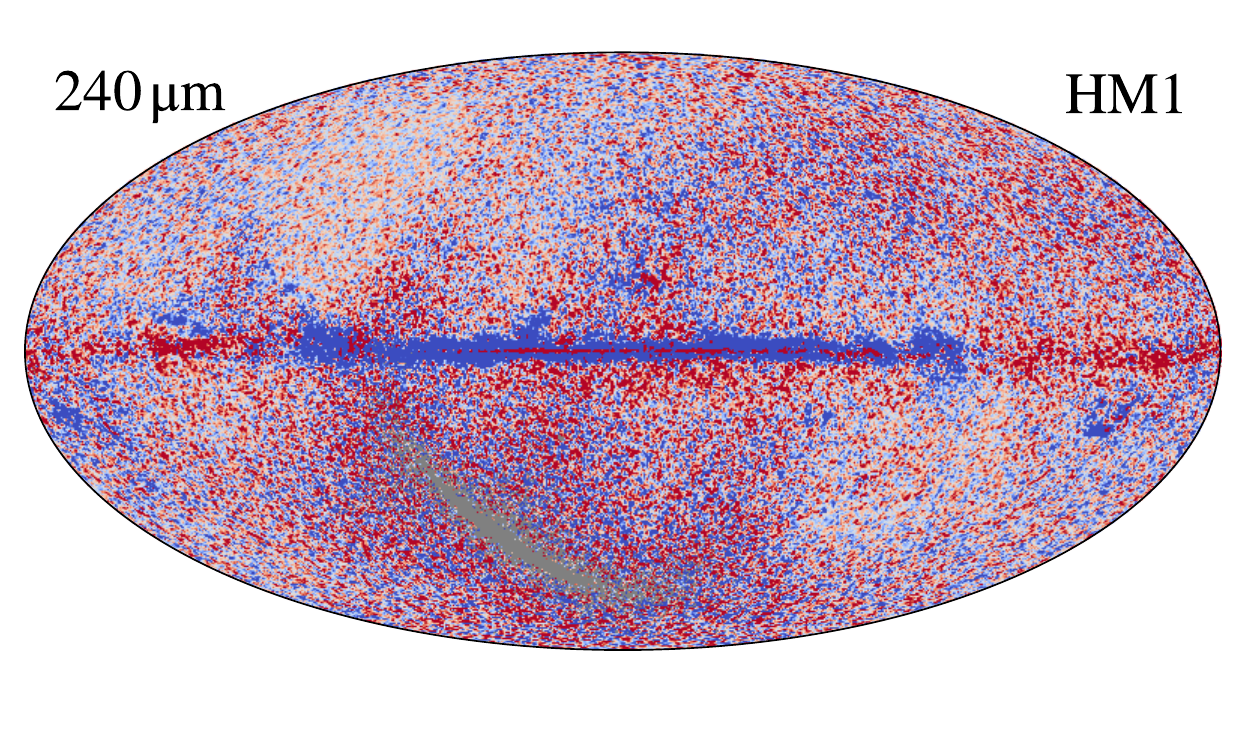}%
    \includegraphics[width=0.22\linewidth]{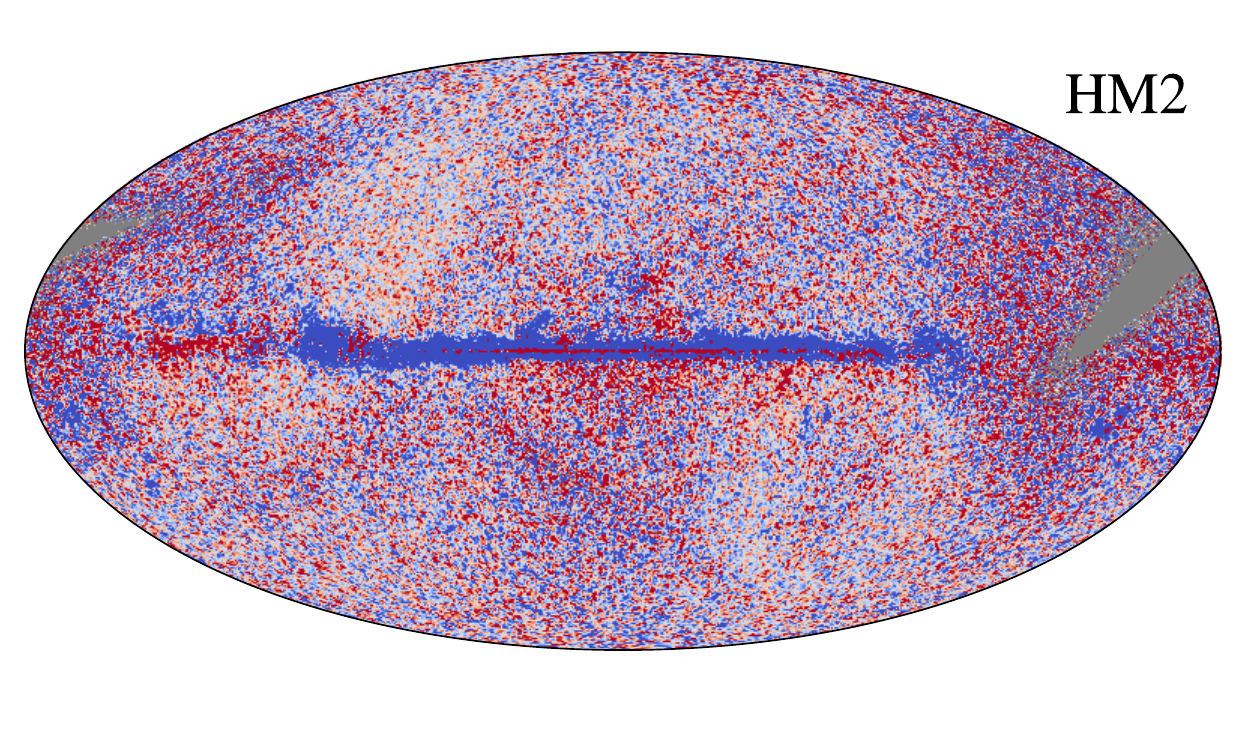}%
    \includegraphics[width=23mm,angle=90]{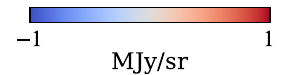}
    \caption{Half-mission data-minus-signal residual maps for each
    channel. The two columns within each wavelength section
    correspond to the first (HM1) and second (HM2) half-mission
    maps, respectively. All bands between 1.25 and 100 $\mu$m have been smoothed with a $15^\prime$ Gaussian kernel,
    and the 140 and 240 $\mu$m channels have been smoothed with a $30^\prime$ FWHM kernel. 
    The gray regions represent unobserved pixels in the half-mission splits.}
    \label{fig:half-mission-res2}
\end{figure*}

\begin{figure*}[t]
    \centering

    \includegraphics[height=2.90cm]{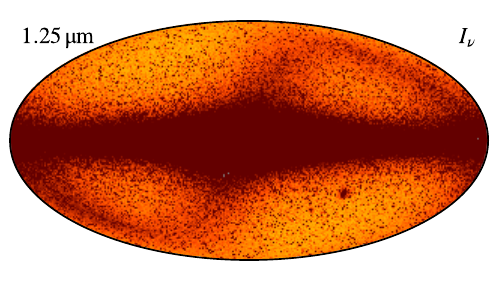}%
    \includegraphics[width=2.90cm,angle=90]{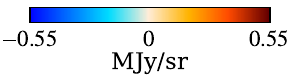}%
    \includegraphics[height=2.90cm]{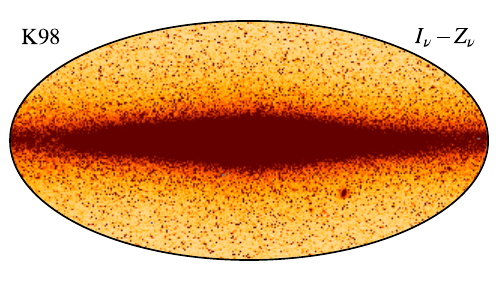}%
    \includegraphics[height=2.90cm]{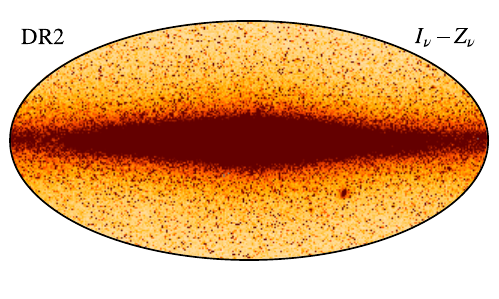}%
    \includegraphics[width=2.90cm,angle=90]{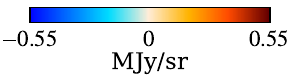}%
    \\

    \includegraphics[height=2.90cm]{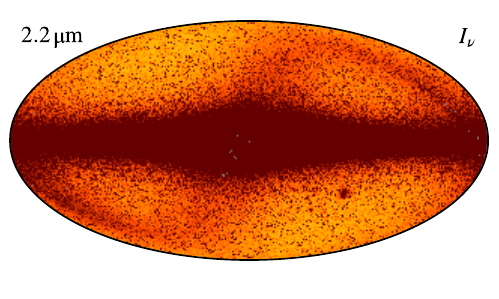}%
    \includegraphics[width=2.90cm,angle=90]{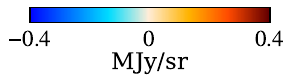}%
    \includegraphics[height=2.90cm]{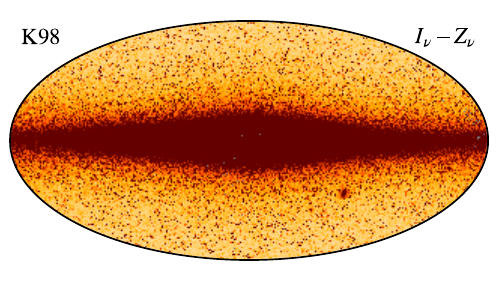}%
    \includegraphics[height=2.90cm]{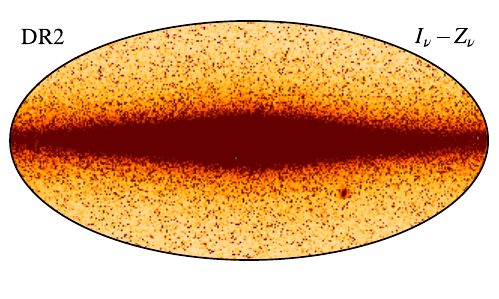}%
    \includegraphics[width=2.90cm,angle=90]{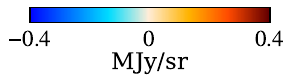}%
    \\

    \includegraphics[height=2.90cm]{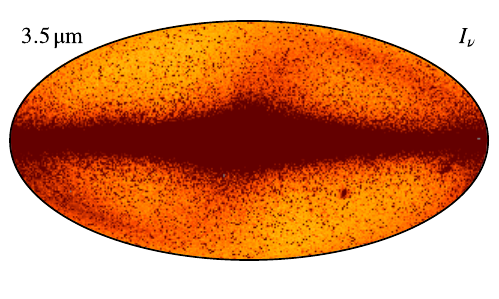}%
    \includegraphics[width=2.90cm,angle=90]{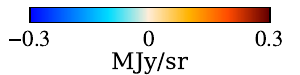}%
    \includegraphics[height=2.90cm]{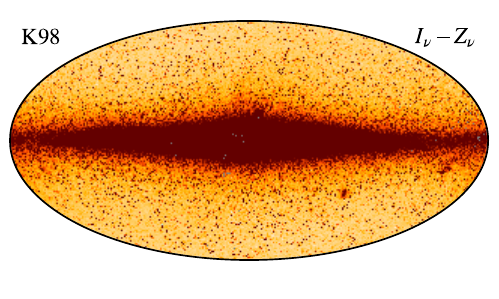}%
    \includegraphics[height=2.90cm]{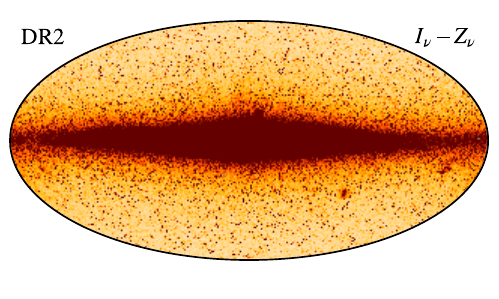}%
    \includegraphics[width=2.90cm,angle=90]{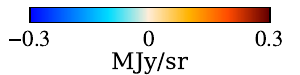}%
      \\

    \includegraphics[height=2.90cm]{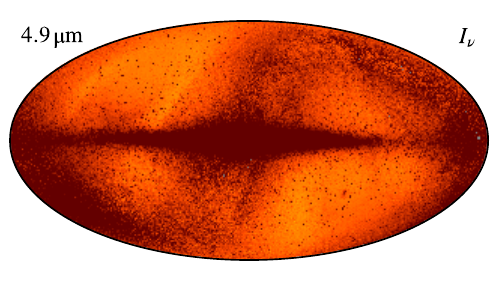}%
    \includegraphics[width=2.90cm,angle=90]{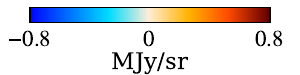}%
    \includegraphics[height=2.90cm]{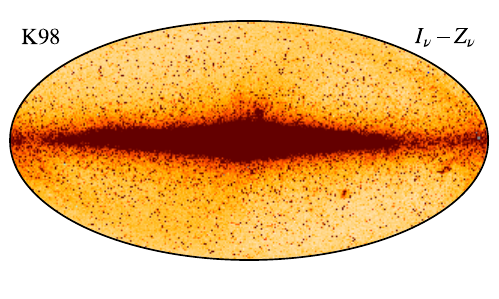}%
    \includegraphics[height=2.90cm]{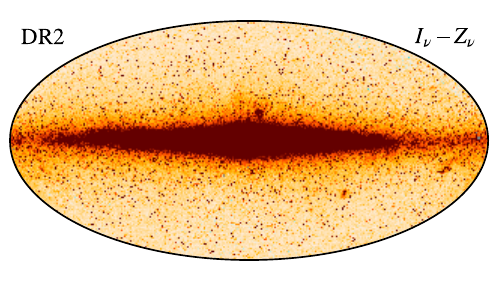}%
    \includegraphics[width=2.90cm,angle=90]{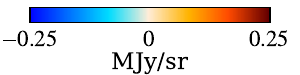}%
      \\

    \includegraphics[height=2.90cm]{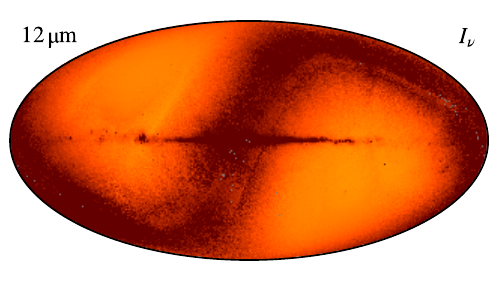}%
    \includegraphics[width=2.90cm,angle=90]{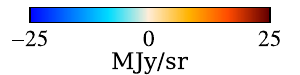}%
    \includegraphics[height=2.90cm]{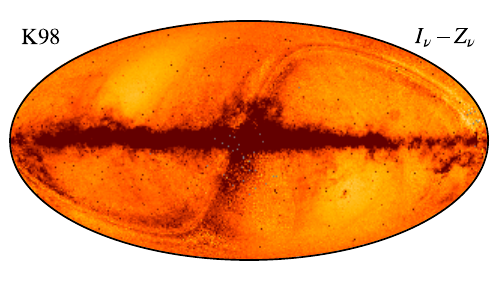}%
    \includegraphics[height=2.90cm]{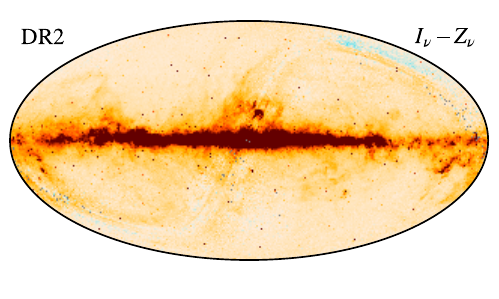}%
    \includegraphics[width=2.90cm,angle=90]{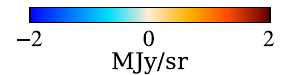}%
      \\

    \includegraphics[height=2.90cm]{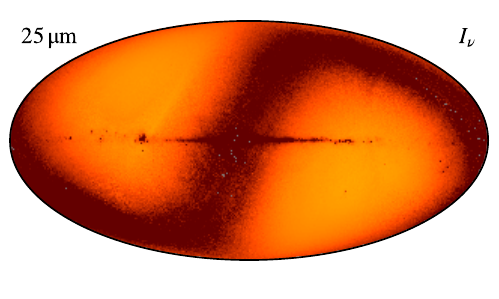}%
    \includegraphics[width=2.90cm,angle=90]{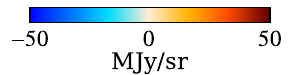}%
    \includegraphics[height=2.90cm]{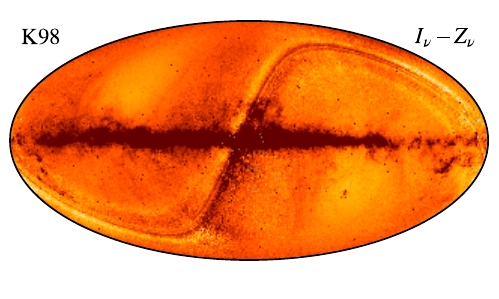}%
    \includegraphics[height=2.90cm]{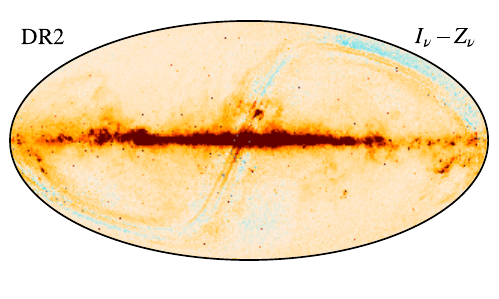}%
    \includegraphics[width=2.90cm,angle=90]{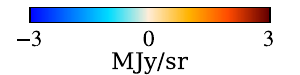}%
      \\

    \includegraphics[height=2.90cm]{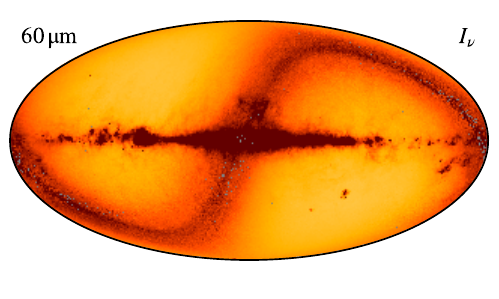}%
    \includegraphics[width=2.90cm,angle=90]{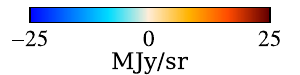}%
    \includegraphics[height=2.90cm]{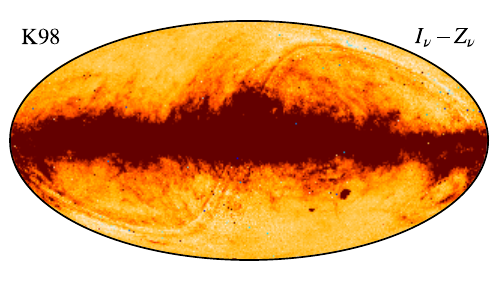}%
    \includegraphics[height=2.90cm]{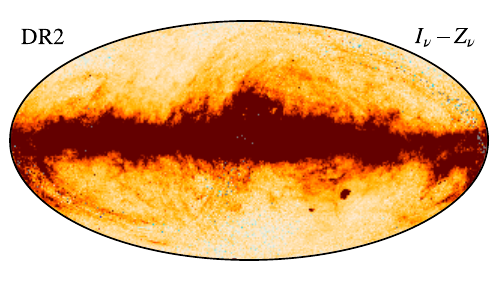}%
    \includegraphics[width=2.90cm,angle=90]{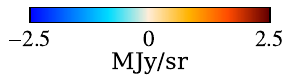}%
      \\
    \caption{\textit{(Left column):} Mission-averaged frequency maps containing ZL after data 
    selection. \textit{(Middle column):} Official DIRBE ZSMA maps. \textit{(Right column):} Our ZL subtracted mission-average 
    maps. All the maps are at our native HEALPix resolution of $N_\mathrm{side} = 512$. Rows show, from top to 
    bottom, DIRBE channels from 1.25 to 60$\,\mu$m.
    }
    \label{fig:dr2-zsma-compare1}
\end{figure*}

\begin{figure*}[t]
    \centering
    \includegraphics[height=2.90cm]{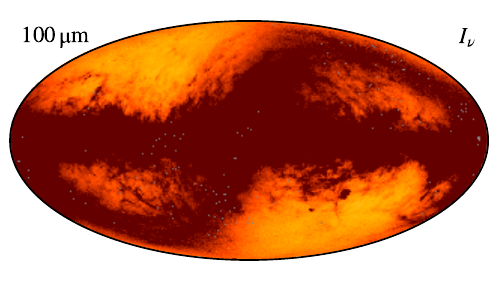}%
    \includegraphics[width=2.90cm,angle=90]{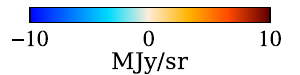}%
    \includegraphics[height=2.90cm]{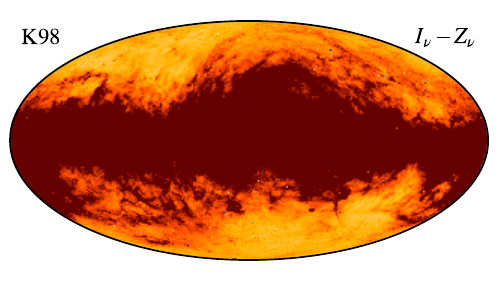}%
    \includegraphics[height=2.90cm]{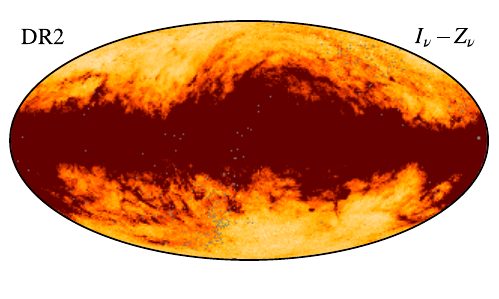}%
    \includegraphics[width=2.90cm,angle=90]{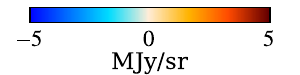}%
    \\

    \includegraphics[height=2.90cm]{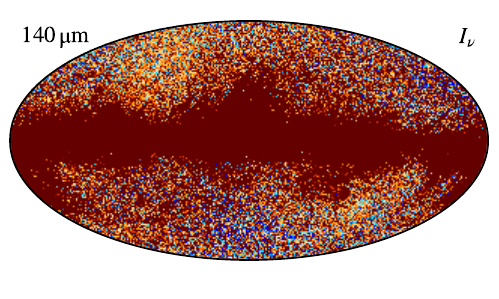}%
    \includegraphics[width=2.90cm,angle=90]{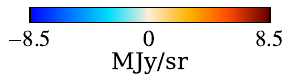}%
    \includegraphics[height=2.90cm]{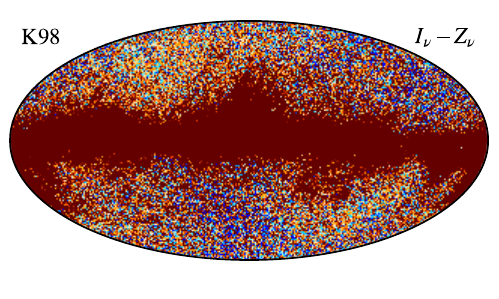}%
    \includegraphics[height=2.90cm]{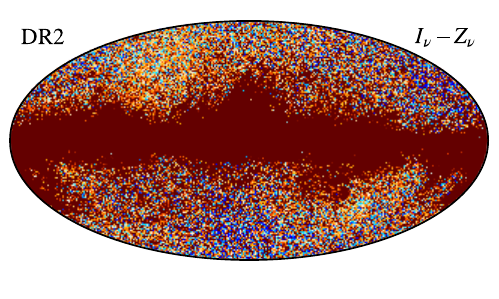}%
    \includegraphics[width=2.90cm,angle=90]{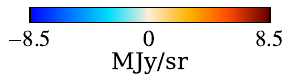}%
      \\

    \includegraphics[height=2.90cm]{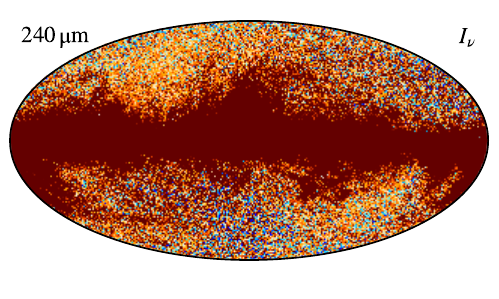}%
    \includegraphics[width=2.90cm,angle=90]{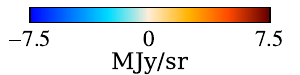}%
    \includegraphics[height=2.90cm]{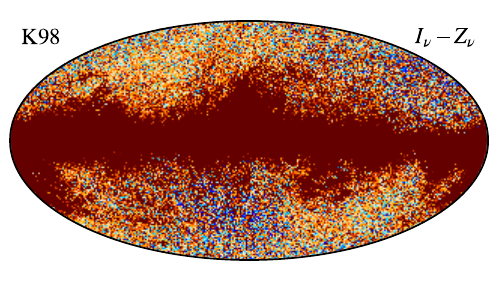}%
    \includegraphics[height=2.90cm]{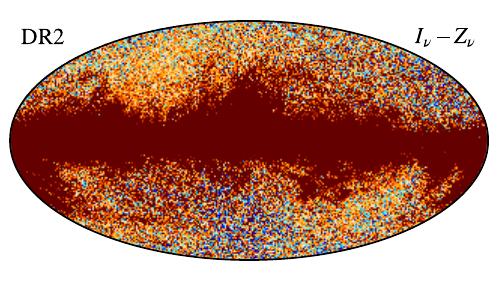}%
    \includegraphics[width=2.90cm,angle=90]{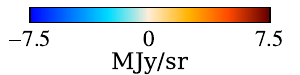}%
      \\

      \caption{Same as Fig.~\ref{fig:dr2-zsma-compare1}, but for the 100--240$\,\mu$m channels. 
    }    
    \label{fig:dr2-zsma-compare2}
\end{figure*}

In general, we see that the differences between DR2 and K98 vary for
many more parameters than predicted by the reported statistical
uncertainties. For instance, the density parameters of dust bands 1 and 2 are 
approximately 67\,\% and 25\,\% lower in the DR2 model than in K98. 
However, at frequencies below 140$\,\mu$m, the emissivities 
of all components are higher in DR2 than in K98, which compensates for the lower densities 
and results in a similar total intensity. 
Figure~\ref{fig:mission-averaged-comp-maps} in
Appendix~\ref{sec:zodi-comps} compares the mission-averaged ZL maps
for both models per component. Here we see visually that in the DR2 maps both the 
smooth cloud and the three dust bands are fainter than in the K98 maps. 
The intensity of the third asteroidal band is particularly weak in our model, and we only conclude with an upper limit.

Figure~\ref{fig:phase_function} compares our new phase function, $\Phi(\Theta)$, 
with those reported by \citet{kelsall1998} and \citet{Hong}. All the models are 
normalized over the interval $\Theta \in [65^\circ,\,125^\circ]$, since this is the interval of 
solar elongation angle that DIRBE was able to observe. However, when comparing these 
it is worth noting that most of the DIRBE constraining power lies between 80 and $110^{\circ}$, 
and the visual impression in this figure does depend on the specific range used for normalization. 
With this caveat in mind, we see that the minimum (which is independent of normalization) 
in our function lies between the Hong et al.\ and K98 minima. The biggest discrepancy is 
seen at very low scattering angles, where our phase function lies about $4\,\sigma$ lower 
than Hong et al. In that respect, it is worth noting that the Hong et al.\ analysis included 
datasets at very low solar elongations. A future \Cosmoglobe\ re-analysis of the current 
model may want to consider including those as well. 

Figure~\ref{fig:zodi-intensity} compares the predicted ZL intensity
as a function of wavelength for three different positions on the sky
at one given day in different colors, as well as the average over the
full sky in black. Solid and dashed lines show the DR2 and K98
predictions, respectively. The bottom panel shows the relative
difference. In general, the two models agree to better than 10\,\% for
most wavelengths, except at the very longest, 
for which the overall amplitudes are very uncertain. 

Figure~\ref{fig:reldiff} shows a similar relative difference, but now 
as a function of position on the sky for the 12 and 25\,$\mu$m
channels. In these channels, for which the ZL emission is the
brightest, the K98 and \cosmoglobe\ DR2 models agree to within
$\lesssim$\,3\,\% over most of the sky. These differences are thus 
similar in magnitude to the relative residuals measured by
\citet{kelsall1998}, which were found to be $\sim$\,2\,\% at
12\,$\mu$m.

\subsection{Goodness of fit}

Next, we consider the absolute goodness of fit of the \cosmoglobe\ DR2
model, and we start by inspecting the half-mission data-minus-model
residual maps for each channel, which are shown in
Fig.~\ref{fig:half-mission-res2}. We note that these maps show total
residuals, and therefore include contributions from ZL, Galactic
foregrounds, and instrumental noise.

\begin{table}
\newdimen\tblskip \tblskip=5pt
\caption{Reduced goodness-of-fit ($\chi_\mathrm{red}^2$) values for one arbitrarily 
selected Gibbs sample, along with the number of TOD samples used to fit the ZL parameters ($N_{\mathrm{samp}}$) 
and the instrument white noise RMS per 8\,Hz sample for each DIRBE channel ($\sigma_0$). 
Note that the 1.25--3.5$\,\mu$m and the 12--60$\,\mu$m channels are analyzed jointly; 
therefore, the values of $N_{\mathrm{samp}}$ and $\chi^2_{\mathrm{red}}$ refer to the combined groups.}
\label{tab:chisq}
\vskip -4mm
\footnotesize
\setbox\tablebox=\vbox{
 \newdimen\digitwidth
 \setbox0=\hbox{\rm 0}
 \digitwidth=\wd0
 \catcode`*=\active
 \def*{\kern\digitwidth}
  \newdimen\dpwidth
  \setbox0=\hbox{.}
  \dpwidth=\wd0
  \catcode`!=\active
  \def!{\kern\dpwidth}
  \halign{\hbox to 2cm{#\leaderfil}\tabskip 2em&
    \hfil$#$\hfil \tabskip 2em&
    \hfil$#$\hfil \tabskip 2em&
    \hfil$#$\hfil \tabskip 0em\cr
\noalign{\doubleline}
\omit\sc $\lambda$ ($\mu\mathrm{m}$)\hfil& N_{\mathrm{samp}} & \sigma_{0} [\mathrm{MJy/sr}] & \chi^2_{\mathrm{red}} \cr
\noalign{\vskip 3pt\hrule\vskip 5pt}
*1.25   & *668 & *0.022 & 1.11 \cr 
*2.2    & \cdots & *0.024 & \cdots \cr 
*3.5    & \cdots & *0.023 & \cdots \cr 
*4.9    & *118 & *0.028 & 1.19 \cr 
*12     & 2052 & *0.109 & 2.17 \cr
*25     & \cdots & *0.209 & \cdots \cr 
*60     & \cdots & *0.369 & \cdots \cr 
100     & *122 & *0.417 & 1.91 \cr 
140     & **31 & 32.030  & 1.02 \cr 
240     & **35 & 18.042  & 1.03 \cr
\noalign{\vskip 5pt\hrule\vskip 5pt}}}
  \endPlancktable
\par
\end{table}

Starting with the near-infrared 1.25--3.5\,$\mu$m channels, we see
that these are primarily dominated by residual starlight emission,
seen in the form of the bright (positive and negative) Galactic plane
and the scattered points source residuals at high Galactic
latitudes. However, sub-dominant contributions from ZL residuals are
also clearly seen in the form of diffuse structures aligned with the
Ecliptic plane and poles. For instance, in the 1.25\,$\mu$m first
half-mission map, denoted by HM1, which corresponds to the first five
months of the mission, we see a faint band that follows the Ecliptic
plane. Even brighter Ecliptic residuals are seen in the 12--60\,$\mu$m
channels, and future work should clearly aim at establishing a more
detailed model of ZL in the Ecliptic plane; again, joint analysis with
high-resolution measurements from \IRAS\ and \AKARI\ should prove
extremely useful for this.

As noted in Sect.~\ref{sect:data}, we exclude one week of observations
at both the beginning and end of the DIRBE survey due to instrumental
instabilities. For the 4.9--100$\,\mu$m channels, a whole month is
removed at the end of the survey, and a corresponding larger region of
missing pixels is seen in HM2. The signatures of these missing data
are seen as large gray regions in the two half-mission maps. After
removing those, no sharp scan-aligned edges are seen in the unmasked
region, as they are in the official maps.

The 100 and 140$\,\mu$m channels are clearly dominated by Galactic
residuals. In order to further improve the ZL model for these channels, it is therefore necessary to also improve the thermal dust model at the same time.  

These qualitative observations can be made more quantitative by
considering the reduced $\chi^2$, as optimized within the parameter
estimation algorithm. The best-fit $\chi^2_{\mathrm{red}}$ values are
reported for each channel in the rightmost column of
Table~\ref{tab:chisq}; note that the 1.25--3.5$\,\mu$m and the 12--60$\,\mu$m channels are
processed jointly when fitting the parameters, and
$\chi^2_{\mathrm{red}}$ is therefore only properly defined for the two groups. 
For completeness, the second column lists the number of
TOD samples used to fit the ZL parameters, and the third column lists
the instrumental white noise RMS per TOD sample.

For ideal data, with a perfectly fitting model, we would
expect $\chi^2_{\mathrm{red}}\sim 1$, with a standard deviation given
by $\sqrt{2/N_{\mathrm{samp}}}\sim0.005$. Clearly, the current model
is not a perfect match to the observed data, and significant
deviations are observed. Still, the values are for several channels
very close to unity, in particular for the 140 and 240$\,\mu$m bands.

The biggest outliers are  the 12--60$\,\mu$m channels, which
have the highest signal-to-noise ratio to ZL emission. In this case,
the reduced $\chi^2$ is 2.17, which indicates that the white noise RMS
accounts for only about 40\,\% of the total variation seen in the residual
TOD; the rest is most naturally explained in terms of ZL modeling
errors. However, in this respect it is worth noting that our estimates
of the noise level, as reported by \citet{CG02_01}, are
significantly lower than those used for previous ZL analysis with
DIRBE observations. For instance, \citet{Robinson2013} adopted
uncertainties of 0.80 and 0.93\,MJy/sr per sample at 12 and
25$\,\mu$m, respectively, which are 8 and 5 times larger than our
estimates. Despite assuming such large uncertainties, they obtained a
reduced $\chi^2$ of 1.30 with their extended model. In contrast, if we
had adopted the same white noise estimates, we would have obtained a
reduced $\chi^2$ of $\sim$\,$0.1$, which immediately would suggest
significant noise RMS over-estimation. In other words, the current
\cosmoglobe\ DR2 analysis has resulted both in significantly lower
white noise uncertainties and lower absolute residuals for the main ZL
channels.

\subsection{Comparison of ZSMA maps}

A primary goal of the parametric ZL model presented here is to
establish a set of zodiacal light subtracted mission averaged (ZSMA)
frequency maps for each DIRBE channel that can be used for
cosmological and astrophysical analysis. These maps are simply derived
by subtracting the ZL component from the TOD, and bin the rest
into pixelized sky maps.

Figures~\ref{fig:dr2-zsma-compare1} and \ref{fig:dr2-zsma-compare2}
compare the K98 ZSMA maps (second column) with the new
\cosmoglobe\ DR2 ZSMA maps (third column), as well as with the
corresponding non-ZL-subtracted maps (first column). 
The 12--60\,$\mu$m DR2 ZSMA maps present negative 
values along the Ecliptic plane, indicating that the actual zodiacal light signal is lower than 
what is predicted by the model in use. Despite this, here we clearly
see that the \cosmoglobe\ approach results in significantly lower ZL residuals at
all channels below 140\,$\mu$m, while for the two longest wavelength
channels the two maps appear visually very similar. The improvements
are particularly striking between 4.9 and 60$\,\mu$m: in this frequency range zodiacal
emission is the strongest, yet the DR2 ZSMA maps are for the first time dominated by Galactic
signal at high latitudes rather than by ZL residuals. In summary, even though further work is needed 
to address the over-subtraction in channels 12--60\,$\mu$m, this analysis represents a 
significant advance in zodiacal light modeling for DIRBE, particularly in the mid-infrared.

\begin{figure}[t]
    \centering
    \includegraphics[width=\linewidth]{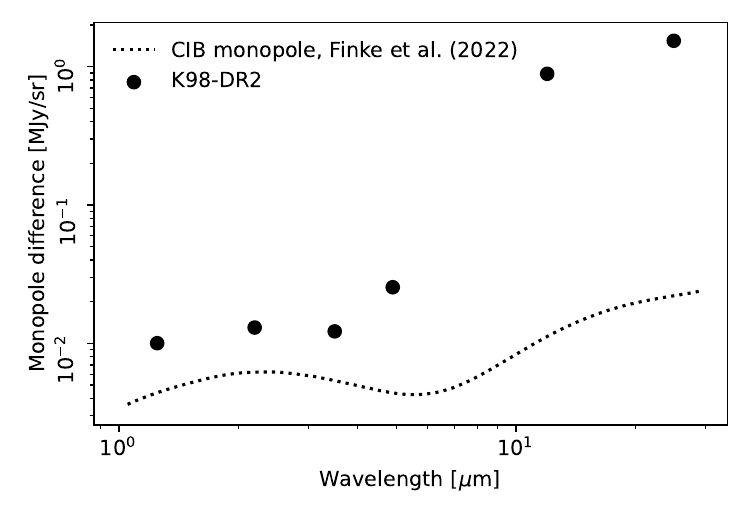}
    \caption{Monopole difference between the official K98 and 
    \cosmoglobe\ DR2 ZSMA maps, evaluated as the average of the 
    difference between the maps at 1.25--25$\,\mu$m, as shown in the second 
    and third columns of Fig.~\ref{fig:dr2-zsma-compare1}, outside the DR2 analysis masks.}
    \label{fig:zsma_mean}
\end{figure}

\begin{figure}[t]
    \centering
    \includegraphics[width=0.98\linewidth]{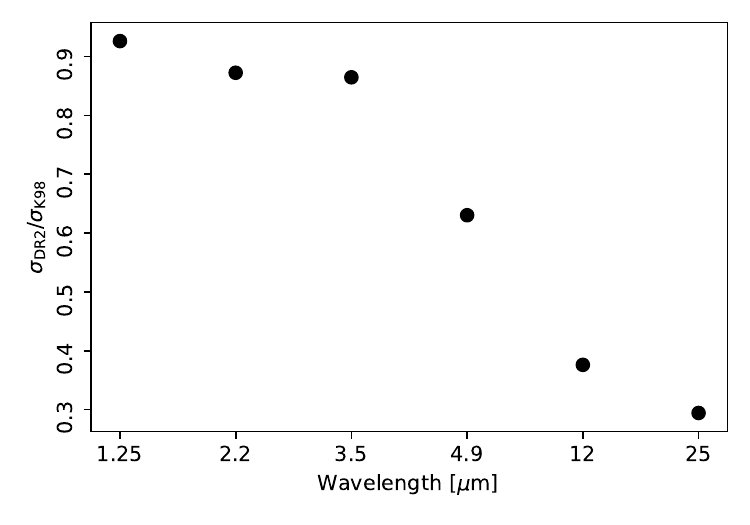}
    \caption{RMS ratio between the \cosmoglobe\ DR2 and K98 ZSMA maps,
      $\sigma_{\mathrm{DR2}}/\sigma_\mathrm{K98}$, as evaluated
      outside the DR2 analysis masks.}
    \label{fig:zsma_rms}
\end{figure}

One of the primary scientific goals of DIRBE was
to measure the CIB monopole spectrum, which essentially corresponds to
the zero-level of the ZSMA maps shown in
Figs.~\ref{fig:dr2-zsma-compare1} and \ref{fig:dr2-zsma-compare2}. It
is therefore of great interest to measure the difference in
zero-levels between the K98 and DR2 maps, shown in
Fig.~\ref{fig:zsma_mean} for the 1.25--25 $\mu$m channels, evaluated
outside the same confidence masks as described in
Sect.~\ref{sec:masks}. For reference, the best-fit theoretical model
of \citet{finke2022} is plotted as a dotted line. First, we note that
the difference is defined as K98-minus-DR2, which is always
positive. In other words, the new analysis appears to have removed
more ZL than the K98 analysis. In fact, the measured monopole
difference is actually also higher than the theoretical CIB monopole
spectrum itself, suggesting that the modifications are in fact
$\mathcal{O}(1)$ or larger as measured relatively to the main science
target of DIRBE; at 12 and 25$\,\mu$m, the differences are almost two orders of
magnitude. For further discussion regarding the scientific
implications of these differences in terms of the CIB monopole, we
refer the interested reader to \citet{CG02_03}.

Figure~\ref{fig:zsma_rms} shows the results from a similar
calculation, but this time in the form of the RMS ratio of the ZSMA
maps evaluated outside the same masks. Here we see that the DR2 maps
have lower RMS than the K98 maps in all the channels between 1.25 and 25\,$\mu$m, which once
again indicates that more ZL emission has been removed from the new maps.

\section{Conclusions}
\label{sec:conclusions}

We have presented a new Bayesian framework for modeling zodiacal
light from time-ordered observations, and applied this to 
DIRBE. We find that the resulting ZL model differs
significantly from the K98 model, both in terms of mean parameter
values and uncertainties, and the new model results in much cleaner
ZL-subtracted frequency maps. We strongly recommend that any future
cosmological or astrophysical analysis that involves DIRBE frequency
maps should use the \cosmoglobe\ DR2 maps.

These large improvements stem from two foundational features of the
\cosmoglobe\ framework. First, the new analysis used external
observations from \Planck\,, WISE, and \GAIA\ to model Galactic signals,
and thereby breaks, or at least minimizes, overall parameter
degeneracies. Second, the current framework fits all parameters
simultaneously, both those of astrophysical and instrumental
nature. This is achieved through a process called Gibbs sampling,
which maps out the full joint posterior distribution by iterating
through a large number of conditional distributions.

While this Gibbs sampling approach is a key ingredient in the global
process, and it is difficult to envision how the full framework would
work without it, this method is also associated with significant
numerical challenges. In particular, it is well known that Gibbs
sampling struggles with probing strongly degenerate distributions, and
the ZL posterior distribution in question in this paper shows
precisely this. This distribution does not only exhibit strong
degeneracies, but also a huge number of local extrema in which a
Markov chain can easily get trapped.

Another issue regarding the current model is physical plausibility. In
this first application of the Bayesian end-to-end framework to ZL
modelling, an important goal was to understand how well the current
data constrain our parametric model. As a result, only weak uniform
priors are imposed on the main fitted parameters. On the one hand,
this does ensure that the final residuals are minimized, which is good
for astrophysical and cosmological applications of the resulting ZSMA
maps, but it also means that some parameters may have likely drifted
into non-physical parameter values.

Overcoming all of these challenges will most likely involve at least
two main tasks. First, by analyzing additional and complementary data,
for instance from high-resolution experiments such as \AKARI\ and \IRAS\, 
jointly with the DIRBE measurements, many of the current strong
degeneracies are likely to be effectively broken. In particular the
structure of the asteroidal bands, for which DIRBE's low angular
resolution is a severe limitation, should become much better probed
with these data. Still, even with more data there are likely to be
several degeneracies present that are difficult to predict in
advance. Implementing better and faster sampling algorithms should
therefore be a high-priority topic for future work, for instance using
ideas from simulated annealing and/or Hamiltonian sampling.
Furthermore, future analyses may also include SPHEREx data, 
which will allow testing of the ZL model at wavelengths closer to the 
visible and contribute to ongoing efforts toward developing an optical 
zodiacal light model (e.g., \citet{skysurf11}).

In addition, it is by no means obvious that a low-dimensional 3D model
such as K98 with only $\mathcal{O}(10^2)$ free parameters can describe
such a rich and dynamical system as ZL to the required precision for
next-generation cosmology. Future work should therefore also consider
conceptual generalizations of the entire framework. One step in this
direction was taken by \citet{Robinson2013}, who introduced a
different set of basis components than K98; indeed, the new fit
presented in this paper appears to have moved toward that model within
the K98 framework. More drastic approaches would be to consider
perturbative Taylor expansion models around the existing components,
or even fully voxelized 3D models. Obviously, these would be highly
degenerate when using only near-Earth observations, but perhaps
in combination with data from Solar System probes such as Juno or New
Horizons, new insights may be gained.

In conclusion, it is clear that the ZL model presented in this paper
suffers from many shortcomings of both modeling and algorithmic
origin, and these must be addressed through future extensions of
the \cosmoglobe\ framework. At the same time, and despite all these
shortcomings, it is equally clear that this model redefines the
state of the art of ZL modeling for the DIRBE data, reducing residuals
by orders of magnitude in the mid-infrared regime. Furthermore, this
work has established an effective computational framework that can
be extended to other datasets with relatively minor effort, and that
will allow equally large steps forward to be made. All in all, this
work has established a new reference for the analysis of past, present
and future infrared experiments, building directly on a long line of
algorithmic breakthroughs initially developed for the CMB field.

\begin{acknowledgements}
  We thank Richard Arendt, Tony Banday, Johannes Eskilt, Dale Fixsen,
  Ken Ganga, Paul Goldsmith, Shuji Matsuura, Sven Wedemeyer, Janet
  Weiland and Edward Wright for useful suggestions and guidance.
  The current work has received funding from the
  European Union’s Horizon 2020 research and innovation programme
  under grant agreement numbers 819478 (ERC; \textsc{Cosmoglobe}),
  772253 (ERC; \textsc{bits2cosmology}), 101165647 (ERC;
  \textsc{Origins}), 101141621 (ERC; \textsc{Commander}), and 101007633 (MSCA;
  \textsc{CMBInflate}). This article reflects the views of the authors only. The funding
  body is not responsible for any use that may be made of the
  information contained therein. This research is
  also funded by the Research Council of Norway under grant agreement
  numbers 344934 (YRT; \textsc{CosmoglobeHD}) and 351037 (FRIPRO;
  \textsc{LiteBIRD-Norway}). Some of the results in this paper have been
  derived using healpy \citep{Zonca2019} and the HEALPix
  \citep{healpix} packages.  We acknowledge the use of the Legacy
  Archive for Microwave Background Data Analysis (LAMBDA), part of the
  High Energy Astrophysics Science Archive Center
  (HEASARC). HEASARC/LAMBDA is a service of the Astrophysics Science
  Division at the NASA Goddard Space Flight Center. This publication
  makes use of data products from the Wide-field Infrared Survey
  Explorer, which is a joint project of the University of California,
  Los Angeles, and the Jet Propulsion Laboratory/California Institute
  of Technology, funded by the National Aeronautics and Space
  Administration. This work has made use of data from the European
  Space Agency (ESA) mission {\it Gaia}
  (\url{https://www.cosmos.esa.int/gaia}), processed by the {\it Gaia}
  Data Processing and Analysis Consortium (DPAC,
  \url{https://www.cosmos.esa.int/web/gaia/dpac/consortium}). Funding
  for the DPAC has been provided by national institutions, in
  particular the institutions participating in the {\it Gaia}
  Multilateral Agreement.
  We acknowledge the use of data provided by the Centre d'Analyse de 
  Données Etendues (CADE), a service of IRAP-UPS/CNRS (http://cade.irap.omp.eu, \citealt{paradis:2012}). 
  This paper and related research have been conducted during and with the support 
  of the Italian national inter-university PhD programme in Space Science and Technology. 
  Work on this article was produced while attending the PhD program in PhD in Space 
  Science and Technology at the University of Trento, Cycle XXXIX, with the support 
  of a scholarship financed by the Ministerial Decree no. 118 of 2nd March 2023, based on the NRRP - 
  funded by the European Union - NextGenerationEU - Mission 4 "Education and Research", Component 1 
  "Enhancement of the offer of educational services: from nurseries to universities” - Investment 4.1 
  “Extension of the number of research doctorates and innovative doctorates for public administration 
  and cultural heritage” - CUP E66E23000110001.
\end{acknowledgements}

\bibliographystyle{aa}
\bibliography{../../common/CG_bibliography,references,../../common/Planck_bib}

\appendix
\onecolumn

\section{Component-wise zodiacal light maps and number density cross-sections}
\label{sec:zodi-comps}

\begin{figure*}[t]
    \centering
    \vspace{0.5cm}
    \resizebox{\textwidth}{!}{
    \includegraphics{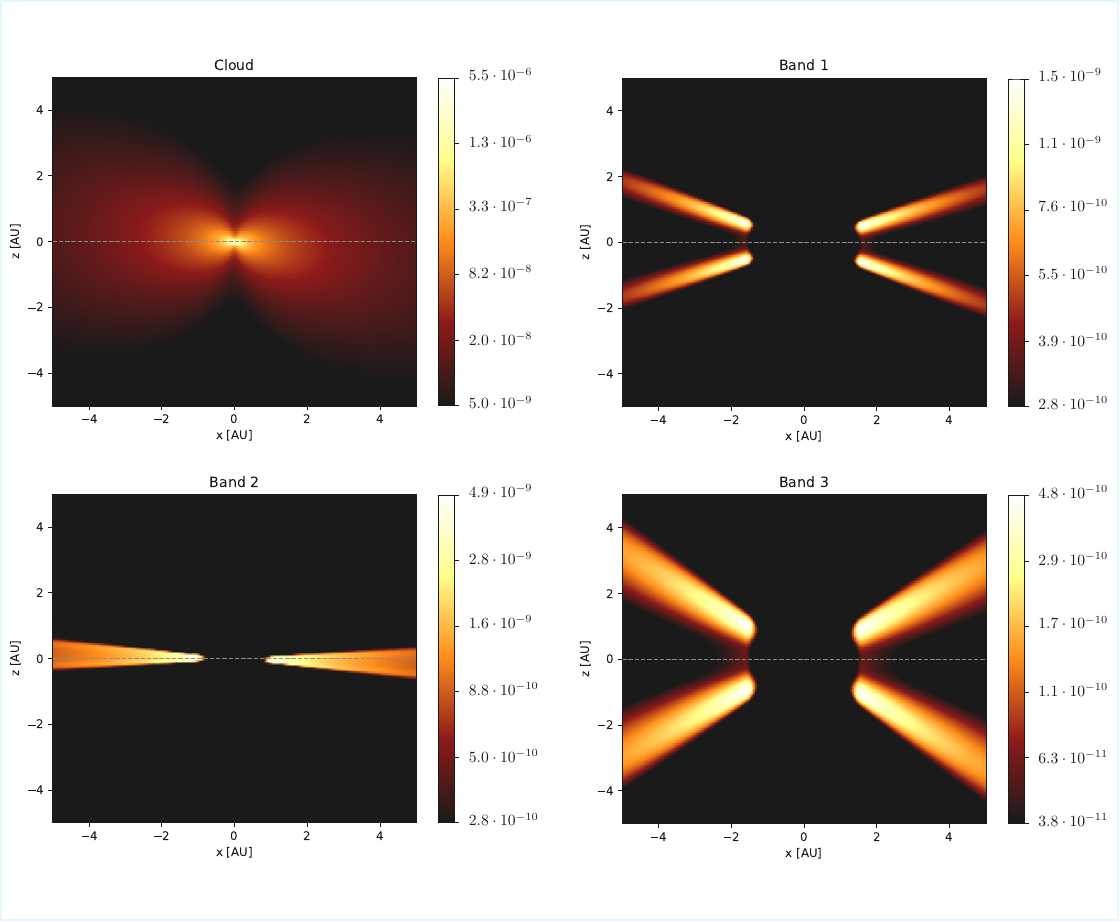}
    }\\
    \caption{Visualization of the IPD number density of the four fitted zodiacal 
    components in our model. The number densities are shown as a cross-section of the 
    Solar System in the xz-plane. \textit{(Top left):} The smooth cloud. \textit{(Top right):} 
    Dust band 1. \textit{(Bottom left):} Dust band 2. \textit{(Bottom right):} Dust band 3. 
    The gray dotted line represents the Ecliptic plane and helps illustrate the variations in the components symmetry planes.}
    \label{fig:ipd-number-density}
\end{figure*}

\begin{figure*}[t]
    \centering
    \resizebox{0.91\textwidth}{!}{%
    \includegraphics[height=1cm]{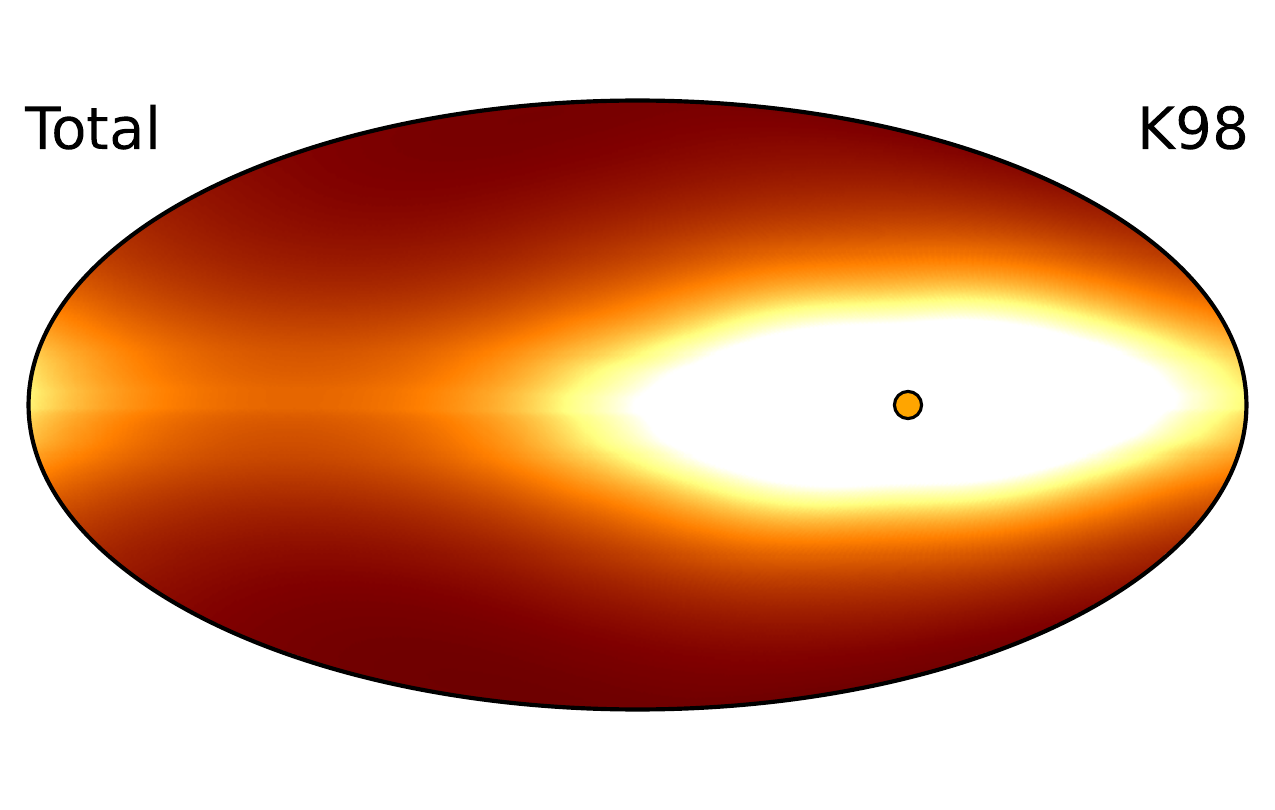}%
    \hspace{2.3pt}
    \includegraphics[height=1cm]{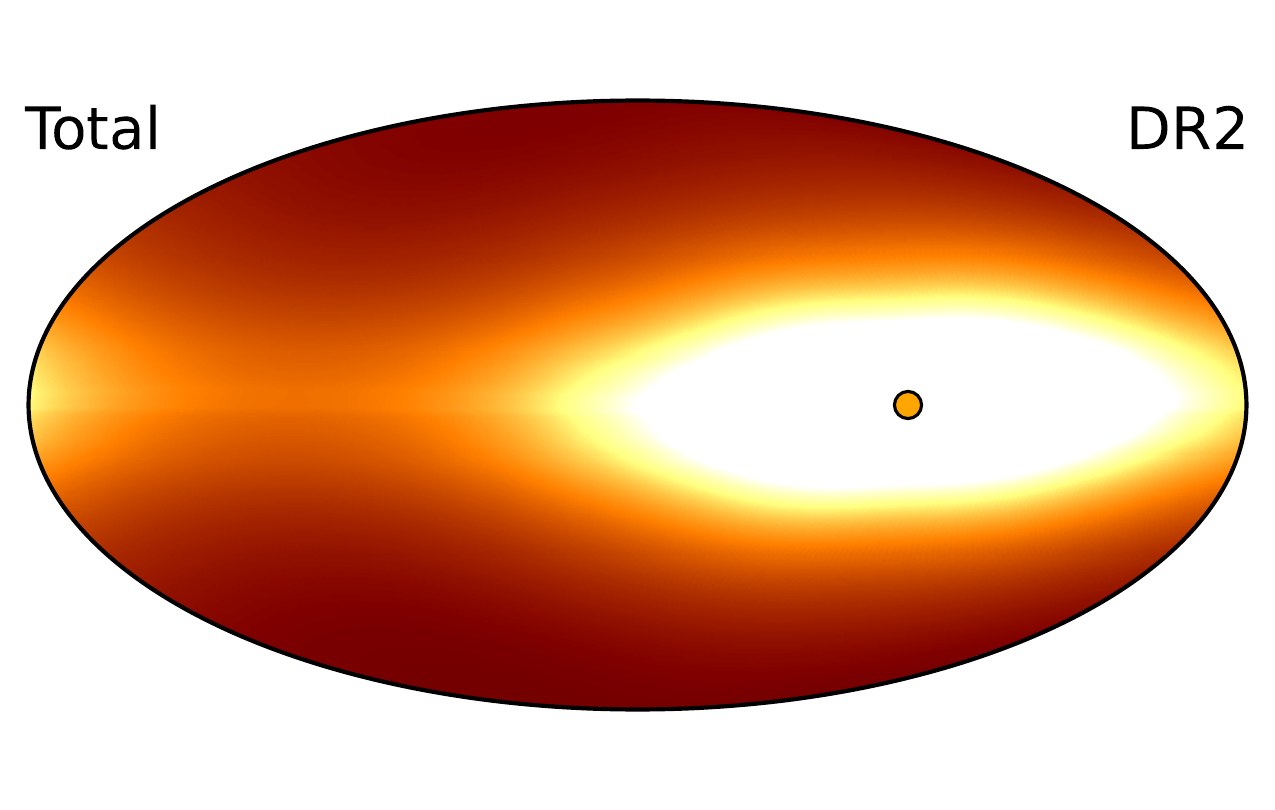}%
    \hspace{2.3pt}
    \includegraphics[width=1cm,angle=90]{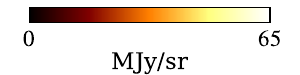}%
    }\\
    \resizebox{0.91\textwidth}{!}{%
    \includegraphics[height=1cm]{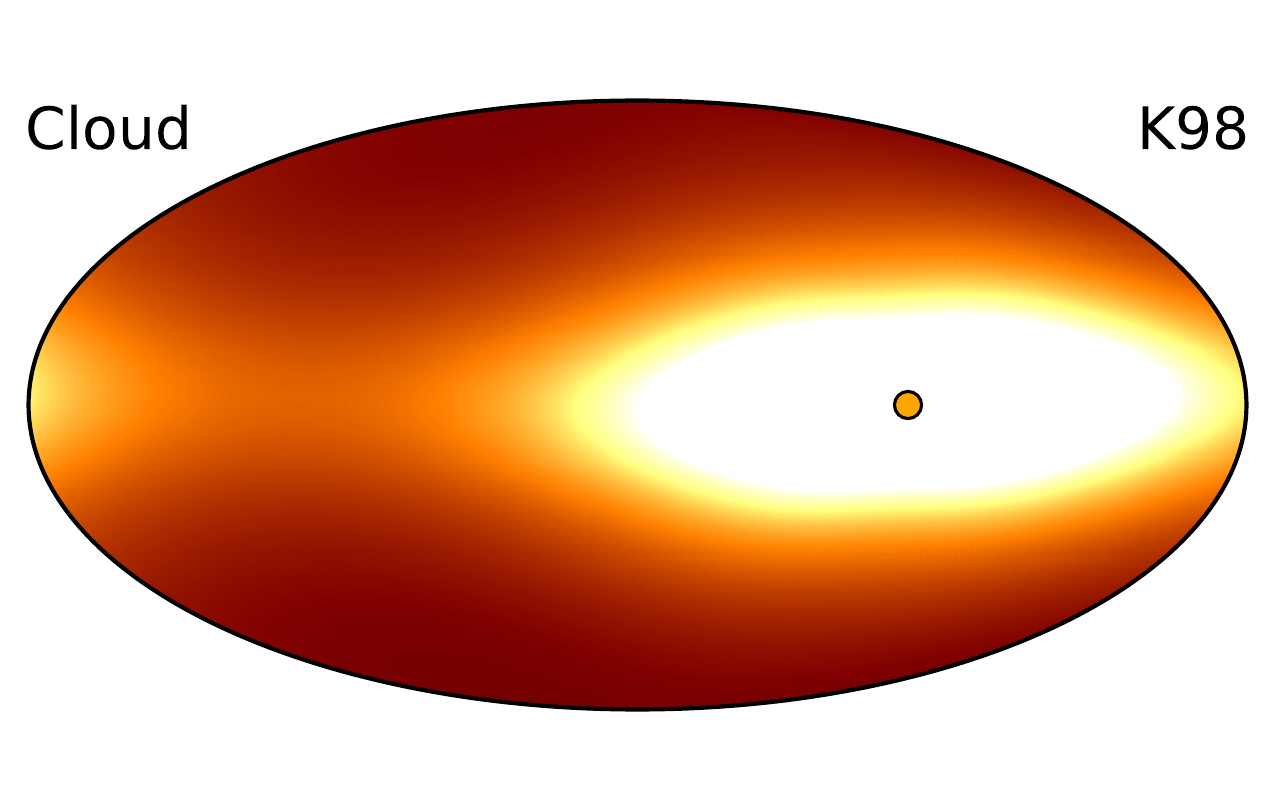}%
    \hspace{2.3pt}
    \includegraphics[height=1cm]{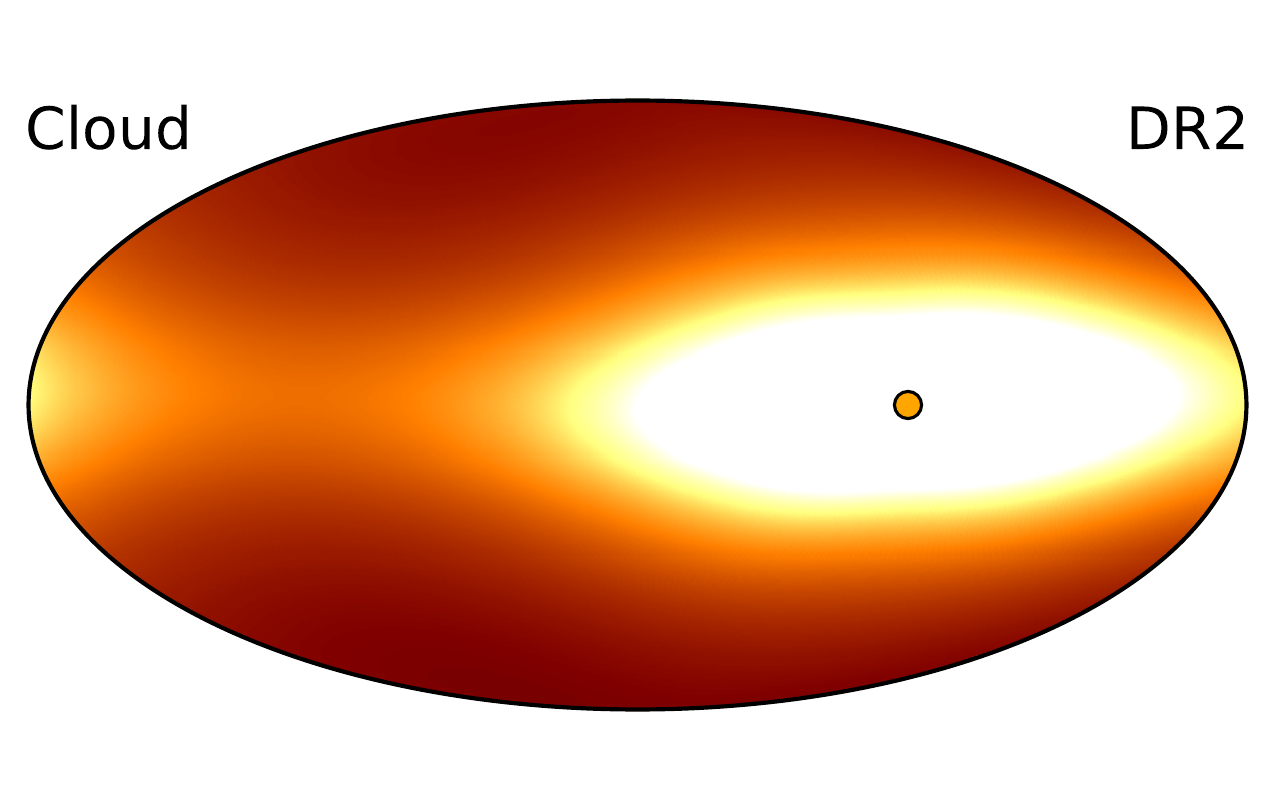}%
    \hspace{2.3pt}
    \includegraphics[width=1cm,angle=90]{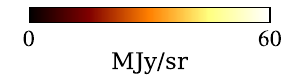}%
    }\\
    \resizebox{0.91\textwidth}{!}{%
    \includegraphics[height=1cm]{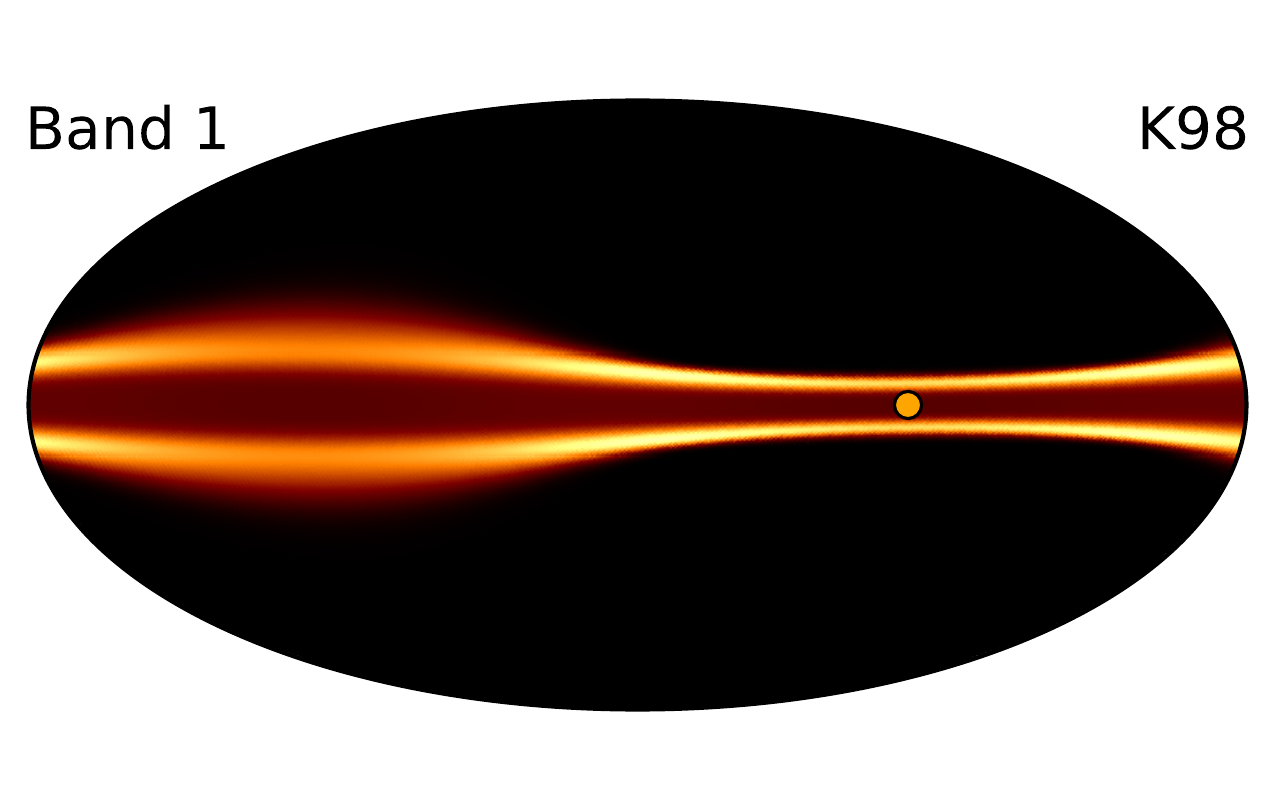}%
    \hspace{2.3pt}
    \includegraphics[height=1cm]{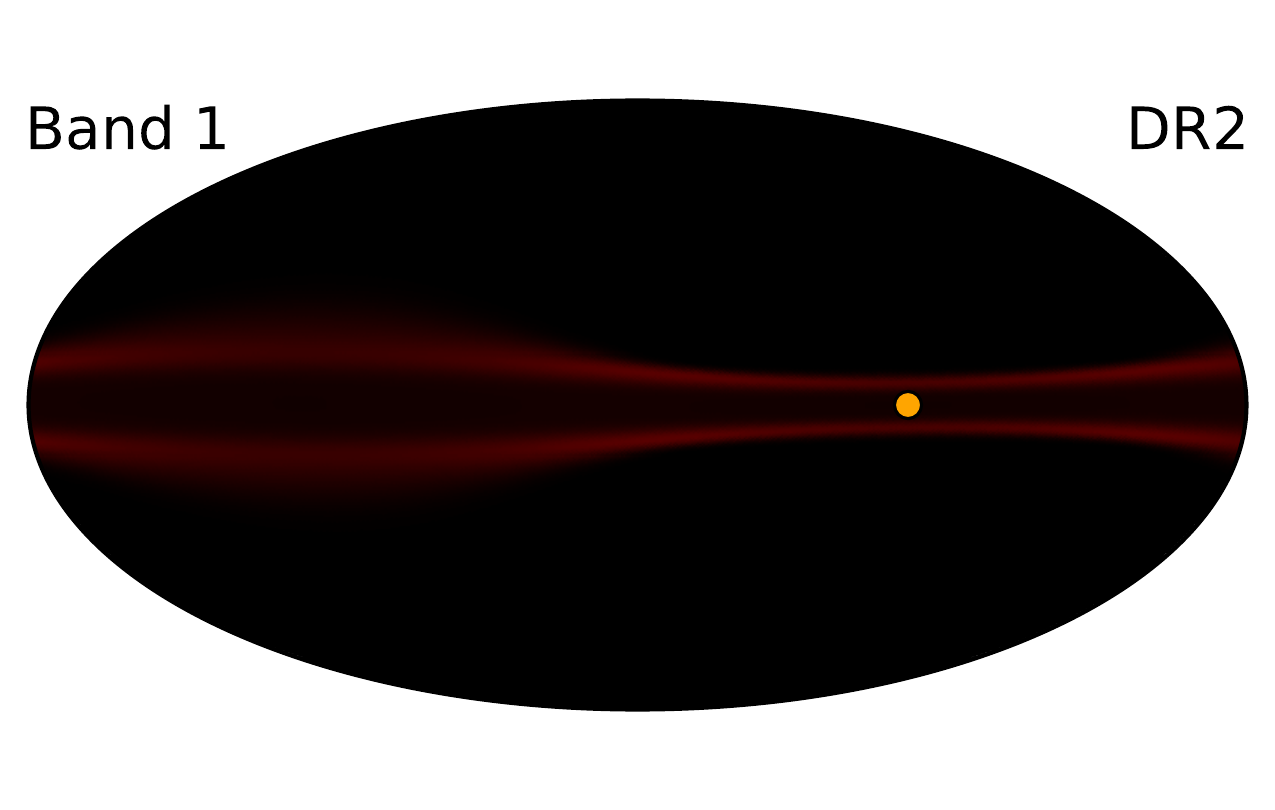}%
    \hspace{2.3pt}
    \includegraphics[width=1cm,angle=90]{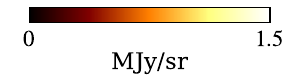}%
    }\\
    \resizebox{0.91\textwidth}{!}{%
    \includegraphics[height=1cm]{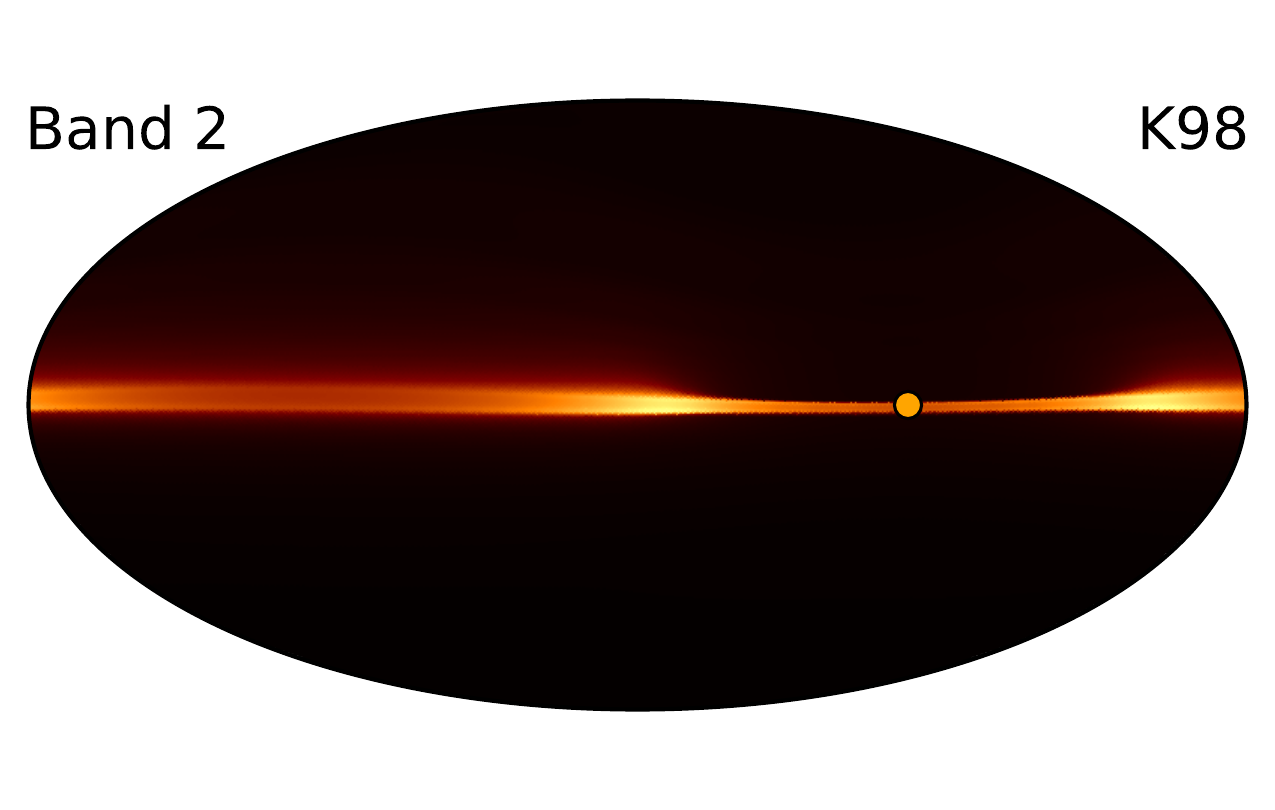}%
    \hspace{2.3pt}
    \includegraphics[height=1cm]{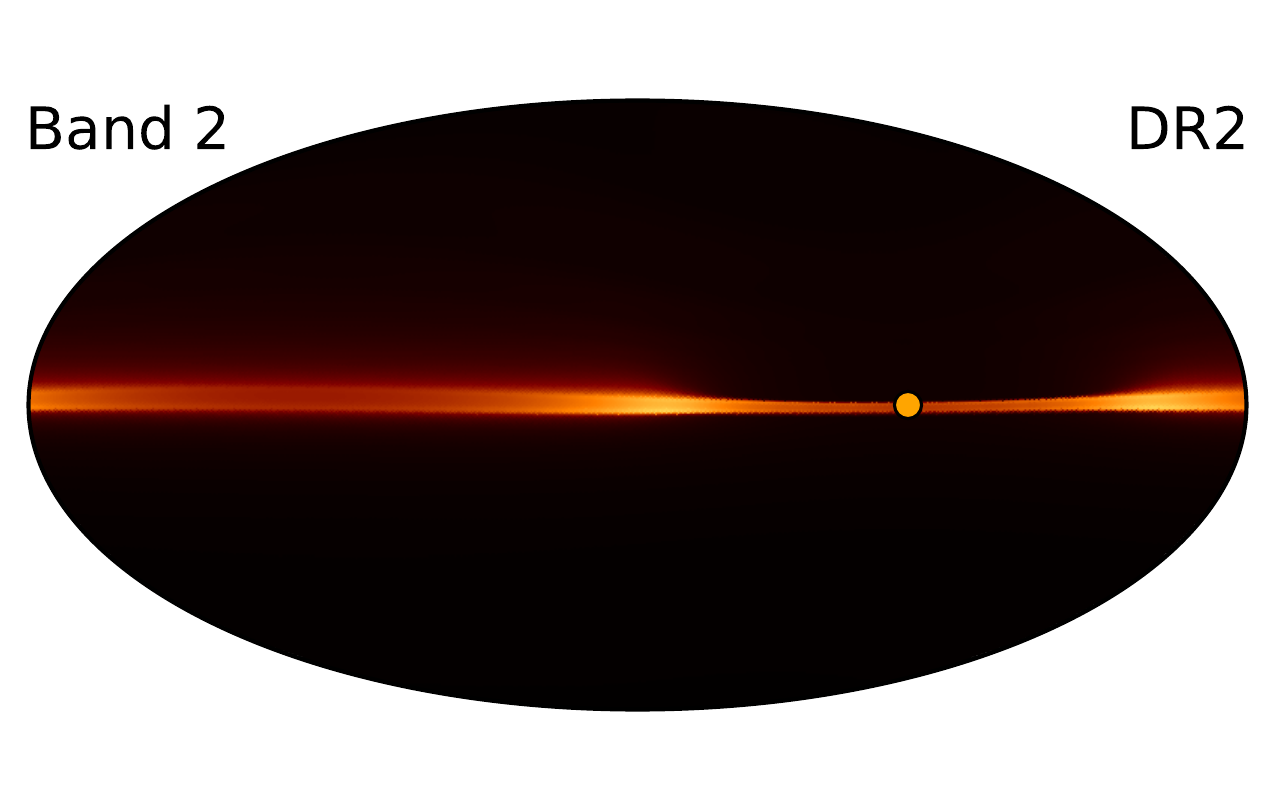}%
    \hspace{2.3pt}
    \includegraphics[width=1cm,angle=90]{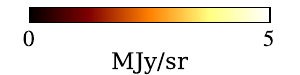}%
    }\\
    \resizebox{0.91\textwidth}{!}{%
    \includegraphics[height=1cm]{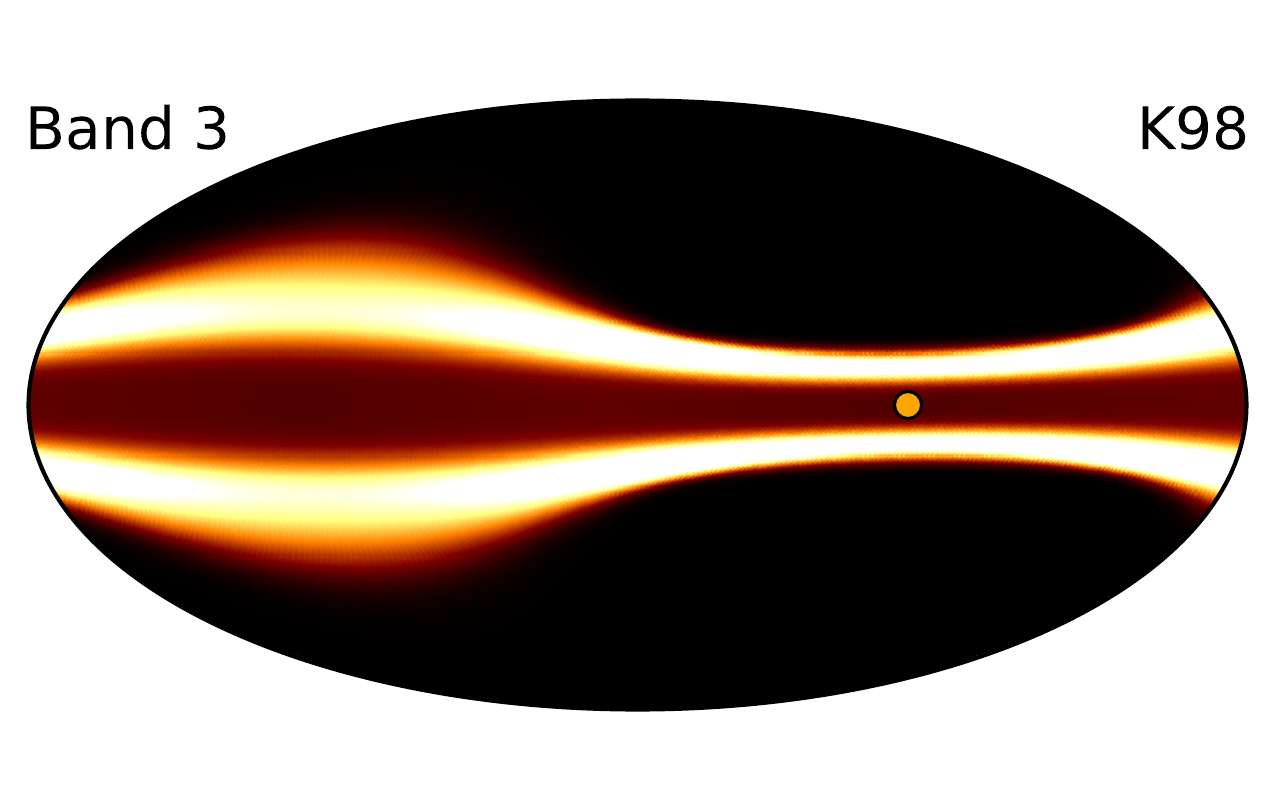}%
    \hspace{2.3pt}
    \includegraphics[height=1cm]{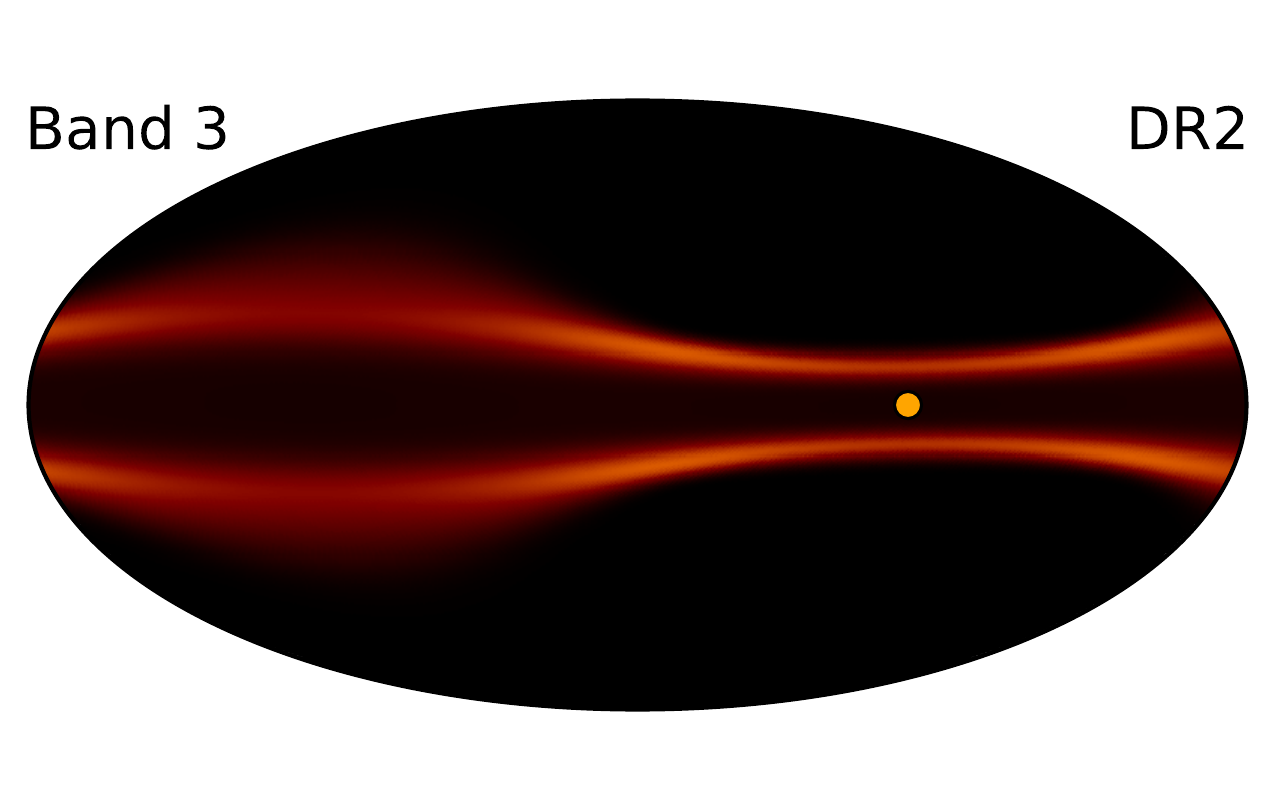}%
    \hspace{2.3pt}
    \includegraphics[width=1cm,angle=90]{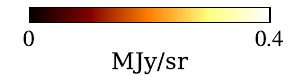}%
    }\\
    \caption{Full-sky total and component-wise ZL maps (January 1, 2024) at $25\mu$m made with ZodiPy. 
    \textit{(Left column:)} The K98 model. \textit{(Right column:)} Best-fit \Cosmoglobe\ ZL model. 
    Rows list the zodiacal components, from top to bottom, 1) total ZL emission, including the 
    circumsolar ring and Earth-trailing feauture; 2) smooth cloud; 3) dust band 1; 4) 
    dust band 2; 5) dust band 3. The maps are in Ecliptic coordinates, with the Sun marked as 
    an orange dot.}
    \label{fig:mission-averaged-inst-maps}
\end{figure*}

\begin{figure*}[hbt]
    \centering
    \resizebox{0.91\textwidth}{!}{%
    \includegraphics[height=1cm]{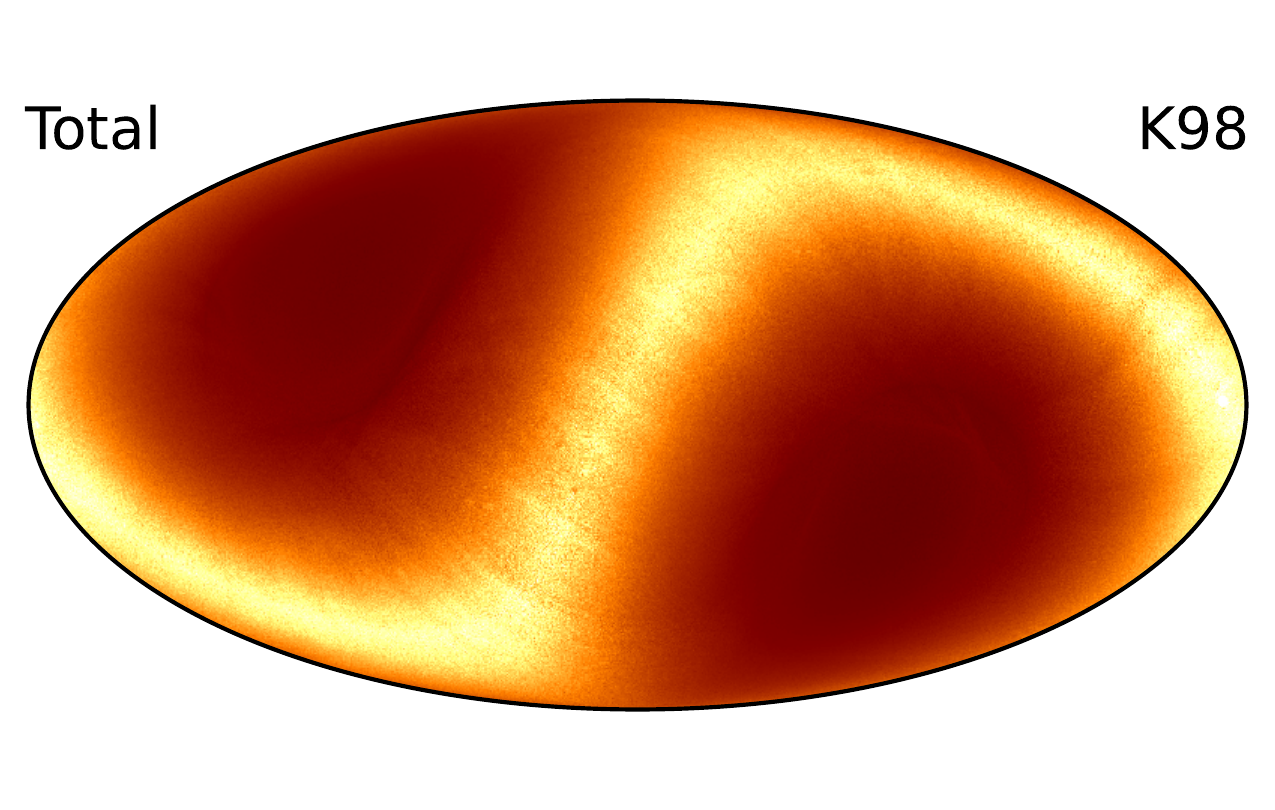}%
    \hspace{2.3pt}
    \includegraphics[height=1cm]{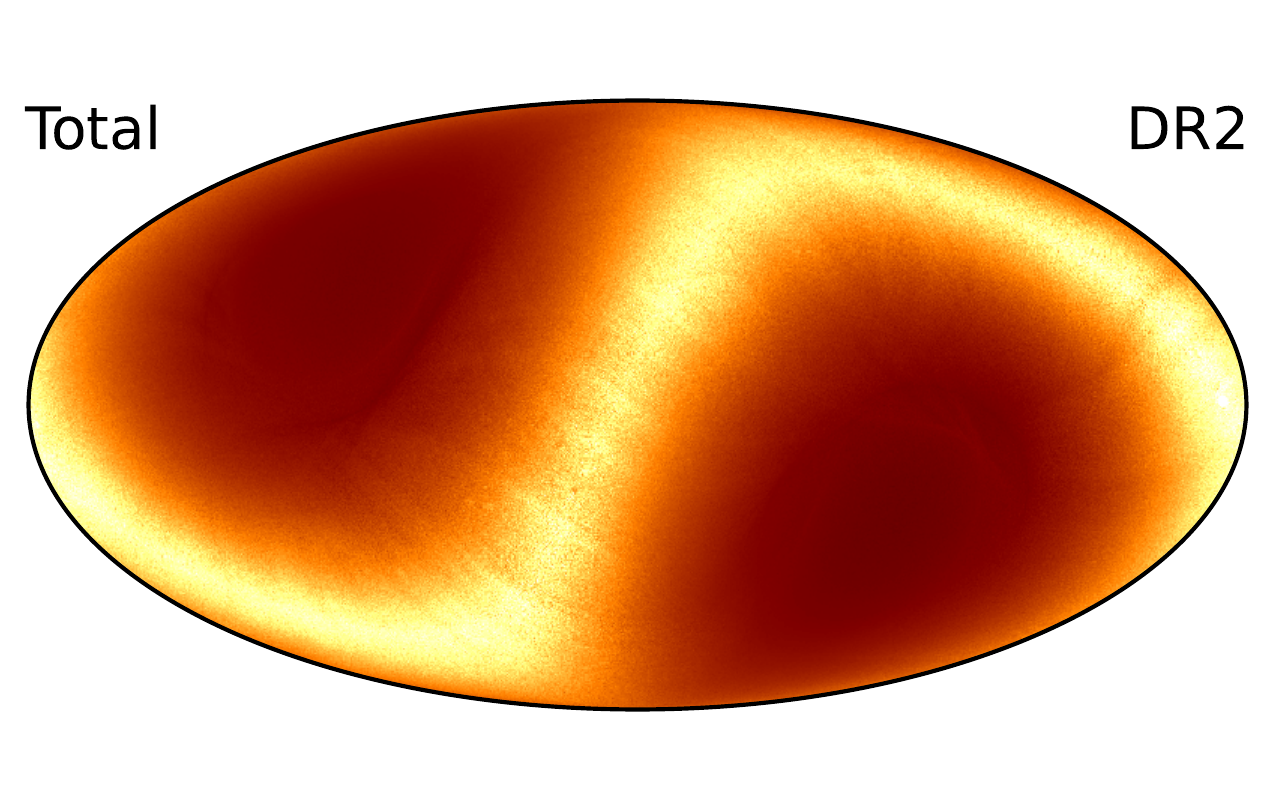}%
    \hspace{2.3pt}
    \includegraphics[width=1cm,angle=90]{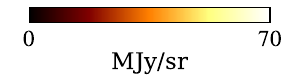}%
    }\\
    \resizebox{0.91\textwidth}{!}{%
    \includegraphics[height=1cm]{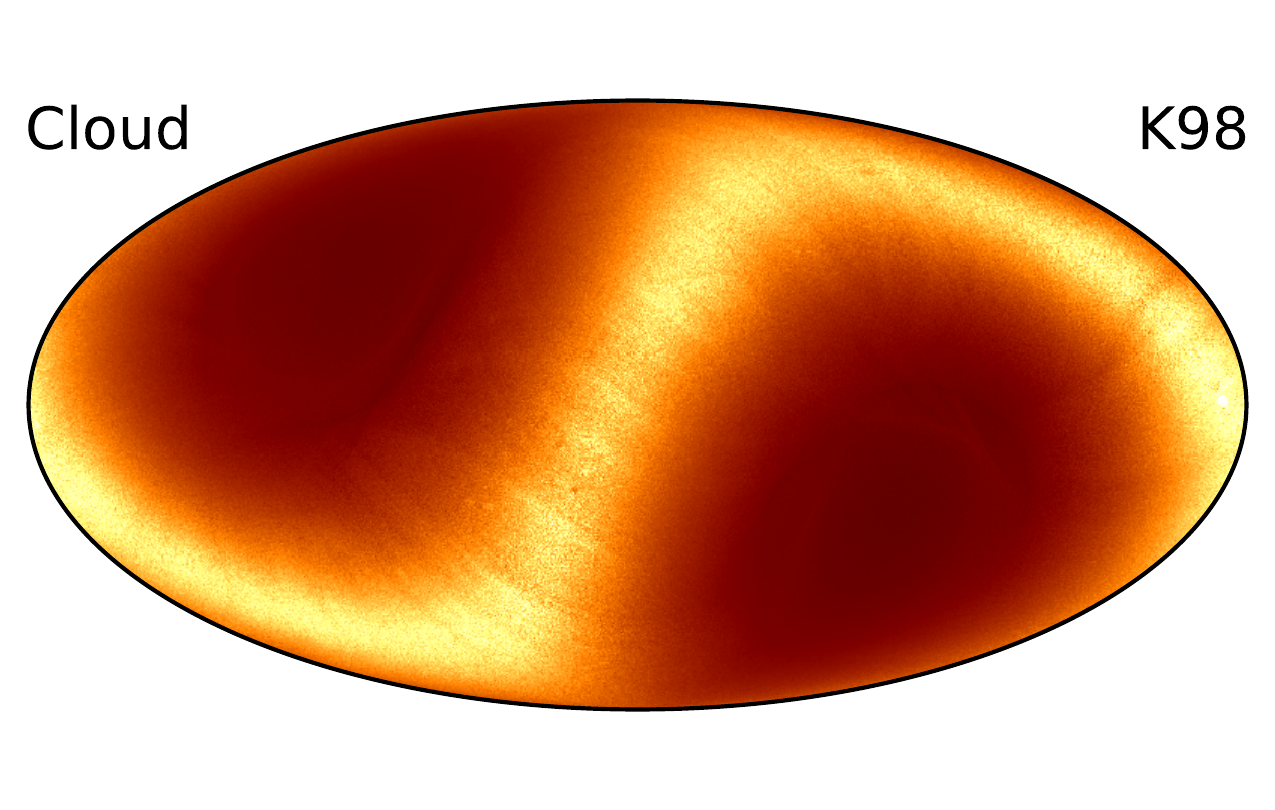}%
    \hspace{2.3pt}
    \includegraphics[height=1cm]{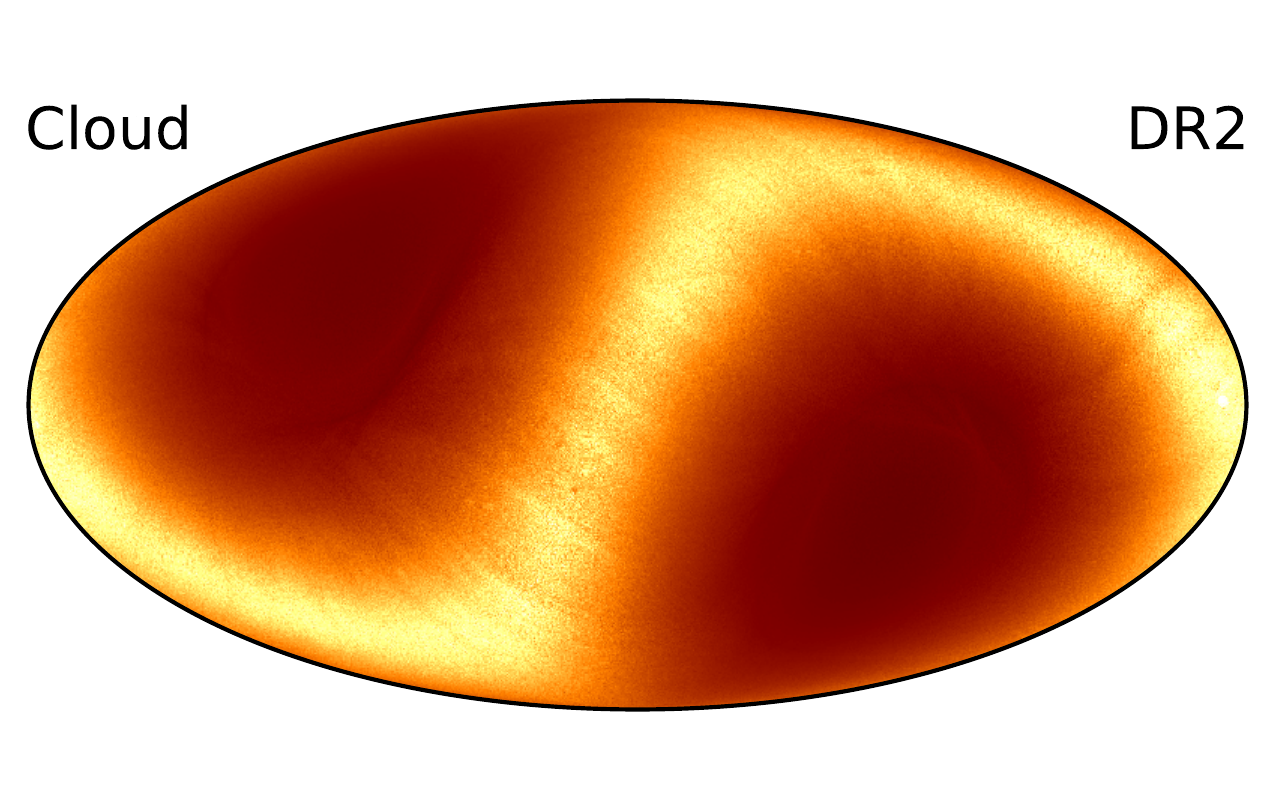}%
    \hspace{2.3pt}
    \includegraphics[width=1cm,angle=90]{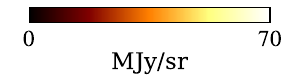}%
    }\\
    \resizebox{0.91\textwidth}{!}{%
    \includegraphics[height=1cm]{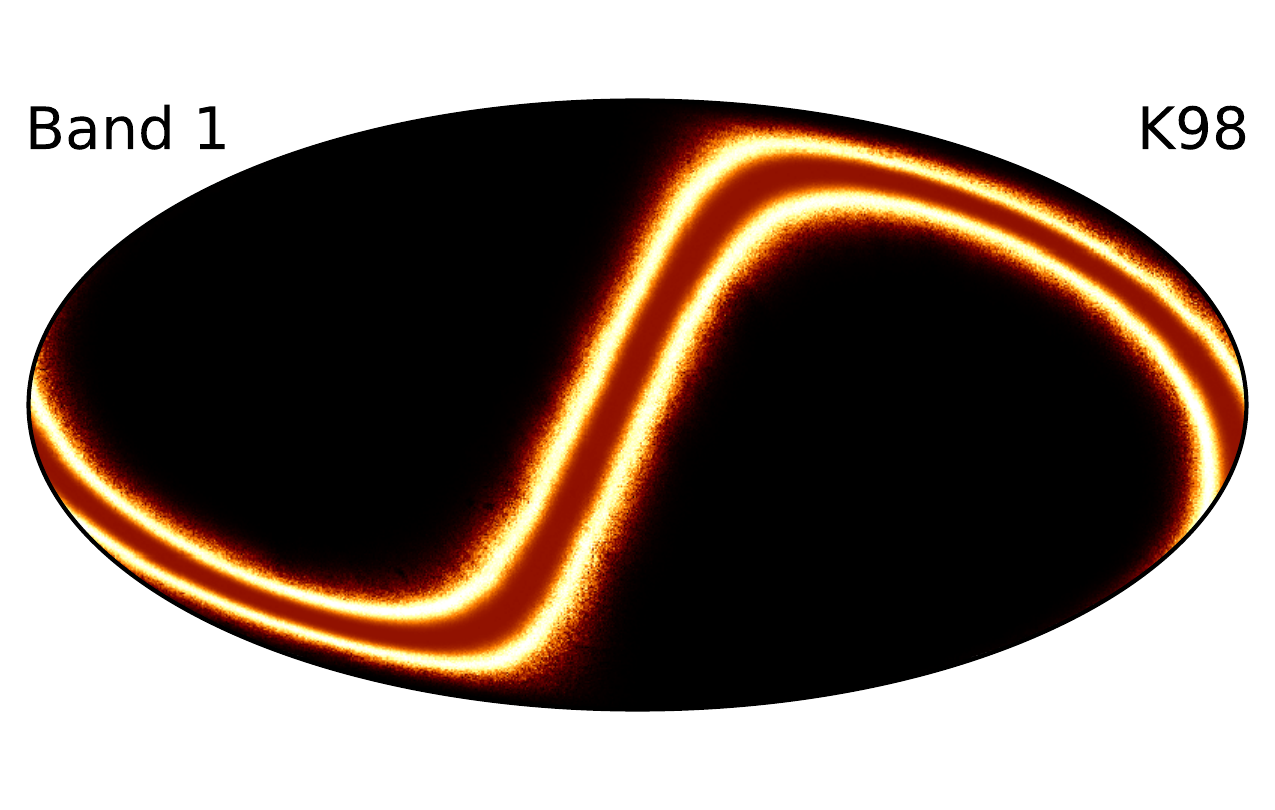}%
    \hspace{2.3pt}
    \includegraphics[height=1cm]{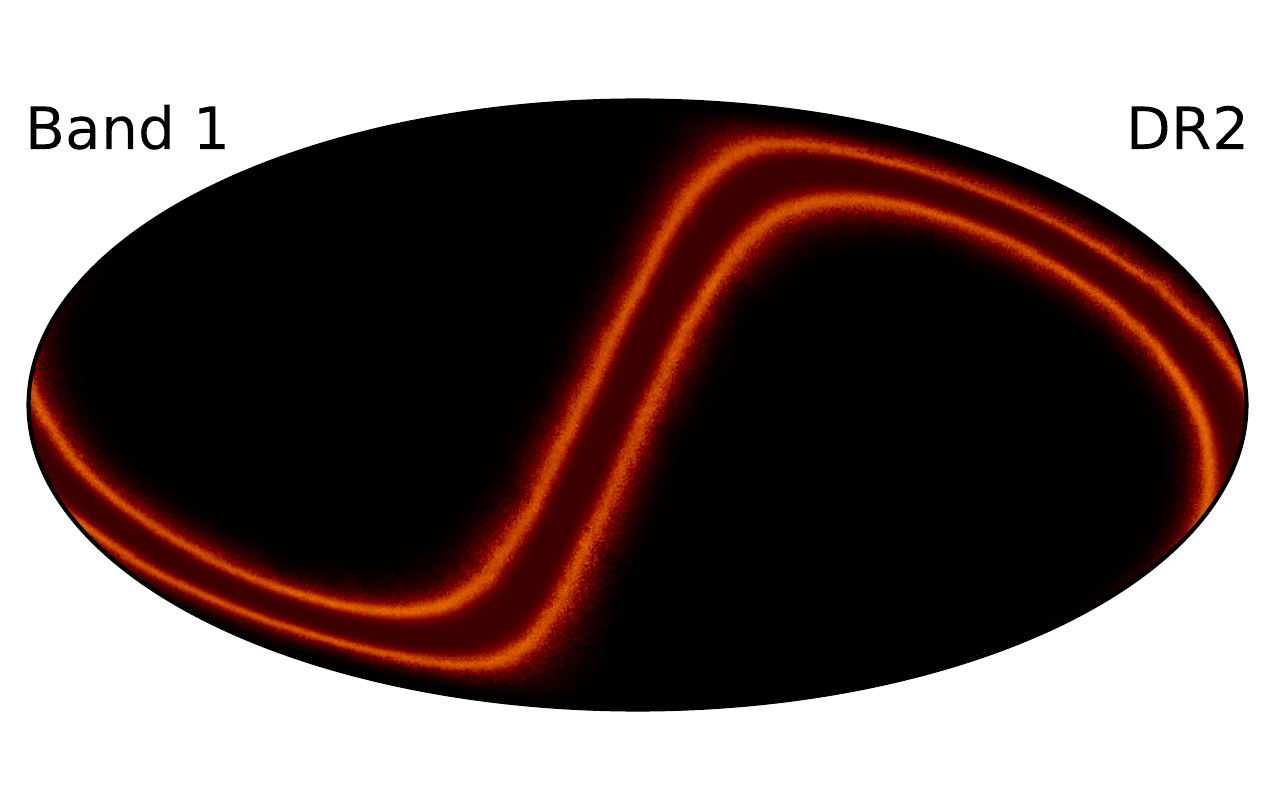}%
    \hspace{2.3pt}
    \includegraphics[width=1cm,angle=90]{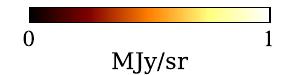}%
    }\\
    \resizebox{0.91\textwidth}{!}{%
    \includegraphics[height=1cm]{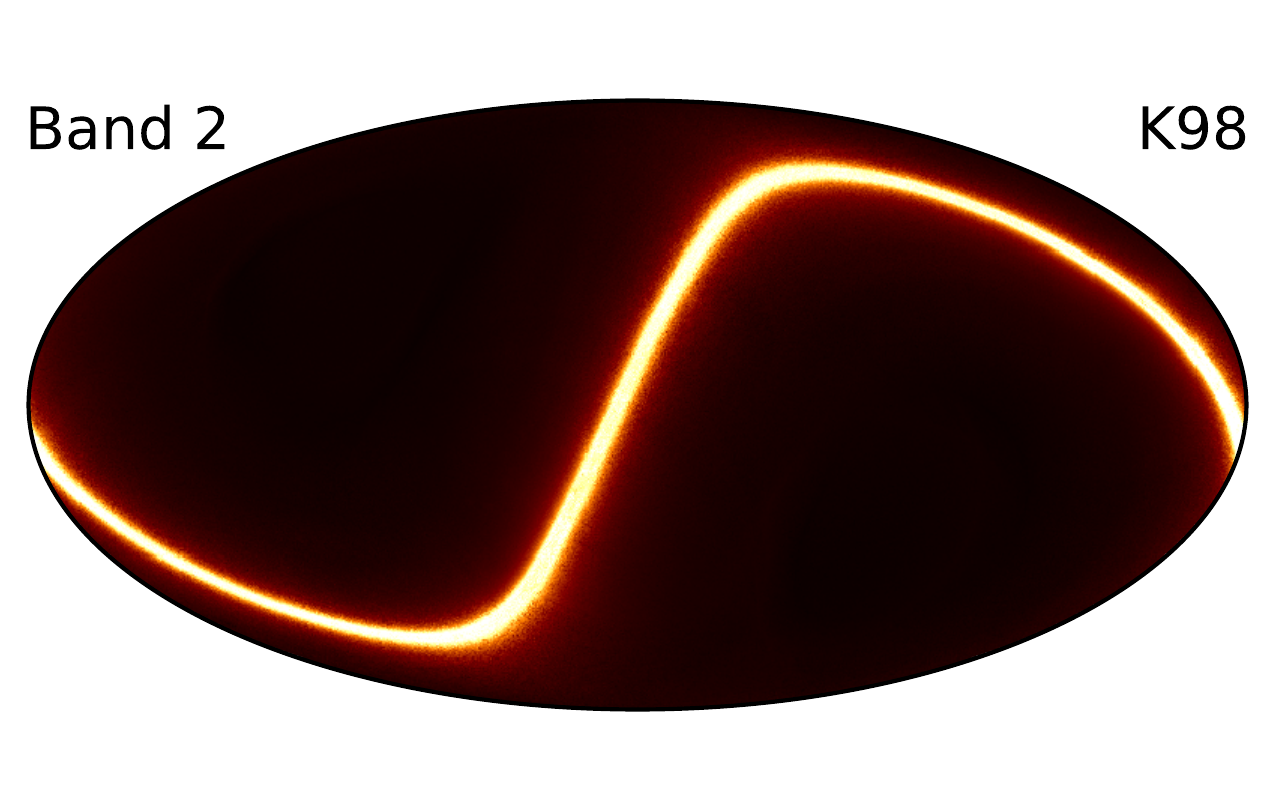}%
    \hspace{2.3pt}
    \includegraphics[height=1cm]{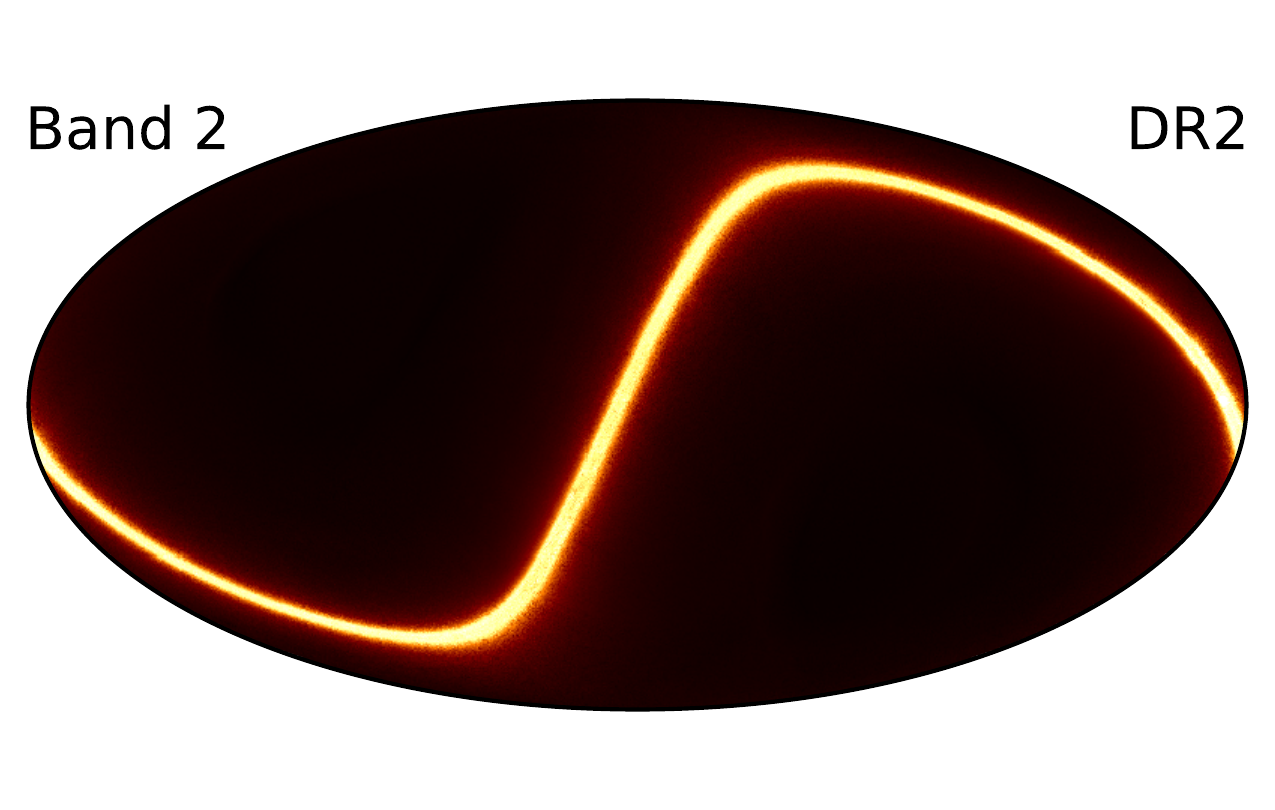}%
    \hspace{2.3pt}
    \includegraphics[width=1cm,angle=90]{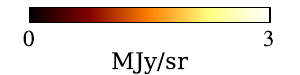}%
    }\\
    \resizebox{0.91\textwidth}{!}{%
    \includegraphics[height=1cm]{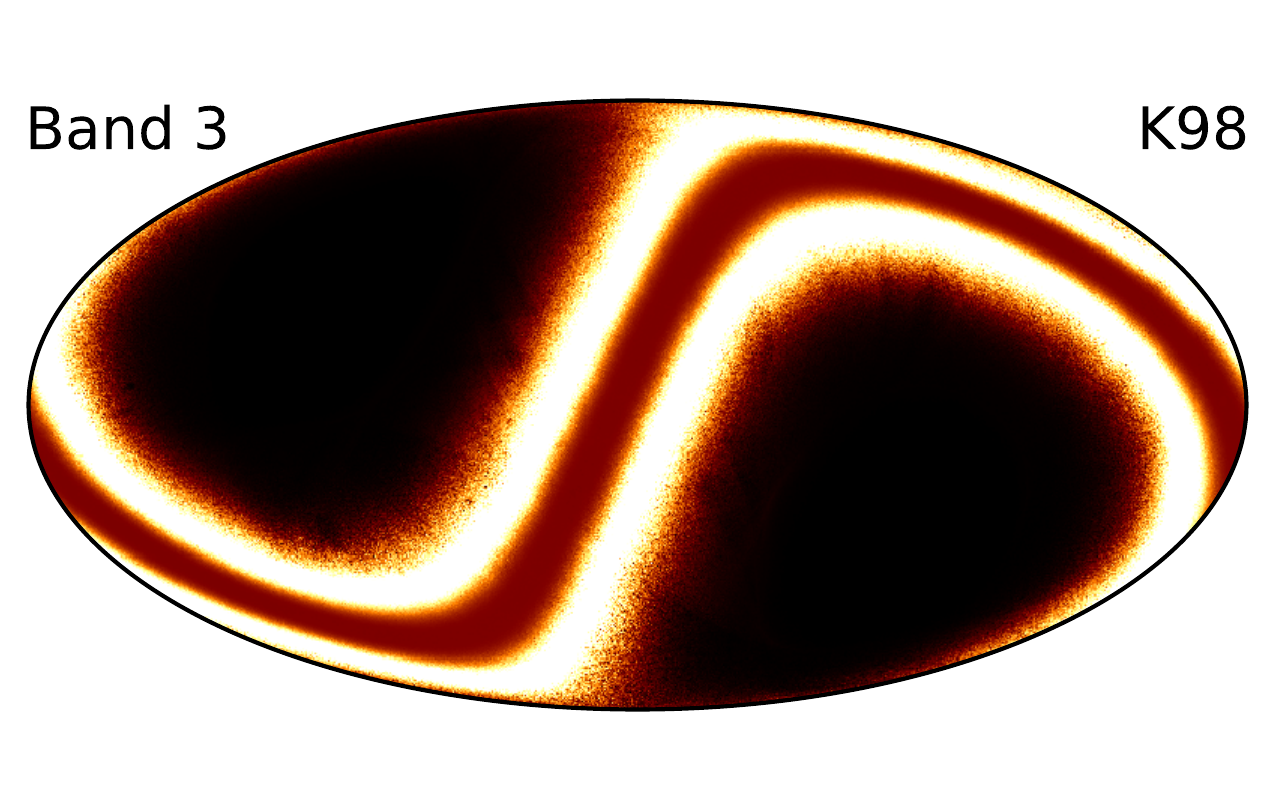}%
    \hspace{2.3pt}
    \includegraphics[height=1cm]{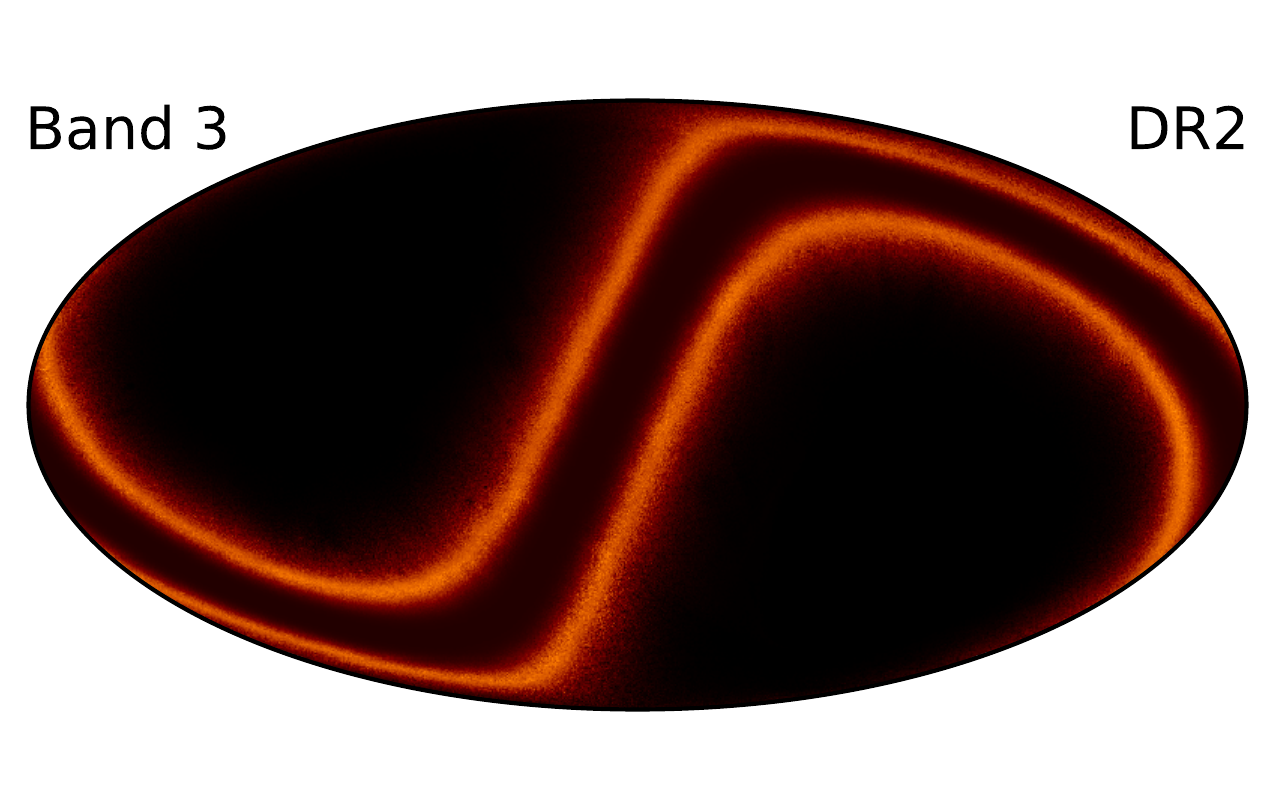}%
    \hspace{2.3pt}
    \includegraphics[width=1cm,angle=90]{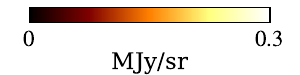}%
    }\\
    \caption{Mission-averaged component-wise ZL maps at $25\mu$m made with ZodiPy. 
    \textit{(Left column:)} The K98 model. \textit{(Right column:)} Best-fit \Cosmoglobe\ ZL model.
    Rows list the zodiacal components, from top to bottom, 1) total ZL emission, including the 
    circumsolar ring and Earth-trailing feauture; 2) smooth cloud; 3) dust band 1; 4) 
    dust band 2; 5) dust band 3. The maps are in Galactic coordinates.}
    \label{fig:mission-averaged-comp-maps}
\end{figure*}

In this Appendix we present maps visualizating of our best-fit ZL 
model. These figures can help illustrate the physical properties 
of the model and help validate how physical our models are. 
First, Fig.~\ref{fig:ipd-number-density} shows cross-sections of the 
Solar System in the xz-plane, illustrating the number density of the four 
components fitted in our zodiacal light model. Then, Fig.~\ref{fig:mission-averaged-inst-maps} 
displays instantaneous all-sky maps of both the total signal and the component 
contributions of the DR2 and K98 ZL models. Lastly, mission-averaged maps of the 
same quantities are presented in Fig.~\ref{fig:mission-averaged-comp-maps}.

\clearpage
\section{Interplanetary dust parameter atlas}
\label{sec:param-atlas}
\begin{figure*}[hbt]
    \centering
    \resizebox{0.82\textwidth}{!}{%
    \includegraphics[height=1cm]{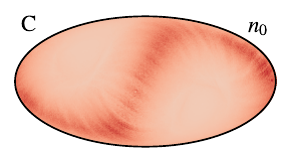}%
    \includegraphics[height=1cm]{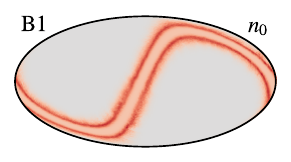}%
    \includegraphics[height=1cm]{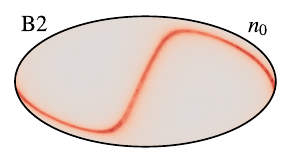}%
    \includegraphics[height=1cm]{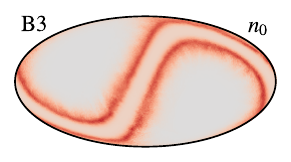}%
    }\\
    \resizebox{0.82\textwidth}{!}{%
    \includegraphics[height=1cm]{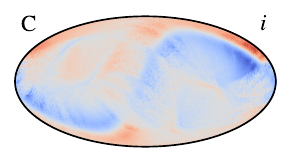}%
    \includegraphics[height=1cm]{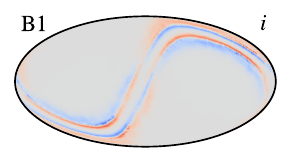}%
    \includegraphics[height=1cm]{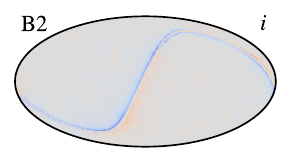}%
    \includegraphics[height=1cm]{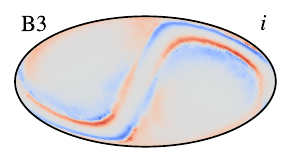}%
    }\\
    \resizebox{0.82\textwidth}{!}{%
    \includegraphics[height=1cm]{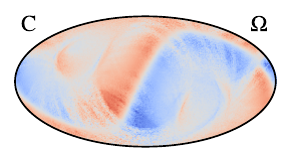}%
    \includegraphics[height=1cm]{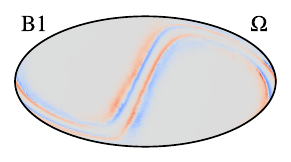}%
    \includegraphics[height=1cm]{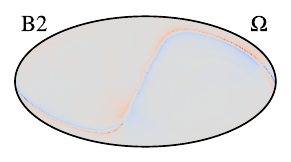}%
    \includegraphics[height=1cm]{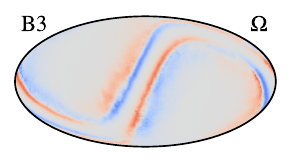}%
    }\\
    \resizebox{0.82\textwidth}{!}{%
    \includegraphics[height=1cm]{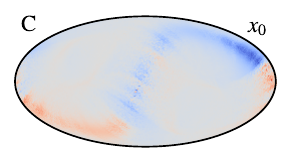}%
    \includegraphics[height=1cm]{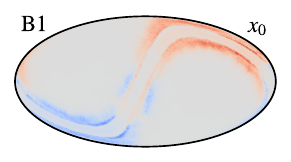}%
    \includegraphics[height=1cm]{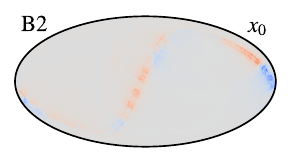}%
    \includegraphics[height=1cm]{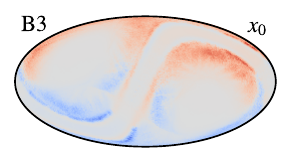}%
    }\\
    \resizebox{0.82\textwidth}{!}{%
    \includegraphics[height=1cm]{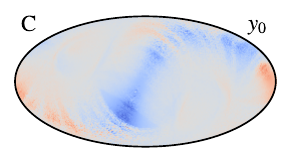}%
    \includegraphics[height=1cm]{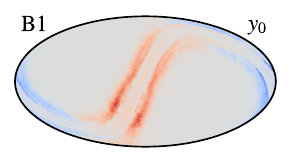}%
    \includegraphics[height=1cm]{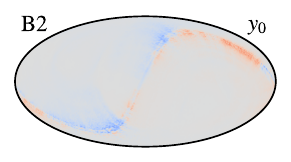}%
    \includegraphics[height=1cm]{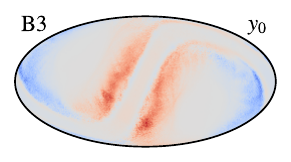}%
    }\\
    \resizebox{0.82\textwidth}{!}{%
    \includegraphics[height=1cm]{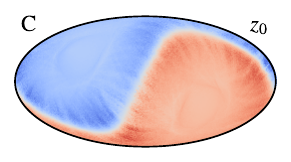}%
    \includegraphics[height=1cm]{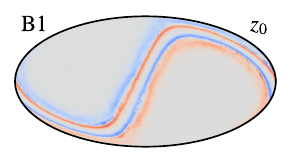}%
    \includegraphics[height=1cm]{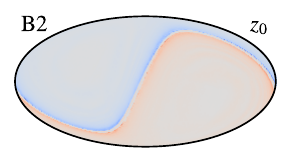}%
    \includegraphics[height=1cm]{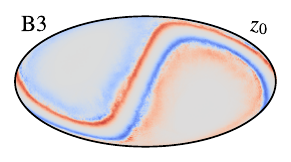}%
    }\\
    \resizebox{0.82\textwidth}{!}{%
    \includegraphics[height=1cm]{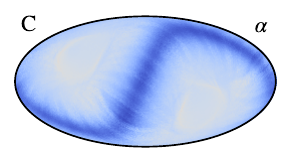}%
    \includegraphics[height=1cm]{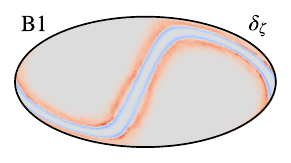}%
    \includegraphics[height=1cm]{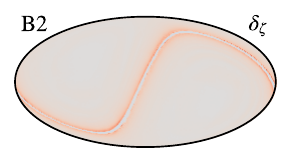}%
    \includegraphics[height=1cm]{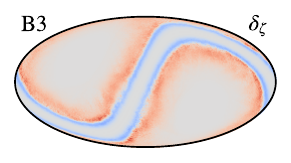}%
    }\\
    \resizebox{0.82\textwidth}{!}{%
    \includegraphics[height=1cm]{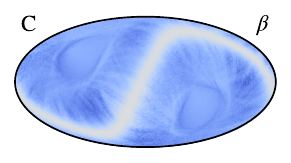}%
    \includegraphics[height=1cm]{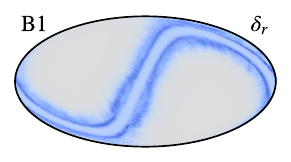}%
    \includegraphics[height=1cm]{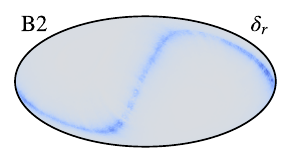}%
    \includegraphics[height=1cm]{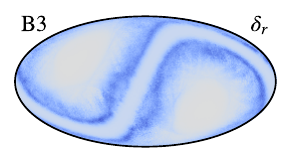}%
    }\\
    \resizebox{0.82\textwidth}{!}{%
    \includegraphics[height=1cm]{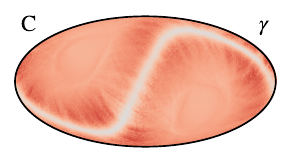}%
    \includegraphics[height=1cm]{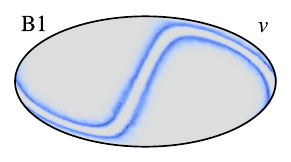}%
    \includegraphics[height=1cm]{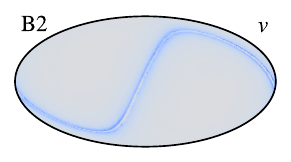}%
    \includegraphics[height=1cm]{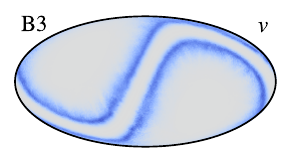}%
    }\\
    \resizebox{0.82\textwidth}{!}{%
    \includegraphics[height=1cm]{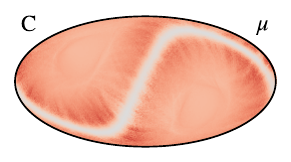}%
    \includegraphics[height=1cm]{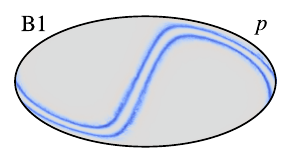}%
    \includegraphics[height=1cm]{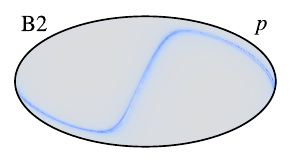}%
    \includegraphics[height=1cm]{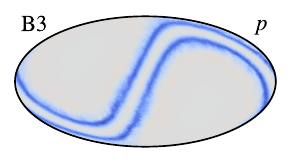}%
    }\\
   
    \caption{ZL parameter atlas showing the difference between increasing 
    and lowering each ZL model parameter by $5\%$ in the form of %
    mission-averaged ZL maps. Columns list, from left to right parameters of
    1) the smooth cloud; 2) dust band 1; 3) dust band 2; and 4) dust band 3.}
    \label{fig:atlas1}
\end{figure*}

\begin{figure*}
    \centering
    \resizebox{0.82\textwidth}{!}{%
    \includegraphics[height=1cm]{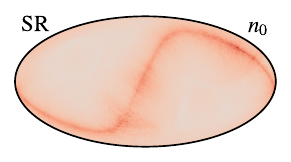}%
    \includegraphics[height=1cm]{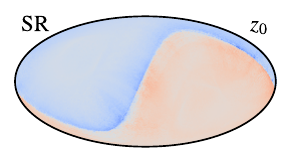}%
    \includegraphics[height=1cm]{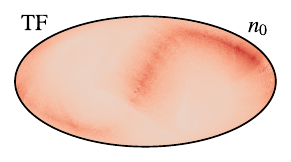}%
    \includegraphics[height=1cm]{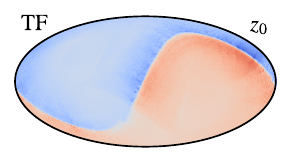}%
    }\\   
    \resizebox{0.82\textwidth}{!}{%
    \includegraphics[height=1cm]{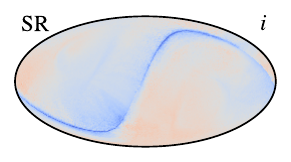}%
    \includegraphics[height=1cm]{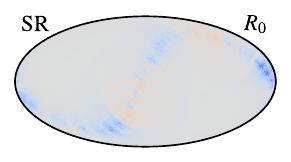}%
    \includegraphics[height=1cm]{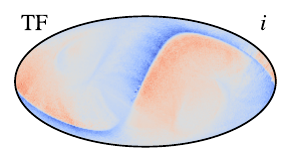}%
    \includegraphics[height=1cm]{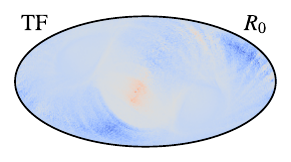}%
    }\\
    \resizebox{0.82\textwidth}{!}{%
    \includegraphics[height=1cm]{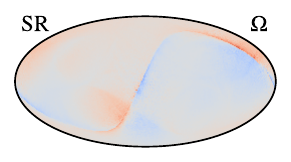}%
    \includegraphics[height=1cm]{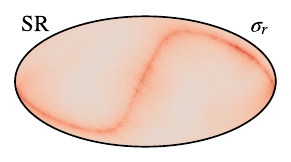}%
    \includegraphics[height=1cm]{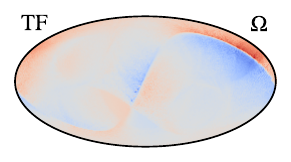}%
    \includegraphics[height=1cm]{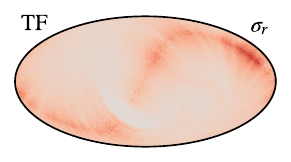}%
    }\\
    \resizebox{0.82\textwidth}{!}{%
    \includegraphics[height=1cm]{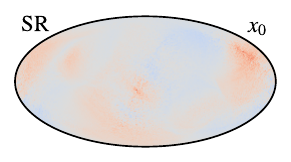}%
    \includegraphics[height=1cm]{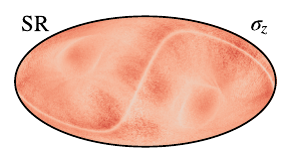}%
    \includegraphics[height=1cm]{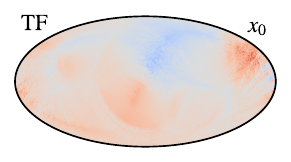}%
    \includegraphics[height=1cm]{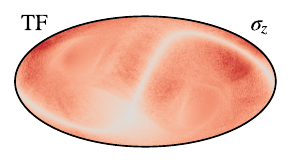}%
    }\\
    \resizebox{0.82\textwidth}{!}{%
    \includegraphics[height=1cm]{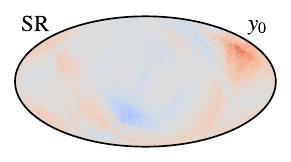}%
    \includegraphics[height=1cm]{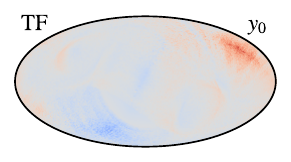}%
    \includegraphics[height=1cm]{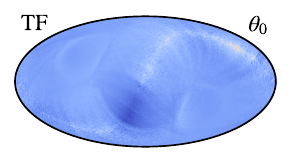}%
    \includegraphics[height=1cm]{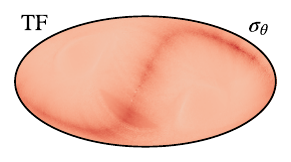}%
    }\\
    \resizebox{0.41\textwidth}{!}{%
    \includegraphics[height=1cm]{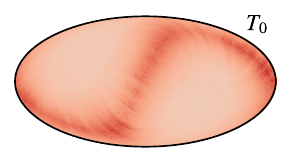}%
    \includegraphics[height=1cm]{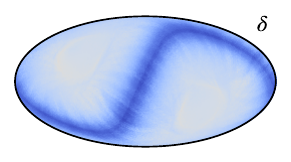}%
    }
    \caption{ZL parameter atlas showing the difference between increasing 
    and lowering each ZL model parameter by $5\%$ in the form of %
    mission-averaged ZL maps. Columns list the circumsolar ring and Earth-trailing feature components.}
    \label{fig:atlas2}
\end{figure*}

In this Appendix, we present an atlas of mission-averaged ZL parameter maps.
Each map shown in Figs.~\ref{fig:atlas1} and ~\ref{fig:atlas2} 
represents the effect of changing one ZL model parameter by $\pm 5\%$ while holding the other fixed. %


\end{document}